\documentclass[12pt, letterpaper]{article}
\usepackage{amsmath,amssymb,mathrsfs,amsthm}
\usepackage{microtype}
\usepackage[T1]{fontenc}

\usepackage{times}

\usepackage{setspace}

\usepackage{xcolor}
\definecolor{BLUE}{RGB}{0,0,100}
\usepackage[margin=1in]{geometry}

\usepackage[colorlinks=true,linkcolor=blue,citecolor=blue,urlcolor=BLUE,backref=page]{hyperref}
\let \backreforig \backref
\renewcommand*{\backref}[1]{[\backreforig{#1}]}

\usepackage{comment}
\usepackage{graphicx}
\usepackage{enumitem}

\usepackage{fnpct}
\usepackage{subcaption}
\usepackage{booktabs}
\usepackage{multirow}

\usepackage[title,titletoc]{appendix}

\newif\ifshowextra
\showextrafalse

\theoremstyle{plain}
\newtheorem{theorem}{Theorem}\newtheorem{proposition}[theorem]{Proposition}
\newtheorem{corollary}{Corollary}[theorem]

\newtheorem{lemma}{Lemma}
\newtheorem*{lemma*}{Lemma}

\newtheorem*{claim*}{Claim}

\newtheorem{assumption}{Assumption}

\theoremstyle{definition}
\newtheorem{remark}{Remark}
\newtheorem*{remark*}{Remark}
\newtheorem{example}{Example}

\usepackage{natbib}
\DeclareMathOperator{\E}{\mathbf{E}}

\DeclareMathOperator*{\argmax}{arg\,max}

\renewcommand{\d}{\mathop{}\!\mathrm{d}}

\newcommand{\lbar}{\underline}
\newcommand{\reals}{\mathbb{R}}

\DeclareSymbolFont{operators}   {OT1}{lmr} {m}{n}
\DeclareMathOperator{\MPS}{\mathrm{MPS}}

\newtheorem*{example*}{Example}
\newtheorem*{corollary*}{Corollary}

\DeclareMathOperator{\erfi}{erfi}

\newcommand{\unif}{\mathrm{Unif}}
\newcommand{\T}{[\lbar\theta,\bar\theta]}

\NewDocumentCommand{\app}{s m}{\IfBooleanTF{#1}
    {Appendix~\ref{#2}}{Appendix~\ref{#2}}}

\newcommand{\supple}{\app}

\newtheorem*{condition*}{\conditionnumber}
\providecommand{\conditionnumber}{}
\makeatletter
\newenvironment{condition}[1]
 {\renewcommand{\conditionnumber}{Condition (#1)}\begin{condition*}\protected@edef\@currentlabel{#1}}
 {\end{condition*}
 }
\makeatother

\newtheorem*{Condition*}{\conditionnumber}
\providecommand{\conditionnumber}{}


\makeatletter
\newcommand{\nameditem}[2]{\item[(#1)]\def\@currentlabel{#1}\label{#2}}
\makeatother

\title{
    Incentivizing Agents through Ratings
}
\author{
    Peiran Xiao\thanks{Department of Economics, University of Southern California. Email: \href{mailto:peiran.xiao@usc.edu}{{peiran.xiao@usc.edu}}.
    \smallskip
    \\   
    I am indebted to Bart Lipman, Juan Ortner, Krishna Dasaratha, and Chiara Margaria for their invaluable help and support throughout this project. 
    I thank Nageeb Ali, Ian Ball, Nina Bobkova, Tilman Börgers, Tan Gan, Kira Goldner, Marina Halac, Xiao Lin, Yijun Liu, Albert Ma, Teddy Mekonnen, Benny Moldovanu, Dilip Mookherjee, Andy Newman, Axel Niemeyer, Jawwad Noor, Marco Ottaviani, Harry Pei, Alex Smolin, Teck Yong Tan, Mark Whitmeyer, Kun Zhang, Dihan Zou, and participants at various conferences and seminars for their helpful comments. 
    }
}

\date{\today  \bigskip \medskip  \\
\href{https://files.peiranxiao.com/InfoDesign/ratings.pdf}{\normalsize [\color{blue}{Link to the latest version}]}
}

\begin{document}
\maketitle

\onehalfspacing

\begin{abstract}
I study the optimal design of ratings to motivate an agent's investment in
quality when transfers are unavailable. The principal designs a rating scheme that maps quality to a (possibly stochastic) signal.
The agent privately knows his ability and chooses a quality level.
A competitive market then offers the agent a wage equal to his expected quality given the signal.
When restricted to deterministic ratings, lower censorship is optimal if the ability density is single-peaked, and pass/fail tests are optimal as well if the density is increasing.
When randomization is allowed, lower censorship remains optimal if the ability density is log-concave or increasing, and pass/fail tests remain optimal if the density is increasing.
By contrast, every optimal rating scheme involves randomization if the density
is decreasing and sufficiently log-convex---roughly speaking, if intermediate ability is
scarce relative to high and low ability.
\end{abstract}


\newpage

\section{Introduction}
In many economic applications, a principal seeks to motivate agents' investment in quality, but monetary transfers between them are infeasible.
In these situations, the principal can instead incentivize agents through a rating scheme (or disclosure policy) that reveals information about their endogenous quality to the market.
When the market infers agents' quality from this information and rewards them
accordingly, ratings can provide reputational incentives.

For example, consider a school whose students invest in their quality (human capital). To encourage such investment, the school designs a grading rule that discloses information about students' endogenous quality to the job market.
Similarly, a regulatory certifier concerned with consumer welfare can motivate firms to invest in product quality through quality certification.\footnote{
    Schools may want to incentivize quality investment to improve placement outcomes, foster human capital formation, or maximize tuition revenue (\citealp{Zubrickas2015, OnuchicRay2023}).
    Regulatory or NGO certifiers care about product quality because of consumer welfare or quality spillovers (\citealp{Zapechelnyuk2020,BizzottoHarstad2023,Vatter2025}), as in restaurant hygiene ratings, Medicare Star Ratings, and certifications for energy efficiency or product safety.
}
In these examples, the market pays the agent the expected value of his quality conditional on the rating result.
By contrast, transfers between the principal and agent contingent on the agent's quality or rating result are often infeasible in practice.

A variety of rating schemes are used in these environments to motivate agents.
A common scheme is the \emph{pass/fail} test, as in qualifying exams, job-training certificates, and product-safety certifications (e.g., UL certification).
Another prevalent scheme is \emph{lower censorship}, which reveals quality if and only if it reaches a minimum standard.
For example, many schools release precise scores above a passing grade, and certifiers
use quality assurance, which effectively keeps products below the standard off the
market \citep{Zapechelnyuk2020}.

Ratings can also be \emph{stochastic}.
In practice, certification may involve randomization in sampling, inspection, or disclosure.
For example, quality assurance can be combined with randomized disclosure: products below the minimum standard remain off the market, while detailed assessments of products above the standard are published randomly.
An exam may also use a random passing grade whose realization is unknown to students ex ante.
Such randomness can be a deliberate incentive instrument rather than arbitrary noise.
However, because strengthening incentives for some agents may weaken incentives for others, it is unclear whether randomization is valuable overall.
In particular, can randomization improve upon pass/fail tests or lower censorship?

In this paper, I study the optimal design of rating schemes to motivate agent investment in quality when transfers are unavailable and ratings are the only instrument.
The principal designs a rating scheme (à la Blackwell) that stochastically maps the agent's quality to a signal. The agent privately knows his ability, which determines his cost of investment, and chooses a quality level. A competitive market observes the realized signal and offers the agent a wage equal to its expectation of his quality given the signal.

Full revelation of quality is a natural benchmark, since every marginal investment in quality is observed and rewarded by the market. 
However, a minimum standard can provide stronger incentives for some agents: because lower-ability agents may prefer not to meet the standard, higher-ability agents are willing to invest more to distinguish themselves from those who fall below. 
Moreover, randomization can reallocate incentives across agents and thus provide some agents with stronger incentives at the expense of others.

To study the optimal rating scheme, I study an equivalent problem of designing a direct mechanism consisting of a quality schedule and an {interim} wage schedule---that is, the agent's interim expectation of the market wage.
Unlike in standard principal-agent models, the agent's wage is offered by the market and equals his expected quality given the signal generated by the rating.
Consequently, the mechanism design problem is subject to a feasibility condition that the interim wage must be induced by some rating scheme, which turns out to be equivalent to the wage being majorized by the quality schedule in quantile space.

I first consider optimal deterministic ratings as a benchmark.
A deterministic rating scheme either fully reveals quality or pools some qualities to the same signal.
In the latter case, among the qualities that are pooled to the same signal, only the lowest one will be chosen by the agent in equilibrium.\footnote{
This argument hinges on the assumption that quality can be chosen deterministically 
and no longer holds if the agent's investment determines quality stochastically.
}
Thus, the market perfectly infers the agent's quality from the signal in equilibrium, so the interim wage always equals the quality, and the feasibility condition is always satisfied.

When the principal maximizes expected quality, lower censorship is optimal among deterministic ratings if the ability density is single-peaked \citep{Zapechelnyuk2020}. 
I further show that when the density is increasing, pass/fail tests are also optimal, as they implement the same bang-bang quality schedule as lower censorship, under which every type either fails or just meets the standard.
When the density is decreasing, lower censorship is optimal with a minimum standard that every
type reaches in equilibrium.
Intuitively, a minimum standard trades off exclusion of low types, who would rather
fail, against stronger incentives for high types, who invest more to distinguish
themselves from those who fail. When high types are relatively abundant, it is optimal to set a high standard that requires even the highest type to invest more than under full revelation, thereby raising the quality of high types at the cost of excluding low types. By contrast, when low types are relatively abundant, excluding them becomes too costly, so the optimal minimum standard is achievable by the lowest type in equilibrium.

Then, I allow for stochastic rating schemes and provide sufficient conditions under which deterministic and stochastic ratings are optimal, respectively.
When stochastic ratings are allowed, the interim wage can differ from quality, and the feasibility condition becomes nontrivial.
For quality maximization, I show that randomization has no value for many common distributions: lower censorship is optimal if the ability density is log-concave, and both pass/fail tests and lower censorship are optimal if the density is increasing. 
\footnote{
A weaker sufficient condition for the former is that the density is single-peaked with increasing elasticity on its decreasing region, which is satisfied by many common unimodal distributions.
A weaker sufficient condition for the latter is that there exists a type whose density is above (below) the average density between it and every lower (higher) type.
}
These results may help explain the prevalence of simple deterministic rating schemes.

By contrast, every optimal rating scheme involves randomization when the density is decreasing and sufficiently log-convex.
Roughly, this condition says that intermediate ability is scarce relative to high and low ability.
To see the intuition, consider a variant of lower censorship that fails those who do not meet the standard; among those who do, it reveals quality with a quality-dependent probability and otherwise gives a ``pass.''
In practice, this can be interpreted as a certification that gives every compliant product an approval seal and publishes a detailed assessment with some probability.
This scheme strengthens incentives at both ends of the ability distribution.
On the extensive margin, the prospect of being pooled with higher types via the pass signal raises the payoff to just meeting the threshold, so the principal can impose a higher threshold without excluding more low types.
On the intensive margin, a U-shaped revelation probability can strengthen incentives for high and low types through carrot and stick: a low type invests in part to be pooled, and a high type invests in part to avoid being pooled.
Consequently, randomization raises quality among low and high types at the expense of intermediate types, so it increases expected quality when intermediate types are relatively scarce.

I also extend the analysis from quality maximization to a class of
type-dependent principal preferences, including weighted sums of expected
quality and agents' payoffs. For deterministic ratings, the preceding results
carry over once the ability density is replaced by a virtual density that
incorporates these preferences. For stochastic ratings, the results are more nuanced because
the majorization constraint continues to depend on the actual ability density.

The main contribution is to provide conditions under which (i) simple ratings are optimal among deterministic ratings, (ii) they remain optimal once stochastic ratings are allowed, and (iii) deterministic ratings are strictly dominated by stochastic ratings.
These results show how the ability distribution shapes the value of
using information to redistribute investment incentives.
I also provide a sufficiency theorem for certifying candidate
stochastic mechanisms and apply it to an explicitly constructed optimal
stochastic rating scheme.
These results extend beyond the deterministic benchmarks analyzed by
\citet{Zapechelnyuk2020,RodinaFarragut2020}.

The analysis combines the interim approach to information design
\citep{SaeediShourideh2020,DovalSmolin2024}
with optimal control methods.
The interim approach reduces stochastic rating design to a mechanism
design problem with a majorization constraint.
To analyze the resulting problem, I develop sufficiency theorems for
optimal control problems with monotonicity, equality, and state constraints,
allowing for jump discontinuities in the monotone variable.
\footnote{The approach addresses three technical difficulties.
First, because transfers are unavailable, the usual Myersonian
reduction to a virtual-surplus maximization problem
(see also \citealp{Toikka2011}) does not apply directly.
Second, classical optimal control formulations used in mechanism
design (e.g., \citealp{GuesnerieLaffont1984}) do not apply, since they assume
absolute continuity of the monotone variable and use its derivative as a control.
Third, the majorization constraint is a pure state constraint; when it binds
in deterministic rating design, the envelope condition becomes a pure state
equality constraint.
}
The approach is related to the Lagrangian methods developed in the delegation literature by
\citet{AmadorBagwell2013,AmadorBagwell2022}, but it incorporates a majorization
constraint arising from the feasibility condition and provides sufficient
conditions for the optimality of possibly discontinuous allocations.
\footnote{In contemporaneous work, \citet{AmadorBagwellCarpizo2026}
extend the Lagrangian approach to provide sufficient conditions
for the optimality of possibly discontinuous allocations in
delegation without participation constraints.
}
Specifically, the feasibility condition links the quality and wage
schedules through a majorization constraint, which enters the
optimal-control formulation as a pure state constraint.

Methodologically, the paper contributes to the literature on optimization under
majorization constraints \citep{KleinerMoldovanuStrack2021} by studying a
setting in which the majorization set is itself endogenous.
The designer jointly chooses two functions---the quality schedule and the
interim wage schedule---subject to incentive constraints and a majorization
constraint requiring that the former majorize the latter in quantile space.
As a result, even when the objective is linear, the optimal interim wage
schedule need not be an extreme point of the majorization set induced
by the optimal quality schedule.

\paragraph*{Literature Review.}
This paper contributes to the literature on optimal rating design to incentivize agents when they are rewarded according to market beliefs on their endogenous quality (\citet{AlbanoLizzeri2001,SaeediShourideh2020,SaeediShourideh2026,Zapechelnyuk2020,RodinaFarragut2020,BoleslavskyKim2021,Vatter2025,CamboniCarnehl2025}).
The paper most closely related to mine is \citet{Zapechelnyuk2020}, who studies optimal \emph{deterministic} rating design when the principal essentially maximizes expected quality.
\footnote{
    Under a different specification of the cost function, \citet{RodinaFarragut2020} characterize the properties of effort-maximizing deterministic rating when the ability distribution is ``sufficiently'' concave, convex, or single-peaked.
    They also provide an example in which randomization is valuable.
}
Recent work has also studied optimal stochastic rating design in pure
moral-hazard environments without private types:  
\citet{BoleslavskyKim2021} assume that effort shifts the distribution of quality,
while \citet{SaeediShourideh2026} assume that effort generates a nondegenerate random quality.
\footnote{
  An earlier version \citep{SaeediShourideh2020} studies Pareto-optimal rating schemes in
a deterministic quality-choice setting with privately informed
sellers. The designer maximizes a weighted average of sellers'
payoffs, with type-dependent Pareto weights.
}
In contrast, I study a setting in which the agent privately knows his ability and chooses quality deterministically.

A closely related literature studies optimal rating design with career concerns, in which the market values the agent's \emph{exogenous} ability rather than his endogenous quality (e.g., \citet{Rodina2020, HornerLambert2021,Goel2025,Xiao2026}).
\citet{HornerLambert2021} study a dynamic career-concerns model in which agents do not privately know their ability and show that the effort-maximizing rating is a linear function of past observations.
In a static model with private abilities, \citet{Xiao2026} shows that the effort-maximizing rating scheme always overrates low-ability agents by partially pooling them with higher-ability agents.

The results on optimal deterministic rating design also contribute to the literature on
optimal delegation with an outside option (\cite{AmadorBagwell2022,KartikKleinerVanWeelden2021,Saran2026}).
Relative to \cite{AmadorBagwell2022}, my conditions admit bang-bang allocations as optimal solutions.
\footnote{
Bang-bang allocations are generally not optimal under the sufficient conditions in \cite{AmadorBagwell2022}.
See also \cite{HalacYared2022} for the optimality of bang-bang incentive schemes.
} 
Relative to \cite{KartikKleinerVanWeelden2021}, I allow for state-dependent preferences of the principal.
More importantly, unlike in the deterministic case, stochastic rating design is \emph{not}
equivalent to stochastic delegation.
\footnote{
In stochastic delegation, the agent selects a lottery over actions; the
mechanism then randomizes the action itself, and the realized action directly
affects the agent's payoff. In stochastic rating design, the agent chooses a
deterministic action (quality); the rating scheme then randomizes the signal (and hence the market wage) generated by the chosen quality.
}

More broadly, this paper contributes to the literature on quality disclosure and certification \citep[e.g.,][]{DranoveJin2010, HarbaughRasmusen2018, DeMarzoKremerSkrzypacz2019, AliHaghpanahLin2022}.
Much of this literature takes quality as given and studies how information acquisition, selective disclosure, and certification affect market beliefs and outcomes. Here, by contrast, quality is endogenous, and the principal designs a rating scheme to motivate investment in quality.
 
\section{The Model}
\label{model}

\subsection{Setup}

The model has three players: a principal, an agent, and a competitive
market.

\paragraph*{Agent.} The agent has a private type
\(\theta\in\Theta=[\underline\theta,\bar\theta]\), where
\(0<\underline\theta<\bar\theta\), distributed according to a commonly known distribution \(F\) with a continuously differentiable density \(f>0\).
The agent chooses a quality level $q\geq0$ at a cost $c(q)/\theta$.
Assume that $c\colon\reals_+\to\reals_+$ is twice continuously
differentiable, with $c'(q)>0$ and $c''(q)>0$ for all $q>0$, $c(0)=c'(0)=0$, and $\lim_{q\to\infty}c'(q)>\bar\theta$.
Define $q_{\max}$ as the unique positive solution to $c(q) =\bar\theta q$.
Without loss of generality, restrict the agent's quality choice to a compact set \( Q=[0,q_{\max}]\).

\paragraph*{Principal.} 
The principal's payoff is \(v(q,\theta)\).
I focus on the case $v(q,\theta)=q$, in which the principal maximizes expected quality.
In general, \(v\colon Q\times\Theta\to\reals\) is twice continuously
differentiable and satisfies $v(0,\theta)=0$, $v_q(0,\theta)>0$, and $v_{qq}(q,\theta)\leq 0$
for every \((q,\theta)\in Q\times\Theta\). 
The principal prefers a higher quality than the agent would choose if the market observes his quality (see Assumption~\ref{DB}).

The principal does not observe the agent's type and cannot use transfers. Instead, she publicly commits to a rating scheme $\pi\colon Q\to\Delta(S)$ that stochastically maps quality to a signal $s\in S$.
Formally, \(S\) is a Polish space, \(\pi\) is a Borel Markov kernel, and \(\pi(\cdot\mid q)\) is the distribution of the realized signal conditional on quality \(q\).
Assume without loss of generality that \(S=Q\sqcup\{\varnothing\}\), where \(\varnothing\) is a null result that represents {nondisclosure} (i.e., the test yields no verifiable information), and each nonnull result is an unbiased estimate of quality, i.e., $s = \E[q\mid s]$ for all $s\in Q$.
\footnote{
Additional signal labels, such as \(\mathsf{pass}\) and \(\mathsf{fail}\), can be added for expositional convenience.
}

Whether the principal directly observes quality is immaterial, as long as the rating scheme takes quality as an input. If she observes \(q\), the rating scheme is a disclosure policy that garbles $q$; otherwise, it is a test that generates a score based on $q$. In either case, the scheme conveys information about quality to the market, thereby creating incentives for the agent.

The agent can choose whether to participate in the rating scheme (i.e., take the test). If he participates, the market observes a signal \(s \sim \pi(\cdot \mid q)\). If he does not, the market observes the null result \(\varnothing\).

\paragraph*{Competitive market.} 
The competitive market values the agent's quality.
After observing \(s\), it offers the agent a wage equal to the posterior mean
$\omega(s)=\E[q\mid s]$.
\footnote{
This can be microfounded by competition between two risk-neutral
employers (or buyers): the employers simultaneously submit wage offers, and the agent accepts the higher offer, with ties broken randomly. An employer who hires an agent of quality \(q\) at wage \(\omega\)
receives payoff \(q-\omega\), while the other employer receives zero.
}

\paragraph*{Timing.}
First, the agent privately knows his
type \(\theta\). Then, the principal publicly commits to a rating scheme
\(\pi\colon Q\to\Delta (S)\). 
Next, the agent chooses a quality level \(q\) and whether to take the test. Finally, the market observes $s\sim \pi(\cdot\mid q)$, or the null result $\varnothing$ if he does not take the test, and offers a wage $\omega(s) = \E[{q}\mid s]$.

\paragraph*{Equilibrium.}
The equilibrium concept is Perfect Bayesian Equilibrium (PBE).
After observing a signal \(s\), the market forms a posterior \(\mu_s\in\Delta(Q)\) using Bayes' rule
whenever possible and offers \(\omega(s)=\E_{\mu_s}[q]\). 

If the agent takes the test $\pi\colon Q\to\Delta(S)$, define his \emph{interim wage} $\hat w\colon Q \to Q$ by
\[
    \hat w_\pi(q)=\E_{s\sim\pi(\cdot \mid q)} [\omega(s) ].
\]
The agent's expected payoff is \(\hat w_\pi(q)-c(q)/\theta\).
Multiplying by \(\theta>0\), which does not affect his choice, gives the
scaled payoff 
\(u(q,\theta)=\theta\hat w_\pi(q)-c(q).\)
If he does not participate, the market observes \(s=\varnothing\) and offers \(\omega(\varnothing)\), and his optimal quality choice is \(q=0\), since his wage is independent of quality.

It is optimal for the principal to induce \(\omega(\varnothing)=0\), which
minimizes the agent's outside option. 
The principal can choose a rating scheme
that either reserves \(\varnothing\) for nonparticipation, or reserves it for a set of qualities containing zero and maps them to \(\varnothing\) with probability $1$.
Consequently, if \(\varnothing\) occurs on
path, sequential rationality and Bayes' rule imply \(\mu_{\varnothing}=\delta_0\) and \(\omega(\varnothing)=0\).  
\footnote{
If \(\varnothing\) is off-path, because every agent can obtain it at zero cost, sequential rationality implies that \(\omega(\varnothing)\) cannot exceed any agent's (unscaled) equilibrium payoff. 
Hence, setting \(\mu_{\varnothing}=\delta_0\) and \(\omega(\varnothing)=0\) is without loss and preserves every agent's equilibrium choice.
\label{ftn:varnothing}
}
Hence, nonparticipation yields a zero payoff.

\paragraph*{Downward bias.}
Under full revelation, the market observes quality perfectly and therefore $\hat w_\pi(q) = q$ for any $q\in Q$. Let
\[
    q_f(\theta)
    :=
    \argmax_{q\in Q}
     \{\theta q-c(q) \} = c'^{-1}(\theta)
\]
denote the agent's optimal quality choice under full revelation. 
The following assumption imposes that the principal prefers a higher quality than the quality chosen by the agent under full revelation.
\footnote{
\cite{AmadorBagwell2022} impose a stronger assumption
that the principal prefers a higher quality than the highest quality the agent would choose rather than opt out.
}

\begin{assumption}[Underinvestment under revelation]
\label{DB}
\(v_q(q_f(\theta),\theta)> 0\) for all \(\theta\in\Theta\).
\end{assumption}
If the principal instead prefers a lower quality, randomization can be valuable because adding pure noise to the rating scheme can discourage investment by reducing the marginal return to investment to $\hat w'_\pi(q)<1$.
I therefore focus on the more interesting case in which full revelation induces too little quality from the principal's perspective, where it is not obvious whether and how randomization can provide stronger incentives overall.

Assumption~\ref{DB} is satisfied trivially when the principal maximizes
expected quality. Examples include a school that maximizes average human capital
\citep{Zubrickas2015}, average placement outcomes, or consumer surplus \citep{Zapechelnyuk2020}. 
The assumption also accommodates objectives that partially internalize the agent's cost in Section~\ref{lineardelegation}.

Unless indicated explicitly, ``increasing'' and  ``decreasing'' refer  to weakly increasing and weakly decreasing, respectively.

\subsection{Direct Mechanism and Feasibility}
\label{preliminaries}

A quality schedule \(q\colon\Theta\to Q\) is \emph{implementable}
if there exist a rating scheme \(\pi\colon Q\to\Delta(S)\) and a PBE
of the induced game such that \(q(\theta)\) and the interim wage
\(
\hat w_\pi(x)
:=
\E_{s\sim\pi(\cdot\mid x)}[\omega(s)]
\)
under that PBE satisfy   
\[
    q(\theta)
    \in
    \argmax_{x\in Q}
    \left\{
        \theta\hat w_\pi(x)-c(x)
    \right\},
    \quad
    \text{for all }\theta\in\Theta.
\]

Consider a direct mechanism 
\((q(\theta),w(\theta))\), where
\(w\colon\Theta\to Q\) is an interim wage schedule. Unlike a transfer
schedule chosen directly by the principal, \(w\) must arise from the
market's beliefs induced by a rating scheme.

Say that \((q(\theta),w(\theta))\) is \emph{feasible} if
there exist a rating scheme
    $\pi\colon Q\to\Delta(S)$
and associated Bayes-consistent market beliefs such that 
\[
    w(\theta) = \hat w_\pi(q(\theta))
    = \E_{s\sim \pi(\cdot \mid q(\theta))}[\omega(s)], \quad \mbox{ for all $\theta\in\Theta$}.
\]

\begin{lemma}
\label{lemma:implementation}
A quality schedule \(q\colon\Theta\to Q\) is implementable if and only
if there exists an interim wage schedule \(w\colon\Theta\to Q\) such
that the direct mechanism \((q(\theta),w(\theta))\) is incentive compatible,
individually rational, and feasible.
\end{lemma}

Lemma~\ref{lemma:implementation} allows us to focus on direct mechanisms $(q(\theta),w(\theta))$.  
The rating scheme is an implementation that does not require the agent's quality to be observable by the principal, as long as it is taken as input by the rating scheme.

Define the interim utility of type \(\theta\) by $U(\theta)
    =
    \theta w(\theta)-c(q(\theta))$.
\begin{lemma}\label{lem:IC}
    A mechanism $(q(\theta), w(\theta))$ is incentive-compatible and individually rational if and only if 
    \begin{itemize}
        \item $ U(\lbar \theta) = \lbar \theta w(\underline \theta) - c(q(\underline\theta))\geq0$ \textnormal{(IR)}, 
        \item  
    $ U(\theta) = \int_{\underline \theta}^\theta w(x) \d x + U(\lbar\theta)$ \textnormal{(IC-Env)},\item $w(\theta)$ is increasing \textnormal{(IC-Mon)}.
    \end{itemize}
    Incentive compatibility also implies that $q(\theta)$ is (strictly) increasing if and only if $w(\theta)$ is (strictly) increasing.
\end{lemma}

For an incentive-compatible mechanism $(q(\theta), w(\theta))$, 
both $q$ and $w$ are increasing.
Assume without loss of generality that both $q$ and $w$ are right-continuous on \([\lbar\theta,\bar\theta)\) and left-continuous at $\bar\theta$.
The following lemma provides the necessary and sufficient condition for feasibility.

\begin{lemma}\label{lem:feasibility}
   An incentive-compatible mechanism $(q(\theta),w(\theta))$ is feasible if and only if $w(\theta)$ is majorized by $q(\theta)$ in quantile space, that is,
    \begin{equation*}
        \int_{ \theta}^{\bar \theta} w(x) \d F(x) \leq  \int_{ \theta}^{ \bar\theta} q(x) \d F(x),  \mbox{ for all  $\theta\in\T$}
    \end{equation*}
    with equality at $\theta=\lbar \theta$.\end{lemma}

\begin{remark}
Together with the revelation principle and Lemma~\ref{lemma:implementation}, this lemma implies that 
eliciting the agent's private information by offering a menu of tests has no value because it does not enlarge the set of implementable outcomes.
\end{remark}

Let $\MPS(q)$ denote the set of increasing functions $g\colon\Theta\to Q$ majorized by $q$ in quantile space.
The
condition in Lemma~\ref{lem:feasibility} is equivalent to
\(w\in\MPS(q)\).
The key subtlety is that quality is endogenous: \(\MPS(q)\) is
defined relative to the prescribed equilibrium quality schedule \(q\).
Although \(\MPS(q)\) has many extreme points, incentive
compatibility rules out all extreme points except possibly \(q\) itself: if \(w\in\operatorname{ext}\MPS(q)\) and \((q,w)\) is
incentive compatible, then \(w=q\). 
\footnote{
An extreme point \(w\) is obtained by averaging \(q\) over a collection
of pooling intervals. Incentive compatibility requires \(q\) to be
constant on every interval pooled by \(w\). Averaging therefore leaves
\(q\) unchanged, so \(w=q\).
}

The proof in the appendix uses the extreme-point method in \citet{KleinerMoldovanuStrack2021}.
\citet[Lemma~1 and Theorem~1]{SaeediShourideh2020} provide a similar characterization. 
The result also echoes Border's theorem in terms of both mathematical structure and role: Border's theorem allows one to optimize over interim rather than ex post allocations, while this result reduces the problem to optimizing over $(q(\theta),w(\theta))$ rather than $\pi$.

\subsection{Discussion of Assumptions}
\label{assumptions:discussion}

\paragraph*{The role of transfers.}
If the principal could make transfers contingent on quality or the rating result, or offer a menu of test-fee structures, the design of ratings becomes irrelevant because transfers can provide incentives in place of $w(\theta)$. \footnote{
    The perfect substitutability between the rating scheme and the transfer scheme is also noted in \cite{AlbanoLizzeri2001}  and \cite{PollrichStrausz2024}. 
    The case with a certification fee is discussed in \app*{transfer}.
}

\paragraph*{Commitment to the rating scheme.}
Adherence to the committed rating scheme is sequentially optimal.
To see this, once quality has been chosen, the principal’s payoff \(v(q,\theta)\) is independent of the realized signal and the wage offered
by the market.
Therefore, she has no incentives to tamper with the rating scheme.
\footnote{
If the principal’s payoff is instead the market wage, which equals quality in expectation, she may have incentives to manipulate the signal. Nevertheless, the rating scheme remains \emph{credible} in the sense of \citet{LinLiu2024}: the principal cannot profitably reassign signals while preserving their marginal distribution.
}

\section{The Deterministic Benchmark}
\label{deterministic}
\subsection{Principal's Problem}
In this section, I restrict attention to \emph{deterministic} rating schemes $\pi\colon Q\to S$.
Deterministic rating schemes either fully reveal the quality or pool a set of qualities into a single score.
When quality is fully revealed, the market learns the quality.
When multiple qualities are mapped to the same score $s$, the lowest quality that generates $s$ strictly dominates all other qualities, so only the lowest quality will be chosen, and the market also learns the quality in equilibrium \citep[see also][Claim~1]{Zapechelnyuk2020}.
Therefore, the equilibrium interim wage schedule always equals the quality schedule, and the majorization condition holds with equality.
\begin{lemma}\label{lemma:w=q}
    When $\pi$ is deterministic, $w(\theta)=q(\theta)$. \end{lemma}

Therefore, finding the optimal rating scheme $\pi\colon Q\to S$ is equivalent to finding the optimal quality schedule $q\colon \Theta\to Q$. The principal's problem becomes
\footnote{
    As shown in \cite{Zapechelnyuk2020}, the principal's problem \ref{eqn:D} is equivalent to a delegation problem (à la \cite{Holmstrom1984}) with an \emph{outside option}:
designing an incentive-compatible quality schedule \(q(\theta)\) is equivalent to choosing a delegation set $\{q(\theta): \theta\in\Theta\}$. 
}
\allowdisplaybreaks
\begin{align*}
\max_{q\colon \Theta\to Q}\quad  &\int_{\underline \theta}^{\bar \theta} {v}(q(\theta), \theta) \d F(\theta)
\tag*{[D]} 
\label{eqn:D}\\
\text{subject to}\quad 
    & \underline{U} = \lbar\theta q(\underline\theta) - c(q(\underline\theta))\geq0 \tag*{(IR\textsuperscript{D})}\\
    & q(\theta) \mbox{ increasing},  \tag*{(IC-Mon\textsuperscript{D})}\\
    & \theta q(\theta) - c(q(\theta) ) =  \int_{\underline \theta}^\theta q(x) \d x + \underline U, \mbox{ for all $\theta\in\T$.}  \tag*{(IC-Env\textsuperscript{D})} 
\end{align*}

An incentive-compatible quality schedule $q(\theta)$ consists of pooling regions, where $q(\theta)$ is constant, and fully revealing regions, where $q(\theta) = q_f(\theta)$, with at most countably many jump discontinuities \citep{MelumadShibano1991}. In particular, $q(\theta)$ can jump at a type $\theta$ who is indifferent between opting out ($q=0$) and a positive quality $q_i(\theta)>0$ characterized by $
 {c(q_i(\theta))}=\theta q_i(\theta)$.
By assumptions on $c(q)$, a unique $q_i(\theta) > q_f(\theta)$ exists for all $\theta\in(0,\bar\theta]$.
In particular, $\lim_{\theta\to0}q_i(\theta)=0$ and $q_i(\bar\theta)=q_{\max}$.

\subsection{Lower Censorship and Pass/fail Tests}
I focus on two simple deterministic ratings: lower censorship and pass/fail tests.

    \emph{Lower censorship} is a deterministic rating scheme $\pi\colon Q\to S$ for which there exists a threshold $q_0$ such that
    \footnote{A fail-or-reveal rating scheme, which outputs $\pi(q) = \mathsf{fail}$ instead of $\varnothing$ when $q<q_0$, implements the same quality schedule.
    A signal $s \in\{\mathsf{pass},\mathsf{fail}\}$ can also be replaced by the signal $s'=\E[q\mid s]$.
    }
    \[
    \pi(q) = \begin{cases}
    q, &\text{if } q\geq q_0,\\
    \varnothing, &\text{otherwise}.
    \end{cases}
    \]
    A \emph{pass/fail test} is a deterministic rating scheme $\pi\colon Q\to S$ for which there exists a threshold $q_0$ such that
    \[
    \pi(q) = \begin{cases}
    \mathsf{pass}, &\text{if } q\geq q_0,\\
    \mathsf{fail}, &\text{otherwise}.
    \end{cases}
    \]
         In both cases, $q_0$ is called a \emph{minimum standard} or a \emph{threshold}.

A threshold $q_0\in(0,q_{\max})$ induces a cutoff type
 $\theta_0 = \frac{c(q_0)}{q_0}\in(0,\bar\theta)$,  who is indifferent between choosing $q=0$ and $q_i(\theta_0)=q_0$.
 For convenience, let
 $\theta_c(\theta_0)= c'(q_i(\theta_0))$
 denote the type that would choose $q_i(\theta_0)$ under full revelation.
 Strict convexity of $c$ with $c(0)=0$ implies $\theta_c(\theta_0)>\theta_0$.
 For example, if $c(q)=q^2/2$, then $q_f(\theta)=\theta$ and $q_i(\theta_0)= \theta_c(\theta_0) = 2\theta_0$.

Lower censorship with a threshold $q_0\in(0,q_{\max})$ implements a quality schedule consisting of an \emph{exclusion} region, a \emph{bunching} region, and a \emph{fully revealing} region,
given by the following piecewise continuous function:
\footnote{
 By convention, $[x, y)$, $(x, y)$, and $[x, y]$ represent the empty set if $x\geq y$, and define $q(\bar\theta) = q(\bar\theta-)$.
 Any of these regions can be empty: (i) if $\theta_0\leq\lbar \theta$, (ii) if $\theta_c(\theta_0)\leq \lbar \theta$, and (iii) if $\theta_c(\theta_0)\geq \bar \theta$.
\label{footnote:empty}
}
\begin{equation*}\label{eqn:lowercensorship}
    q(\theta ) = 
    \begin{cases}
    0   , &\text{if }\theta \in [\underline \theta, \theta_0)\\
    q_i(\theta_0) , &\text{if } \theta \in [  \theta_0, \theta_c(\theta_0))\\
    q_f(\theta)   , &\text{if } \theta \in [\theta_c(\theta_0), \bar\theta]
   \end{cases}
\end{equation*}
where $\theta_0=c(q_0)/q_0\in(0,\bar\theta)$.
Analogously, a pass/fail test with a threshold $q_0\in(0,q_{\max})$ implements a quality schedule consisting of an exclusion region $[\lbar \theta, \theta_0)$ where $q(\theta)=0$, and a bunching region $[\theta_0, \bar\theta]$ where $q(\theta)=q_i(\theta_0)$.

\begin{figure}[htb]
    \centering
    \includegraphics[width= 0.9 \textwidth,trim={4cm 0cm 4cm 0cm},clip]{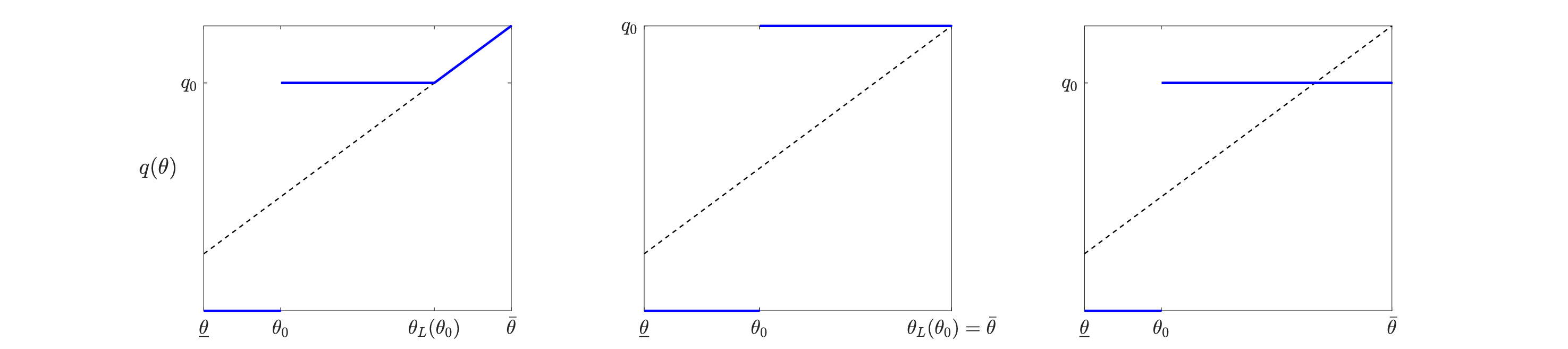}
    \begin{minipage}{0.85\textwidth}
        \footnotesize
        $q(\theta)$ on the left is implemented by lower censorship with a threshold $q_0$; right is implemented by a pass/fail test with a threshold $q_0$;
         center can be implemented by both.
    \end{minipage}
    \caption{$q(\theta)$ implemented by lower censorship and pass/fail tests}
    \label{fig:lcs_passfail}
\end{figure}  

Two observations are useful for subsequent analysis.
First, if $\theta_0 := c(q_0)/q_0 \leq \lbar\theta$, the exclusion region $[\lbar\theta,\theta_0)$ is empty and all types choose $q>0$.
Second, if $q_0\geq q_f(\bar\theta)$, then $\theta_c(\theta_0)\geq\bar\theta$, so the fully revealing region $[\theta_c(\theta_0), \bar\theta]$ is empty: the minimum standard is so high that no one chooses any higher quality in equilibrium.
In that case, lower censorship and a pass/fail test with the same minimum standard $q_0$ implement the same quality schedule.\footnote{
 Even when they implement the same quality schedule, a pass/fail test is \emph{not} a special case of lower censorship, because off-path deviations to $q>q_0$ generate different signals.
}
On the other hand, if $\theta_c(\theta_0)< \bar\theta$, $q$ is continuous at $\theta_c(\theta_0)$.

\subsection{Quality Maximization}

I first focus on the quality maximization case where $v(q,\theta)=q$.

\begin{proposition}\label{prop:quality-maximization}
Suppose that $f$ is single-peaked.
Then, lower censorship is optimal among deterministic rating schemes. In particular,
\begin{itemize} 
    \item if \(f\) is increasing, or if \(\theta_c(\underline\theta)\geq\bar\theta\), then a pass/fail test is also optimal;
    \item if \(f\) is decreasing, then lower censorship with a threshold $q_0^*  = q_i(\lbar\theta)$, which every type reaches in equilibrium, is optimal.
\end{itemize}

   More generally, lower censorship with a minimum standard \(q_0^* = q_i(\theta_0^*)\) is optimal among deterministic rating schemes if there exists some $\theta_0^*\in[\lbar\theta,\bar\theta)$ that satisfies the following conditions:
    \begin{description}[labelindent=1em]
        \nameditem{S}{S0}
        There exists a constant $A^*$ that satisfies
\(
F(\theta)-F(\theta_0^*) \geq A^*(\theta-\theta_0^*)
\)
for all 
\(
\theta\in[0,\theta_c(\theta_0^*)],
\)
with equality at $\theta=\theta_c(\theta_0^*)$.
\footnote{
    That is, $A^* = \frac{F(\theta_c(\theta_0^*))-F(\theta_0^*)}{\theta_c(\theta_0^*)-\theta_0^*}$ is a subgradient of $F$ at $\theta_c(\theta_0^*)$ on $[0,\theta_c(\theta_0^*)]$.
    If $\theta_0^*>\lbar\theta$, then $A^* = f(\theta_0^*)$, and \eqref{S0} is equivalent to $f(\theta_0^*)$ being a subgradient of $F$ at $\theta_0^*$ on $[0,\theta_c(\theta_0^*)]$.
}

        \nameditem{C}{C0}
        $f(\theta)$ is decreasing on
        $[\theta_c(\theta_0^*),\bar\theta]$.
    \end{description} 
    Pass/fail tests are also optimal if, in addition, \(\theta_c(\theta_0^*)\geq\bar\theta\).
 \end{proposition}
\begin{remark}
If \(f\) is single-peaked at a unique $\theta_m \in (\lbar\theta, \bar\theta]$, then $\theta_0^*$ is unique and ensures that $\theta_m\in(\theta_0^*, \theta_c(\theta_0^*))$ lies in the bunching region.
If \(f\) is strictly increasing, then $q_0^*>q_f(\bar\theta)$; if \(f\) is strictly decreasing, then $q_0^*=q_i(\lbar\theta)$. If $f$ is constant and $\theta_c(\lbar\theta)<\bar\theta$, then every $q_0^*\in[q_i(\lbar\theta),q_f(\bar\theta)]$ is optimal.
\end{remark}

Proposition~\ref{prop:quality-maximization} extends
\citet{Zapechelnyuk2020} by weakening the condition under which a pass/fail test is optimal.
\footnote{\citet{Zapechelnyuk2020} shows that if $f$ is single-peaked, lower censorship is optimal among deterministic rating schemes, and that if, in addition, \(\theta_c(\underline\theta)\geq\bar\theta\), a pass/fail test is also optimal.}
The proof uses the optimal control method that parallels the Lagrangian approach of \citet{AmadorBagwell2022}, and it allows the fully revealing region to be empty.

Given the main result, the first claim follows because lower censorship and a pass/fail test with the same threshold \(q_0\) induce the same quality schedule whenever \(q_0\geq q_f(\bar\theta)\), or equivalently, \(\theta_c(\theta_0)\geq\bar\theta\); this holds at the optimum if $f$ is increasing, or if $f$ is single-peaked and \(\theta_c(\underline\theta)\geq\bar\theta\).
The second claim follows because \(\theta_0=\lbar\theta\) is optimal when \(f\) is decreasing, in which case the exclusion region is empty.

\paragraph*{Optimal minimum standard.}
It is equivalent to characterize the optimal cutoff type \(\theta_0\) because \(q_0^* = q_i(\theta_0^*)\).
For lower censorship with a threshold $q_0=q_i(\theta_0)$, let $V(\theta_0)$ denote the expected quality of the implemented quality schedule:
\[
V(\theta_0)
=
\int_{\theta_0}^{\theta_c(\theta_0)}q_i(\theta_0)\d F(\theta) + \int_{\theta_c(\theta_0)}^{\bar\theta}q_f(\theta)\d F(\theta).
\]
Define the average density between
\(\theta_0\) and \(\theta_c(\theta_0)\) by
\footnote{
Extend \(F\) and \(f\) to $\reals_+$ by
\(F(\theta)=f(\theta)=0\) for all \(\theta<\underline\theta\), and
\(F(\theta)=1\), \(f(\theta)=0\) for all \(\theta>\bar\theta\).
}
\begin{equation} \label{eqn:A}
     A(\theta_0)=
     \frac{
            F(\theta_c(\theta_0))-F(\theta_0)
        }{
            \theta_c(\theta_0)-\theta_0
        },
\end{equation}
which satisfies $A(\theta_0^*)=A^*$ in Condition~\eqref{S0}.
We have 
\[
V'(\theta_0)
=
q_i(\theta_0)[A(\theta_0)-f(\theta_0)] \mbox{ for all $\theta_0\in(\lbar\theta,\bar\theta)$.}
\]
If $f$ is single-peaked, so is $V$, and thus the optimal cutoff $\theta_0^* \in\argmax_{t} V(t)$ satisfies
 $A(\theta_0^*) = f(\theta_0^*)$ if  $\theta_0^* >\lbar\theta$, and $A(\theta_0^*) \leq f(\theta_0^*)$ if $\theta_0^* =\lbar\theta$.
    \footnote{
In general, the optimal $\theta_0^*$ is characterized by condition \eqref{S0} and \eqref{C0}.
If $f$ is strictly single-peaked, then $V$ is also strictly single-peaked on $\T$,
and $\theta_0^*\in\T$ is uniquely characterized by $0\in\partial V(\theta_0^*)$, where \(\partial\) denotes the Clarke subdifferential.
}

\begin{example}[Power Distribution]
    \label{ex:power}
    Assume $c(q)=q^2/2$ and \(F(\theta)=\frac{\theta^a-\delta^a}{1-\delta^a}\) on $\Theta=[\delta,1]$, where $a>0$ and $\delta\in(0,1/2)$ (so that $\theta_c(\delta) = 2\delta<1$).

    If $a\leq1$, $f$ is decreasing, so lower censorship with a threshold $2\delta$ is optimal, and 
    $A(\delta)= \frac{F(2\delta)-F(\delta)}{\delta}\leq f(\delta)$.
    If $a\geq1$, $f$ is increasing, lower censorship or a pass/fail test with a threshold $2\theta_0^*$ is optimal, where $\theta_0^*= (1+a)^{-1/a}$ satisfies
    \(
     A(\theta_0^*) =   {f(\theta_0^*)}.
    \)
    As $a\to\infty$, $F$ becomes more convex, and $\theta_0^* \uparrow 1$. \end{example}

\paragraph*{Intuition.} 
\label{intuition}
Assume $c(q)=q^2/2$.
Consider a small perturbation to the optimal quality schedule on $
(\hat\theta-\varepsilon, \hat\theta + \varepsilon)$ in the fully revealing region.
To maintain incentive compatibility, the perturbed quality schedule takes the form of 
\[ \hat q(\theta) = \begin{cases}
   \hat q_L = q_f(\hat\theta-\varepsilon), &\text{ if }\theta\in(\hat\theta-\varepsilon, \hat\theta),\\
  \hat q_H = q_f(\hat\theta+\varepsilon), &\text{ if }\theta\in [\hat\theta,\hat\theta + \varepsilon),
\end{cases}
 \quad \mbox{ for all }\theta\in(\hat\theta-\varepsilon, \hat\theta+\varepsilon).
\]
This schedule is induced by a local threshold at $\hat q_H$ that assigns $[\hat q_L, \hat q_H)$ the same signal as $\hat q_L$, reveals $\hat q_H$, and retains full revelation outside the local interval.
This threshold creates two pooling regions: $(\hat\theta-\varepsilon,\hat\theta)$ and $[\hat\theta, \hat\theta+\varepsilon)$ and leads to a trade-off: On the one hand, it discourages the lower types from investing in quality because they would rather bunch at $\hat q_L$ and not reach $\hat q_H$. On the other hand, it induces the higher types to invest more than they would under full revelation to reach $\hat q_H$ and separate themselves from the lower types.

\begin{figure}[htb]
    \centering
    \includegraphics[width=0.4\textwidth]{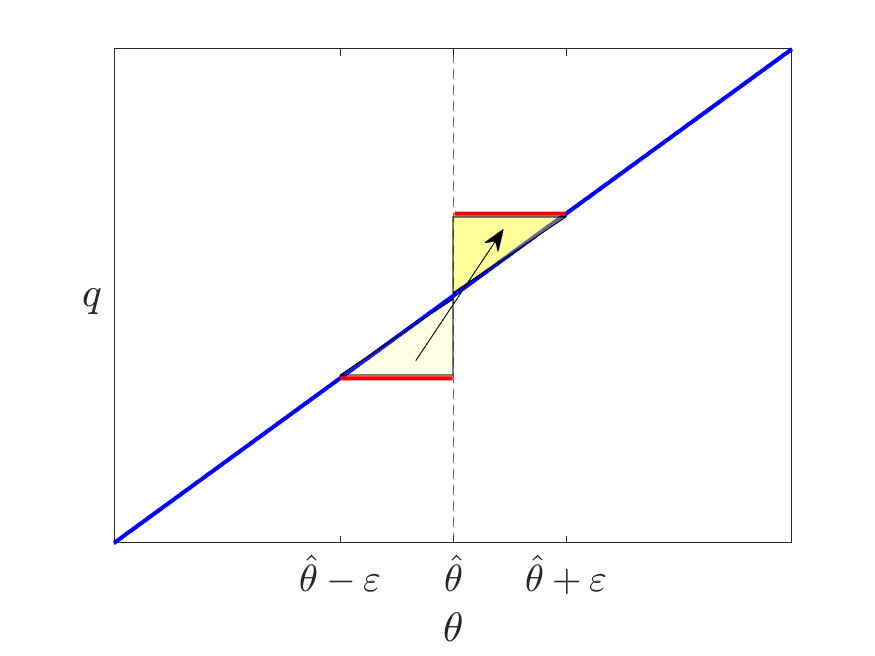}
    \caption{A perturbation to $q_f(\theta)$ in the fully revealing region}
    \label{fig:perturbation}
\end{figure}

Figure~\ref{fig:perturbation} illustrates this trade-off.
The loss in expected quality from lower types is the left triangle, while the gain from higher types is the right triangle.
The two triangles have the same area. 
\footnote{
For general cost functions $c(q)$, the loss and the gain regions still have the same area, although not necessarily triangles, and the endpoints around $\hat\theta$ need not be symmetric.
}
Therefore, shifting the mass from left to right weakly decreases
(increases) expected quality if the density \(f\) is decreasing
(increasing) on \((\hat\theta-\varepsilon,\hat\theta+\varepsilon)\). 
In other words, whenever there are more low types than high types, it is undesirable to reduce the quality of lower types to incentivize higher types.
Consequently, if $f$ is decreasing on $\Theta$, it is optimal to fully reveal quality above a minimum standard $2\lbar\theta$, which induces the highest possible quality from $[\lbar\theta, 2\lbar\theta]$ without incurring any loss from lower types.

In another extreme, if $f$ is increasing on $\Theta$, the mass of the right triangle is larger, so this perturbation increases expected quality, even as $\varepsilon$ becomes large.
Consequently, a bang-bang quality schedule without any fully revealing region is optimal, which can be implemented by a pass/fail test or lower censorship with a high minimum standard.

\section{Optimal Stochastic Ratings}
\label{general}

\subsection{Principal's Problem}
Now I study the optimal rating design when stochastic rating schemes are allowed.
The principal's problem becomes
\allowdisplaybreaks
\begin{align*}
    \max_{q,w\colon \Theta\to Q} \quad  &\int_{\underline \theta}^{\bar \theta} {v}(q(\theta), \theta) \d F(\theta) \tag*{[P]} \label{prob:P} \\
    \text{subject to}\quad 
    & \int_{\underline \theta}^{ \theta} w(x) \d F(x) \geq  \int_{\underline \theta}^{ \theta} q(x) \d F(x), \mbox{ with equality at $\theta=\bar \theta$} \tag*{(MPS)}\\
    & \underline{U} = \lbar \theta w(\underline\theta) - c(q(\underline\theta)) \geq 0 \tag*{(IR)}\\
    & w(\theta) \mbox{ increasing}  \tag*{(IC-Mon)}\\
    & \theta w(\theta) - c(q(\theta) ) =  \int_{\underline \theta}^\theta w(x) \d x + \underline U, \mbox{ for all $\theta\in\T$}.  \tag*{(IC-Env)} 
\end{align*}

Unlike in the deterministic case, (IC-Env) can no longer be reduced to a
majorization constraint, à la
\citet[Section 4.3 on optimal delegation]{KleinerMoldovanuStrack2021}, since \(w(\theta)=q(\theta)\) need not
hold.
Moreover, although (MPS) is itself a majorization constraint, it involves both choice variables \(w(\theta)\) and \(q(\theta)\) simultaneously, and the principal must choose these two functions jointly subject to both the majorization and incentive constraints. Consequently, the method of \citet{KleinerMoldovanuStrack2021} does not apply directly, even when $v(q,\theta)$ is linear.
In particular, the optimal $w$ need not be an extreme point of $\MPS(q)$. Indeed, the optimal rating scheme can be stochastic, with (MPS) slack and both \(w\) and \(q\) strictly increasing on some intervals.

\begin{lemma}[Existence] \label{lem:existence}
    A solution to problem \ref{prob:P} exists.
\end{lemma}
Consequently, whenever an optimal deterministic rating scheme is not optimal among all ratings, the optimal rating scheme must involve randomization.

\subsection{When are deterministic ratings optimal?}
I will first focus on the quality maximization case $v(q,\theta)=q$ and provide sufficient conditions under which deterministic ratings remain optimal.
 
\begin{proposition}\label{prop:optimalstochastic}
     If $f$ is single-peaked and has an increasing elasticity $\epsilon_f= -\frac{\theta f'(\theta)}{f(\theta)}$ on its decreasing part, then lower censorship is optimal among all rating schemes.
    In particular,  
    \begin{itemize}
    \item if $f$ is log-concave, lower censorship is optimal. 
    \item if $f$ is increasing, pass/fail tests and lower censorship are optimal;
    \end{itemize}
    
    More generally, lower censorship and a pass/fail test with a threshold $q_i(\theta_0^*)$ are optimal if Condition~\eqref{S0} holds for some $\theta_0^*\geq\theta_c^{-1}(\bar\theta)$.
    \footnote{
    When $f$ has no flat intervals, this is equivalent to $A(\theta_0^*)$ being a subgradient of $F$ at $\theta_0^*$ on $\T$.
    A sufficient condition is that $f$ is single-peaked and $\theta_c(\lbar\theta)>\bar\theta$.
    }
    This holds if $f(\theta_0^*)$ (or $A(\theta_0^*)$) is a strict subgradient of $F$ at $\theta_0^*$ on $\T$.
\end{proposition}

The proposition implies that randomization has no value if intermediate or high ability is abundant.
The next section provides intuitions on why this matters for the value of randomization.
The proof also uses the optimal control method, but with different multipliers, to establish that $w(\theta)=q(\theta)$ is optimal.

Say that a density $f$ satisfies \emph{increasing elasticity of density} (IED) if its elasticity $\epsilon_f(\theta) = -\frac{\theta f'(\theta)}{f(\theta)}$ is increasing.
Many common unimodal distributions satisfy IED wherever their densities are decreasing, including power, Pareto, and log-normal distributions, even though they may violate standard regularity conditions.
The proposition therefore implies the optimality of lower censorship for a wide range of common distributions, which may help explain the prevalence of pass/fail tests and lower censorship in practice.

If $f$ is decreasing and log-concave, then $\epsilon_f(\theta) = -\theta \frac{f'(\theta)}{f(\theta)}\geq 0$ is increasing.
Because log-concavity also implies single-peakedness, it is a sufficient condition for the optimality of lower censorship among all rating schemes.

If $f$ is increasing, it is single-peaked with an empty decreasing interval, on which IED holds vacuously.
Consequently, lower censorship, which implements a bang-bang quality schedule, is optimal, and pass/fail tests are also optimal since they can implement the same schedule.

\begin{figure}[hbt]
\centering
\includegraphics[width=0.35\textwidth]{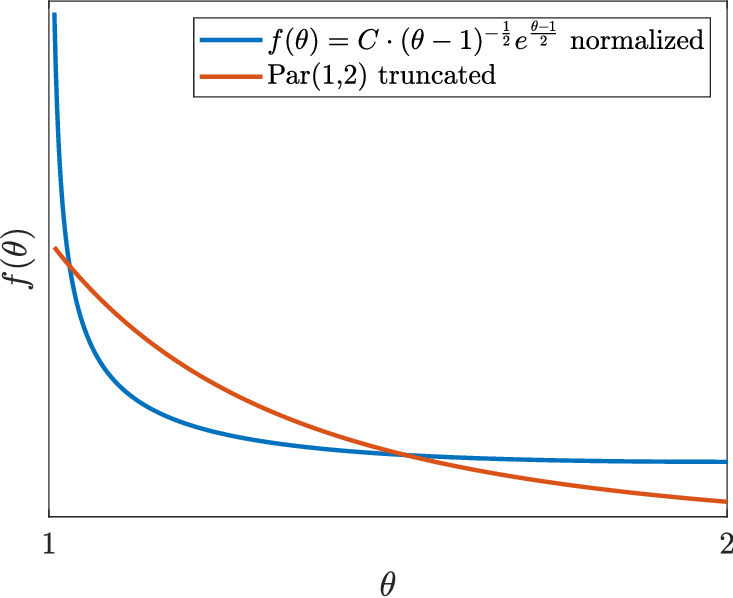}
\caption{A decreasing $f$ more log-convex than Pareto may violate IED}
\label{fig:IED}
\end{figure}

The Pareto (power-law) family, which has a constant elasticity of density, marks the boundary of the IED condition.
\begin{example}[Pareto]
    The Pareto distribution $\text{Par}(a,b)$, truncated at $\bar\theta$, has a strictly decreasing density $f(\theta)=\frac{a b^a \theta^{-(a+1)}}{1-(b/\bar\theta)^a}$ and a \emph{constant} elasticity $\epsilon_f(\theta)= a+1$. 
\end{example}
Consequently, IED fails whenever the density is decreasing and sufficiently log-convex relative to the Pareto distribution. Figure~\ref{fig:IED} illustrates a counterexample with
$f(\theta) = C \cdot (\theta-1)^{-1/2}\exp((\theta-1)/2)$ on $\Theta=[1+\varepsilon,2]$,
where \(C>0\) is a normalizing constant and $\varepsilon>0$ is sufficiently small.
Here, $f$ is strictly decreasing, yet its elasticity is strictly decreasing.
Intuitively, log-convexity of a decreasing density implies that intermediate types are relatively scarce compared with low and high types.

\subsection{When is randomization valuable?}

Now I provide conditions under which stochastic rating schemes are optimal.
The next proposition shows that IED in the fully revealing region is also necessary.
 
\begin{proposition}\label{prop:improve}
    If an optimal deterministic rating scheme induces a quality schedule with a
    nonempty fully revealing interval on which $f$ violates IED, then every
    optimal rating scheme involves randomization.

    A sufficient condition is that $f$ is decreasing, and its elasticity $\epsilon_f(\theta)= -\frac{\theta f'(\theta)}{f(\theta)}$ fails to be increasing on a nonempty subinterval of $(\theta_c(\lbar\theta),\bar\theta)$.
\end{proposition}
\begin{remark}
Theorem~\ref{thm:certificate} in the appendix provides certificates of optimality for candidate rating mechanisms.
\end{remark}
To see how stochastic ratings can outperform deterministic ones, recall that the agent's marginal benefit from raising quality is $\hat w'(q)$, which equals one under full revelation. A stochastic rating scheme can generate $\hat w'(q)>1$ over a range of qualities, thereby providing stronger marginal incentives than full revelation.
This gain is not free: feasibility requires the interim wage to be majorized by quality, so rewarding some types above their quality requires paying others below theirs. Additionally, randomization has value on the extensive margin, strengthening the incentive to meet a higher minimum standard.

To make this concrete, consider a stochastic rating scheme that fails any agent who does not meet a minimum standard $q_0$; among those who meet the standard, it reveals quality with probability $p(q)$ and otherwise issues a ``pass'' signal.
\footnote{
When $q<q_0$, giving a ``fail'' is equivalent to disclosing no verifiable information, as in lower censorship, i.e., $\pi(\cdot\mid q)=\delta_{\varnothing}$.
For the reason discussed in Footnote~\ref{ftn:varnothing}, it is without loss to set $\omega(\mathsf{fail}) = \omega(\varnothing)= 0$.
}
\begin{equation}
    \pi(\cdot\mid q)
    =
    \begin{cases}
        p(q) \delta_q + (1-p(q)) \delta_{\mathsf{pass}}, &\text{if }q\geq q_0,\\
        \delta_{\mathsf{fail}} , &\text{otherwise,} 
    \end{cases}
    \label{eqn:noisytest}
\end{equation}
where $\delta_s$ is the Dirac measure at $s$.
In practice, this can be interpreted as a variant of quality assurance that rejects products falling below the minimum standard, grants every compliant product an approval seal, and publishes a detailed quality assessment with some probability.

In equilibrium, after observing a ``pass'' signal, the market infers the agent's quality to be the average quality $q^m=\E[q\mid \mathsf{pass}]$.
The resulting interim wage is
\begin{equation} \label{eqn:hatw}
        \hat w(q) =
        \begin{cases}
            p(q)q + (1-p(q) )q^m, &\text{if }q\geq q_0,\\
            \omega(\mathsf{fail}) = 0, &\text{otherwise.} 
        \end{cases}
\end{equation}

The scheme strengthens incentives through two channels.
On the extensive margin, it allows the principal to raise the minimum standard without excluding additional low types, since the prospect of being pooled with higher types raises the payoff to just meeting the standard: $\hat w(q) > q$ for all $q \in [q_0, q^m)$ with $p(q)<1$.

On the intensive margin, it raises the marginal return to quality.
To see this, note that for $q\geq q_0$ with $p'(q)(q-q^m)>1-p(q)$, the marginal return is
\[
    \hat w'(q)=p(q)+p'(q)(q-q^m) > 1.
\]
Consider a U-shaped revelation probability $p(q)$.
For low types who choose $q<q^m$, a decreasing revelation probability can raise the marginal return by making them more likely to be pooled with the average quality.
For high types who choose $q>q^m$, an increasing revelation probability can also raise the marginal return by making it more likely that their true quality is revealed.
Thus, a U-shaped revelation probability can strengthen
incentives at both ends of the distribution through a combination of carrot and stick, at the expense of intermediate types.

When the density is decreasing and sufficiently log-convex, intermediate types
are relatively scarce compared to low and high types. 
Therefore, the gains of low and high types outweigh the losses of intermediate types, and this stochastic rating scheme increases expected quality, as shown in the following example.

\begin{example}[Optimal stochastic ratings]
    \label{ex:stochastic}
    Assume that $c(q)=q^2/2$ and
    $\Theta=[\lbar\theta,1]$, where $\lbar\theta=0.01$. Let $
    f(\theta)=C\theta^{-1/2}\exp(\theta/2)$,
    where $C\approx0.46$ is the normalizing constant. 
    The density $f$ and its elasticity $\epsilon_f(\theta) = \frac{1-\theta}{2}$ are both strictly decreasing.

    By Proposition~\ref{prop:quality-maximization},  the optimal deterministic
    rating is lower censorship with a threshold
    $2\lbar\theta=0.02$, which induces $q^D (\theta) = \max\{\theta,2\lbar\theta\}$.    
    By Proposition~\ref{prop:improve}, the optimal rating is stochastic.
    Indeed, the optimal $(q^*(\theta),w^*(\theta))$ is given by 
   \[
q^*(\theta)=
\begin{cases}
q_L, & \theta\in[\lbar\theta,\theta_L],\\
 \frac{(\Lambda-1)\theta^2 f(\theta)}
    {\Lambda\theta f(\theta)+B-\Lambda F(\theta)}, & \theta\in(\theta_L,\theta_H),\\
q_H, & \theta\in[\theta_H,1],
\end{cases}
\quad 
w^*(\theta)
=
\begin{cases}
w_L:=  {q_L^2}/{2\lbar\theta} > q_L,
& \theta\in[\lbar\theta,\theta_L],\\  
w_L
+\int_{\theta_L}^{\theta}
    \frac{ q^*(x)q^{*\prime}(x)}{x}\d x,
& \theta\in(\theta_L,\theta_H),\\ 
q_H,
& \theta\in[\theta_H,1],
\end{cases}
\]
where $\Lambda\approx0.648$, $B\approx-0.03$,
$(\theta_L,q_L)\approx(0.032,0.038)$, $(\theta_H,q_H)\approx(0.845,1.009)$.
   Figure~\ref{fig:q_w} illustrates the optimal mechanism.
   \footnote{
       In \supple*{app:stochastic-certificate}, I construct multipliers to certify the optimality of $(q^*(\theta),w^*(\theta))$ using Theorem~\ref{thm:certificate}.
   }
    \begin{figure}[htb]
    \centering
    \begin{subfigure}[h]{0.45\textwidth}
        \centering
        \includegraphics[width= \textwidth]{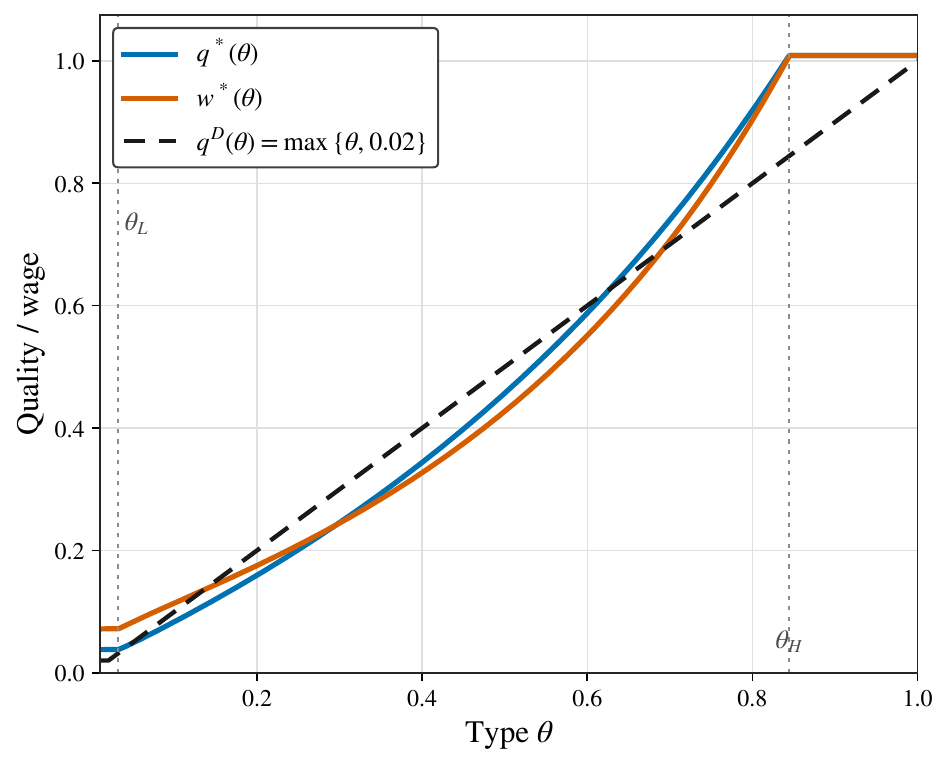}
        \caption{Optimal $(q^*(\theta),w^*(\theta))$ vs.~$q^D(\theta)$}
        \label{fig:q_w}
    \end{subfigure}
    \hspace{20pt}
    \begin{subfigure}[h]{0.45\textwidth}
        \centering
        \includegraphics[width= \textwidth]{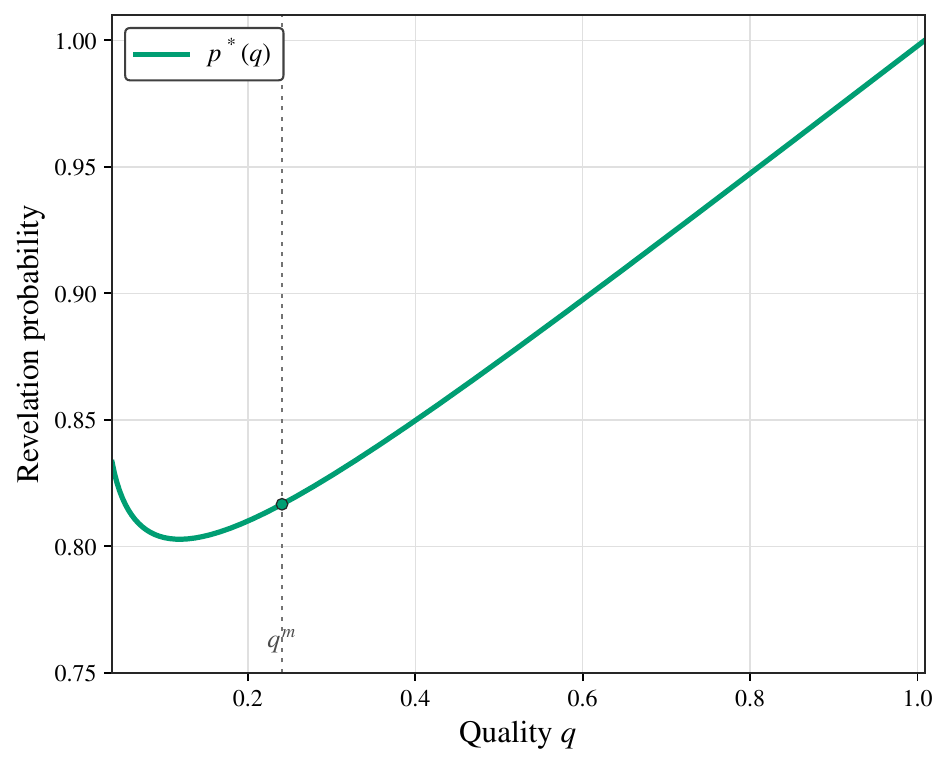}
        \caption{Revelation probability $p^*(q)$}
        \label{fig:p(q)}
    \end{subfigure}
    \caption{Optimal stochastic rating scheme for $f(\theta) = C \cdot \theta^{-1/2} \exp(\theta/2)$}
    \label{fig:improve}
    \end{figure}

   This mechanism can be implemented by a variant of lower censorship with a threshold $q_L>2\lbar\theta$, as in equation~\eqref{eqn:noisytest}, that fails $q<q_L$ and reveals quality $q\geq q_L$ with probability
    \[
        p^*(q)
        =
        \begin{cases}
           \dfrac{\hat w^*(q)-q^m}{q-q^m} , &\text{if }q\in[q_L,q_H],\\
            1, &\text{if }q\geq q_H,
        \end{cases}
    \] 
    where $\hat w^* = w^*\circ q^{*-1}$ is strictly increasing on $(q_L,q_H)$, and $q^m\approx 0.24$ is its interior fixed point that satisfies $\E[q\mid \mathsf{pass}] = \frac{\E[(1-p^*(q))q]}{\E[1-p^*(q)]} = q^m$.
    \footnote{
At $q=q^m$, define $p^*(q^m)=\hat w^{*\prime}(q^m)$.
By construction, this rating scheme induces $\hat w^*(q)$ through
equation~\eqref{eqn:hatw}. Since $\hat w^*$ is strictly increasing,
single-crosses the identity from above at $q^m$ on $(q_L,q_H)$, and
satisfies $\hat w^*(q_H)=q_H$, we have
\[
q<\hat w^*(q)<q^m \;\text{ for all }q\in[q_L,q^m),
\quad
q^m<\hat w^*(q)<q \; \text{ for all }q\in(q^m,q_H).
\]
Therefore, $p^*(q)\in[0,1]$ for every $q\in[q_L,q_H]$, with
$p^*(q_H)=1$.
    }
    Figure~\ref{fig:p(q)} illustrates the associated revelation probability $p^*(q)$.
    The U-shaped probability, together with the higher threshold, induces higher quality than $q^D$ on $[\lbar\theta,q_L]$ and at the top.
\end{example}

Although the optimal stochastic rating scheme increases expected quality by only \(0.37\%\) relative to $q^D$ 
in this example, the gain through randomization can exceed \(10.69\%\) for other ability densities that violate IED.
\footnote{
Let
        $
f(\theta)
=
Z 
[\varepsilon+\exp( -k(\theta-\lbar\theta) )]$, where $\varepsilon=0.005$, $k=100$,
and
$Z \approx66.89$ is the normalizing constant.
For this density, I construct a feasible stochastic rating in \supple*{app:stochastic-certificate} that 
raises expected quality by \(10.69\%\) relative to the optimal
deterministic benchmark \(q^D\).
}

\section{Beyond Quality Maximization}
\label{lineardelegation}

This section extends the preceding analysis beyond quality maximization.
I focus on a type-weighted extension of the linear-delegation
preferences in \citet{KolotilinZapechelnyuk2025}.
\begin{condition}{LD} \label{LD}
    The principal's payoff is $v(q,\theta) = [\beta (\theta)q - \alpha c(q)]\zeta(\theta)$, where $\beta,\zeta\colon \Theta\to \reals_+$ are continuously differentiable and $\alpha\geq0$ is a constant.
     Under Assumption~\ref{DB},
    $\beta(\theta)>\alpha\theta$ for every $\theta\in\Theta$.
\end{condition}

As a simple special case, suppose the principal internalizes a constant fraction $\alpha\in[0,1)$ of the agent's cost: $v(q,\theta)=q-\alpha\frac{c(q)}{\theta}$, which can arise when a school maximizes expected tuition revenue \citep{OnuchicRay2023} or a certifier maximizes a weighted sum of expected quality and firms' profits \citep{BizzottoHarstad2023}.
This objective satisfies Condition~\ref{LD} with $\zeta(\theta)=1/\theta$ and $\beta(\theta)=\theta$.
Similarly, the condition is satisfied when the principal maximizes a weighted sum of
matching efficiency and agents' payoffs \citep{LiQiu2024}.

\subsection{Optimal Deterministic Ratings under Linear Delegation}
When restricted to deterministic ratings, the rating design problem for linear-delegation preferences can be transformed to quality maximization as follows.

First, extend $q$ to $[0,\bar\theta]$ by defining $\theta_0=c(q(\lbar\theta))/q(\lbar\theta)$ and setting $q(\theta)=0$ for all $\theta\in[0,\theta_0)$ and $q(\theta)=q(\lbar\theta)$ for all $\theta\in[\theta_0,\lbar\theta)$.
\footnote{
 If $q(\lbar\theta)=0$, define $\theta_0=0$.
 When $\theta_0\leq\lbar\theta$, no types are excluded.
 }
This extension preserves incentive compatibility on $[0,\lbar\theta]$.
By (IC-Env\textsuperscript{D}), $\theta q(\theta) - c(q(\theta)) = \int_{0}^\theta q(x) \d x$.

Next, define $f_{\zeta}(\theta) = \zeta(\theta) f(\theta)$ on $\Theta$, extend both $f_\zeta$ and $\beta$ to be zero outside $\Theta$, and let $F_{\zeta}(\theta) = \int_0^\theta f_{\zeta}(x) \d x$. Substituting (IC-Env\textsuperscript{D}) into the objective, we have
\begin{align}
    \int_{\lbar\theta}^{\bar\theta} v(q(\theta),\theta) f(\theta) \d \theta 
    &= \int_{0}^{\bar\theta} \left[\beta(\theta) - \alpha \theta \right] q(\theta) f_\zeta(\theta) \d \theta + \alpha \int_{0}^{\bar\theta} \left( \int_{0}^{\theta} q(x) \d x \right) \d F_{\zeta}(\theta) \notag \\
    &= \int_{0}^{\bar\theta} \underbrace{[(\beta(\theta) - \alpha \theta)f_{\zeta}(\theta) + \alpha (F_{\zeta}(\bar\theta)-F_{\zeta}(\theta))]}_{:=\phi(\theta)} q(\theta)  \d\theta  
    \label{eqn:LD}
\end{align}
Thus, the problem is equivalent to quality maximization with a \emph{virtual density} $\phi\colon \reals_+ \to \reals_+$, given by
\footnote{
One can also define the \emph{virtual distribution} $\Phi\colon\reals_+\to\reals_+$ by 
\(
{\Phi}(\theta) = \int_{0}^\theta {\phi} (x) \d x = \int_{0}^{\theta} \beta(x)  \d F_{\zeta}(x) + \alpha\theta (F_{\zeta}(\bar\theta)-F_{\zeta}(\theta)).
\)
Note that $F_\zeta(\bar\theta) = \E[\zeta(\theta)]$ and $\Phi(\bar\theta) = \E[\beta(\theta)\zeta(\theta)]$ are constants.
}
\begin{equation*}
    {\phi}(\theta) =  (\beta(\theta) -\alpha\theta) f_{\zeta}(\theta) + \alpha (F_{\zeta}(\bar\theta)-F_{\zeta}(\theta)).
\end{equation*}

\begin{proposition}\label{prop:lineardelegation}
Suppose that Condition~\ref{LD} holds. If $\phi$ is single-peaked, lower censorship is optimal among deterministic ratings.
Moreover,
\begin{itemize} 
    \item if \(\phi\) is increasing, a pass/fail test is also optimal;
    \item if \(\phi\) is decreasing, lower censorship with a threshold $q_0^* \leq q_i(\lbar\theta)$, which every type meets in equilibrium, is optimal.
\end{itemize}

More generally, lower censorship with a minimum standard \(q_0^* = q_i(\theta_0^*)\) is optimal among deterministic rating schemes if there exists some $\theta_0^*\in[0,\bar\theta)$ that satisfies Conditions~\eqref{S0} and \eqref{C0} with \(\phi\) and $\Phi$ replacing \(f\) and $F$, respectively.
\end{proposition}

The virtual density captures both the principal's type-dependent
valuation of quality and her internalization of the agent's cost.
While \(\beta\) and \(\zeta\) directly reweight types, the effect of
the cost-internalization parameter \(\alpha\) also operates through a
tail term. The following example illustrates the comparative statics.

\begin{example}\label{ex:alpha}
Assume that $v(q,\theta)=q-\alpha\frac{c(q)}{\theta}$, where $\alpha \in [0,1)$. Then, \[
\phi(\theta)
=
(1-\alpha)f(\theta)
+
\alpha \left( F_{\zeta}(\bar\theta)-F_{\zeta}(\theta) \right) 
\]
Assume that $f$ satisfies IED on $\Theta$. Then, $\phi$ is single-peaked, so lower censorship is optimal for every $\alpha \in [0,1)$.
\footnote{
    To see this, when $\epsilon_f$ is increasing, 
\(\phi '(\theta)
=
-\frac{(1-\alpha)f(\theta)}{\theta}
\left(
\epsilon_f(\theta)+\frac{\alpha}{1-\alpha}
\right)\) changes sign at most once, from positive to negative.
In fact, under IED, lower censorship is optimal among \emph{all} rating schemes; the proof is in \app*{app:alpha-all-ratings}.
}
When \(\alpha=0\), the principal maximizes expected
quality, and $\phi=f$.  
As \(\alpha\) increases, the principal internalizes more of the
agent's cost, and the optimal threshold $q_0^*$ of the lower censorship decreases.
\footnote{
When there are multiple optimal thresholds, the smallest and largest optimal thresholds both decrease in $\alpha$.
The proof is provided in the appendix using the single-crossing property.
}

As $\alpha\to1$, $\phi\to\int_\theta^{\bar\theta}f(x)/x \d x$, which is decreasing. In the
limit, the
principal's preference aligns with the agent's, so full revelation
is optimal, which can be implemented by lower censorship with a nonbinding threshold \(q_0\leq q_f(\lbar\theta)\).
\end{example}

\subsection{When Are Stochastic Ratings Optimal}
\label{sec:stochastic_generalpref}
The virtual-density reduction above relies on $w=q$ under deterministic
ratings and therefore does not extend directly to stochastic ratings.
Moreover, the feasibility condition (MPS) continues to depend on the
actual density $f$. Consequently, the preceding results do not carry
over simply by replacing the actual density with the virtual density.
The following proposition provides a sufficient condition under which
every optimal rating scheme involves randomization.

\begin{proposition} \label{prop:optimalrating2}
    Under Condition~\ref{LD}, if an optimal deterministic rating scheme induces a quality schedule with a nonempty fully revealing region in which 
    \[
\rho(\theta)  = \frac{[ v_q(q_f(\theta),\theta) \cdot \theta f(\theta)]'}{f(\theta)}
= \beta(\theta) \zeta(\theta) + \theta \frac{\phi'(\theta)}{f(\theta)}
\]
fails to be decreasing, then every optimal rating scheme involves randomization.
 \end{proposition}

For linear-delegation preferences, randomization can be valuable even when $f$ satisfies IED.
For example, suppose that $v(q,\theta)=q/\theta$: the principal maximizes weighted expected quality and places a higher weight on lower types.
This weight $1/\theta$ affects the objective but not the
majorization constraint (MPS).
In this case, $\rho(\theta) =  {f'(\theta)}/{f(\theta)}$,
which is decreasing if and only if $f$ is log-concave.
Hence, every optimal rating scheme involves randomization whenever $f$ is decreasing but fails to be log-concave on a nonempty interval \((\theta_c(\lbar\theta),\bar\theta)\), as is the case for Pareto and concave power distributions.
Intuitively, stochastic ratings can strengthen lower types' incentives by partially pooling them with higher-quality agents.
When the principal places greater weight on lower types, reallocating incentives toward them becomes more valuable.

Finally, although I focus on linear-delegation preferences, the results in this section can be extended to general $v(q,\theta)$.
Theorem~\ref{thm:certificate} in the appendix provides certificates of optimality for candidate rating mechanisms, whether deterministic or stochastic, for general $v$.
The sufficient condition for the optimal rating scheme to involve randomization in Proposition~\ref{prop:optimalrating2} also applies to general $v$.
 
\section{Conclusion}
\label{conclusion}

This paper studies the optimal design of rating schemes to motivate an agent's investment in quality when transfers are unavailable. I reduce the rating design problem to a mechanism design problem subject to a majorization constraint.
This formulation delivers sufficient conditions under which lower censorship or pass/fail tests are optimal among deterministic rating schemes, conditions under which they remain optimal among all rating schemes, and conditions under which every optimal rating scheme involves randomization.
It also provides a useful framework for constructing and certifying optimal rating schemes more generally.

For many common unimodal distributions, simple deterministic ratings maximize expected quality. When intermediate ability is sufficiently scarce relative to low and high ability, however, stochastic ratings can increase expected quality by reallocating incentives from intermediate types toward both ends of the distribution. Thus, optimal rating design concerns not only how much information to disclose, but also how disclosure allocates incentives across types.

\newpage

\counterwithin*{equation}{section}
\renewcommand{\theequation}{\thesection.\arabic{equation}}

\counterwithin*{theorem}{section}
\renewcommand{\thetheorem}{\thesection.\arabic{theorem}}

\counterwithin*{lemma}{section}
\renewcommand{\thelemma}{\thesection.\arabic{lemma}}

\counterwithin*{observation}{section}
\renewcommand{\theobservation}{\thesection.\arabic{observation}}

\counterwithin*{figure}{section}
\renewcommand{\thefigure}{\thesection.\arabic{figure}}

\counterwithin*{example}{section}
\renewcommand{\theexample}{\thesection.\arabic{example}}

\begin{appendices}

\section{Proofs}

\subsection{Proof of Section \ref{model}}

\begin{proof}[Proof of Lemma~\ref{lemma:implementation}]
Necessity follows from the revelation principle and the definition of feasibility, since $w(\theta) = \hat w_{\pi}(q(\theta))$ for all $\theta\in\Theta$.
\footnote{
Sufficiency would follow from the taxation principle if $\hat w_\pi (q)$ were a quality-contingent transfer schedule,
but here \(\hat w_\pi (q)\) must be induced by a rating scheme and the
market's equilibrium beliefs.}

\medskip \noindent
\emph{Sufficiency.}
Suppose that 
\((q(\theta),w(\theta))\) is incentive compatible, individually rational,
and feasible. Incentive compatibility implies that \(q(\theta)\) is
increasing. 
Let $\mathcal Q:=q(\Theta)$, $\bar q:=q(\bar\theta) = \max\mathcal Q$, and $\lbar q: = q(\lbar\theta) = \min\mathcal Q$.

By feasibility, there exist a rating scheme \(\pi^0\) and  associated
Bayes-consistent beliefs and wages such that $w(\theta) = \E_{s\sim\pi^0(\cdot \mid q(\theta))}[\omega^0(s)]$ for all $\theta\in\Theta$.

Because signal labels are payoff irrelevant, it is without loss to relabel
\(\pi^0\) so that \(\varnothing\) is unused on \(\mathcal Q\).
This relabeling preserves all interim wages and Bayes consistency.
Reserve
\(\varnothing\) for nonparticipation and set $\mu_{\varnothing}=\delta_0$ and $\omega(\varnothing)=0$.

Construct a new rating scheme \(\pi^1\) by
\[
    \pi^1(\cdot\mid x)
    =
    \begin{cases}
        \pi^0(\cdot\mid x),
        & x\in\mathcal Q,\\
        \delta_{\varnothing},
        & x<\lbar q,\\
        \pi^0(\cdot\mid\lbar q),
        & x>\lbar q,\quad x\notin\mathcal Q.
    \end{cases}
\]
Because \(q\) is increasing, its range \(\mathcal Q\) is Borel.
Hence, \(\pi^1\) is a Borel Markov kernel.

Because \(\pi^1=\pi^0\) on \(\mathcal Q\), the equilibrium joint
distribution of quality and signals is unchanged, so the associated beliefs and wages remain
Bayes-consistent.
It is straightforward to verify that \(\pi^1\) induces
\[
\hat w^1(x)
=
\begin{cases}
   w(\theta_x),
    & x\in\mathcal Q,\quad \exists \theta_x \text{ such that } q(\theta_x) = x, \\
    0,
    & x<\lbar q,\\
    w(\lbar\theta),
    & x>\lbar q,\quad x\notin\mathcal Q.
\end{cases}
\]
It is then straightforward to verify that incentive compatibility rules out deviations
within \(\mathcal Q\), while the construction of \(\pi^1\) deters
deviations to \(q\notin\mathcal Q\). Hence, \(\pi^1\) implements \((q(\theta),w(\theta))\) in a PBE.
\end{proof}

\begin{proof}[Proof of Lemma~\ref{lem:feasibility}]
Let
\(\tilde\theta\sim F\) so that \(F(\tilde\theta)\sim\unif[0,1]\). Define the quantiles  $\tilde
q(t)=q(F^{-1}(t))$ and $\tilde w(t)=w(F^{-1}(t))$ for $t\in[0,1]$.
By incentive compatibility, \(\tilde q\) and \(\tilde w\)
are increasing. As in
\citet{KleinerMoldovanuStrack2021}, we use versions that are
right-continuous on \([0,1)\) and left-continuous at \(1\).

The condition in the statement is equivalent to
\begin{equation}
    \int_t^1 \tilde w(r)\,dr
    \leq
    \int_t^1 \tilde q(r)\,dr
    \quad\text{for every }t\in[0,1],
    \qquad
    \int_0^1 \tilde w(r)\,dr
    =
    \int_0^1 \tilde q(r)\,dr.
    \label{eq:quantile-majorization}
\end{equation}
Thus, in the notation of
\citet{KleinerMoldovanuStrack2021}, $\tilde w\in\MPS(\tilde q)$.

\paragraph*{Necessity.}
Suppose that a rating scheme \(\pi\) induces
\((q(\theta),w(\theta))\). Conditional on \(\tilde\theta\), let
   $ \tilde s\sim\pi(q(\tilde\theta))$
denote the signal generated on the equilibrium path. By definition,
$\omega(\tilde s) = \E[q(\tilde\theta)\mid\tilde s]$ and $w(\tilde\theta) =
\E[\omega(\tilde s)\mid\tilde\theta]$.
Hence, for every convex function \(\varphi\),  
conditional Jensen's inequality implies
\[
\begin{aligned}
    \E[\varphi(w(\tilde\theta))]
     =
    \E\!\left[
        \varphi\!\left(
            \E[\omega(\tilde s)\mid\tilde\theta]
        \right)
    \right] 
     \leq
    \E[\varphi(\omega(\tilde s))]  
     =
    \E\!\left[
        \varphi\!\left(
            \E[q(\tilde\theta)\mid\tilde s]
        \right)
    \right] 
     \leq
    \E[\varphi(q(\tilde\theta))].
\end{aligned}
\]
Moreover, iterated expectations imply
$\E[w(\tilde\theta)] = \E[\omega(\tilde s)] = \E[q(\tilde\theta)]$.
Therefore,
$w(\tilde\theta) \preceq_{\mathrm{cx}} q(\tilde\theta)$,
where \(\preceq_{\mathrm{cx}}\) denotes the convex order.

Hence, the integrated-quantile characterization of convex order yields $\tilde w \in \MPS(\tilde q)$ after taking \(t=F(\theta)\).

\paragraph*{Sufficiency.}
Suppose that the mechanism is incentive compatible and satisfies
\eqref{eq:quantile-majorization}. By Lemma~\ref{lem:IC},
\(\tilde w\) is constant on every level set of \(\tilde q\).

By \citet[Proposition~1]{KleinerMoldovanuStrack2021}, there exists a
probability measure \(\lambda\), supported on
\(\operatorname{ext}\MPS(\tilde q)\), such that
\(\tilde w=\int h\,\lambda(\d h)\) in \(L^1([0,1])\).
Take every \(h\) in its canonical version, right-continuous on
\([0,1)\) and left-continuous at \(1\). The canonical evaluation map
is Borel, so Fubini's theorem implies that
\[
    m(t):=\int h(t)\,\lambda(\d h)  = \tilde w(t) \quad \mbox{ a.e.}
\]
Because the extreme points
are uniformly bounded, dominated convergence implies that \(m\) has
the same canonical continuity properties as \(\tilde w\). Hence their
almost-everywhere equality implies equality everywhere:
\[
    \tilde w(t)
    =
    \int h(t)\,\lambda(\d h)
    \quad\text{for every }t\in[0,1].
\]

Let \(T:=F(\tilde\theta)\sim\unif[0,1]\), and independently draw
\(H\sim\lambda\). By
\citet[Theorem~1]{KleinerMoldovanuStrack2021}, each extreme point
averages \(\tilde q\) on its pooling intervals and equals \(\tilde q\)
elsewhere. Consequently,
\[
    \E[\tilde q(T)\mid H,H(T)]
    =
    H(T)
    \quad\text{a.s.}
\]

For \(x\in\tilde q([0,1])\), let \(K(x,\cdot)\) be the uniform
distribution on the quantiles \(t\) satisfying \(\tilde q(t)=x\),
interpreted as a point mass when there is only one such quantile.
Set \(K(x,\cdot)=\delta_0\) outside \(\tilde q([0,1])\).
Under the canonical-version convention, \(K\) is a Borel version of
the conditional distribution of \(T\) given \(\tilde q(T)\).

Set \(X:=\tilde q(T)\). Given quality \(x\), independently draw
\(\widehat T\sim K(x,\cdot)\) and \(H\sim\lambda\), and report
\(Y:=H(\widehat T)\).
By the extreme-point characterization and the fact that \(Q\) is an
interval, \(Y\in Q\). For every Borel set
\(B\subseteq Q\), define
\[
    \pi(B\mid x)
    :=
    \int\!\!\int
        \mathbf 1\{h(t)\in B\}\,
        K(x,\d t)\lambda(\d h),
    \qquad
    \pi(S\setminus Q\mid x)=0.
\]
The measurability of \(K\) and of the canonical evaluation map
\((h,t)\mapsto h(t)\) implies that \(\pi\) is a Borel Markov kernel.

On the equilibrium path,
\[
    (X,\widehat T,H,Y)
    \overset{d}{=}
    \bigl(\tilde q(T),T,H,H(T)\bigr),
    \qquad
    \tilde q(\widehat T)=X
    \quad\text{a.s.}
\]
Consequently,
\[
    \E[X\mid H,Y]
     =
    \E[\tilde q(\widehat T)\mid H,H(\widehat T)]
    =
    Y, \quad 
    \E[X\mid Y]
     =
    \E\!\left[\E[X\mid H,Y]\mid Y\right]
    =
    Y, \quad\text{a.s.}
\]

To select market beliefs pointwise, let
\(y\mapsto\mu_y\in\Delta(Q)\) be a Borel regular conditional
distribution of \(X\) given \(Y=y\), and define
\[
    b(y):=\int_Q x\,\mu_y(\d x),
    \qquad
    N:=\{y\in Q:b(y)\neq y\}.
\]
The function \(b\) is Borel, and the preceding conditional-expectation
identity implies that \(N\) is Borel and
\(\Pr(Y\in N)=0\). Define the modified posterior kernel
\[
    \mu_y^*
    :=
    \begin{cases}
        \mu_y, & y\notin N,\\
        \delta_y, & y\in N.
    \end{cases}
\]
Since \(N\) is Borel and \(y\mapsto\delta_y\) is Borel,
\(y\mapsto\mu_y^*\) is a Borel kernel. Because \(\mu^*\) differs from
\(\mu\) only on the \(Y\)-null set \(N\), it remains a regular
conditional distribution and hence is Bayes-consistent. Take
\(\mu^*\) as the market's selected posterior version. It then induces
\[
    \omega(y)
    =
    \int_Q x\,\mu_y^*(\d x)
    =
    y
    \qquad\text{for every }y\in Q.
\]
For the separate nondisclosure signal \(\varnothing\), which is not
generated by the rating scheme, set $\mu_{\varnothing}=\delta_0$ and 
\(\omega(\varnothing)=0\).  

Finally, for \(x\in q(\Theta)\) and any \(t_x\) satisfying
\(\tilde q(t_x)=x\),
\[
\begin{aligned}
    \hat w_\pi(x)
    =
    \int_Q y\,\pi(\d y\mid x) =
    \int\!\!\int h(t)\,K(x,\d t)\lambda(\d h) =
    \int\tilde w(t)\,K(x,\d t)
    =
    \tilde w(t_x).
\end{aligned}
\]
The last equality follows because \(K(x,\cdot)\) is supported on the
level set of \(\tilde q\) corresponding to \(x\), and \(\tilde w\) is
constant there. Taking \(x=q(\theta)\) and \(t_x=F(\theta)\) gives
\[
    \hat w_\pi(q(\theta))=w(\theta)
    \qquad\text{for every }\theta\in\Theta,
\]
proving feasibility.

\end{proof}

\subsection{Proofs of Sections~\ref{deterministic}}

\begin{proof}[Proof of Lemma~\ref{lemma:w=q}]
Fix a type \(\theta\in\T\), let \(q=q(\theta)\), and let \(s\) denote
the observed signal: \(s=\pi(q)\) if the agent participates in the rating
scheme and \(s=\varnothing\) otherwise. Any feasible choice of participation
and quality that generates \(s\) yields the same wage \(\omega(s)\).
Therefore, the agent must choose the lowest quality that generates \(s\);
otherwise, he could obtain the same wage at a strictly lower cost.

Consequently, all types generating \(s\) in equilibrium choose the same
quality \(q\). In particular, if \(s=\varnothing\), then \(q=0\), since
nonparticipation with $q=0$ generates \(\varnothing\).

Since every type generating \(s\)
chooses quality \(q\) in equilibrium, the market's posterior is degenerate at \(q\). Hence
\(
    \omega(s)
    =q.
\)
Therefore, $w(\theta)=\omega(s)=q(\theta)$. 
\end{proof}

\subsubsection{Proof of Proposition~\ref{prop:quality-maximization}}
 
First, I show the following lemma, which shows that single-peakedness of $f$ implies that a subgradient condition and a concavity condition hold at some $\theta_0^*\in[0,\bar\theta)$.
\begin{lemma}\label{lemma:single-peaked}
    If $f(\theta)$ is single-peaked, then there exists some $\theta_0^*\in[0,\bar\theta)$
    that satisfies the following conditions:
    \begin{description}[labelindent=1em]
        \nameditem{S}{S}
        $F(\theta)-F(\theta_0^*) \geq
        (\theta-\theta_0^*)A(\theta_0^*)$
        for all $\theta\in[0,\theta_c(\theta_0^*)]$,
        with equality at $\theta=\theta_c(\theta_0^*)$.

        \nameditem{C}{C}
        $f(\theta)$ is decreasing on
        $[\theta_c(\theta_0^*),\bar\theta]$.
    \end{description}

    Moreover, if $f(\theta)$ is single-peaked at a unique $\theta_m\in(\lbar\theta,\bar\theta]$, then $\theta_0^* \in (\theta_c^{-1}(\theta_m),\theta_m)$ is unique.
    If $f(\theta)$ is decreasing, then $\theta_0^* = \lbar\theta$ satisfies both conditions.
\end{lemma}
 
\begin{remark}
    In particular, $\theta_0^*$ satisfies the following first-order condition:
\begin{equation}
\label{eqn:FOC}
        {A}(\theta_0^*) := \frac{F(\theta_c(\theta_0^*))-F(\theta_0^*)}{\theta_c(\theta_0^*)-\theta_0^*} \leq f(\theta_0^*),\quad \mbox{ and ${A}(\theta_0^*)=f(\theta_0^*)$ if $\theta_0^*\neq \lbar\theta$}.
\end{equation} 
If there exists some $\theta_0^*\geq \theta_c^{-1}(\bar\theta)$ that satisfies Condition~\eqref{S},
then it also satisfies Condition~\eqref{C} vacuously.
\end{remark}

    Lemma \ref{lemma:single-peaked} is straightforward to verify graphically, as a single-peaked $f$ implies a convex-concave $F$.
    The proof is in \supple*{sec:deferredproofs}.
    \footnote{
            A proof when $c(q)=q^2/2$ (and $\theta_c(\theta_0)=2\theta_0$) is in Footnote 19 in \cite{KartikKleinerVanWeelden2021}.
    }
   Now it suffices to show the second part of Proposition~\ref{prop:quality-maximization} in terms of these two conditions.

\begin{proof}[Proof of Proposition~\ref{prop:quality-maximization}]

Suppose that $v(q,\theta)=q$ and that $f$ is single-peaked.

\subsubsection*{Relaxation of the Problem}
Define the candidate solution $q^*\colon [0,\bar\theta]\to Q$, extended to $[0,\bar\theta]$ while preserving incentive compatibility, by
    \begin{equation} \label{eqn:q*}
    q^*(\theta) =
    \begin{cases}
    0   , &\text{if }\theta \in [0, \theta_0^*)\\
    q_i(\theta_0^*) , &\text{if } \theta \in [\theta_0^*, \theta_c(\theta_0^*)]\\
    q_f(\theta)   , &\text{if } \theta \in [\theta_c(\theta_0^*), \bar\theta]
    \end{cases}        
    \end{equation}
Consider a relaxed problem [R] obtained by replacing (IC-Env\textsuperscript{D}) with
\footnote{
The relaxed constraint (IC-Env\textsuperscript{R}) becomes a convex constraint.
}
\begin{equation}
    \theta q(\theta) - c(q(\theta) ) - \left(  \int_{\underline \theta}^\theta q(x) \d x + \lbar\theta q(\underline\theta) - c(q(\underline\theta))\right) \geq 0 \tag{IC-Env\textsuperscript{R}}
\end{equation}
All other constraints and the objective remain unchanged. 
Because \(q^*\) satisfies the original constraint (IC-Env\textsuperscript{D}), if \(q^*\) solves [R], then it also solves \ref{eqn:D}.

Then, I use the optimal control method to show that $q^*$ solves [R].

\subsubsection*{Setup of the Hamiltonian}
Define $U(\theta) =  \int_{\underline \theta}^\theta q(x) \d x + \underline U$.
Rewrite constraints as 
\begin{align}
    & \theta q(\theta) - c(q(\theta) ) - U(\theta) \geq 0 \tag{IC-Env\textsuperscript{R}} \\
    & \dot U = q(\theta),\quad \mbox{$q$ increasing ($\dot q  \geq 0$ if it exists)}\\ 
    & U(\underline  \theta), q(\underline \theta) \geq  0,\quad  U(\bar \theta), q(\bar \theta) \mbox{ free.}
\end{align}  
Define the Hamiltonian and the Lagrangian by
\begin{align}\label{H}
   H   = q(\theta) f(\theta)  
         + \Gamma (\theta) q(\theta) + \mu(\theta) \dot q (\theta),\quad 
   \mathcal{L}   = H + \gamma(\theta)  [\theta q(\theta) - c(q(\theta) ) -U(\theta)],
\end{align}
where $U,q$ are the state variables;
$\Gamma,\mu$ are the costates associated with $U$ and $q$; and $\gamma$ is the Lagrangian multiplier on (IC-Env\textsuperscript{R}). 

For convenience, the displayed equation treats $\dot q$ as a control variable, so that $\frac{\partial L}{\partial \dot q} = \mu$ whenever $\dot q$ exists. 
More generally, however, this formulation does not require the state variable $q$ to be absolutely continuous: the monotonicity dual condition continues to apply even where $\dot q$ fails to exist---for example, at a jump of $q$ (see \citet[Theorem 3.1]{Hellwig2008}; see also Remark~\ref{rmk:Theorem1} and Theorem~\ref{thm:certificate-D} in \supple*{sec:certificate}). 
The maximum principle implies the following conditions:
\footnote{These conditions are ``almost'' necessary conditions which, together with the concavity condition, ensure sufficiency (see \cite[Chapter 6]{SeierstadSydsaeter1987}; see also \citet[Theorem 10.3.1]{LeonardLong1992}). Since necessity is not needed for this proof, the precise necessary conditions are not stated here.}

\subparagraph*{1. Costate and control equations, dual feasibility, and transversality conditions:} 
\allowdisplaybreaks
\begin{align}
    -&\frac{\partial \mathcal{L}}{\partial q} = - (  f + \gamma  (\theta-c'(q))+\Gamma) = \dot \mu \; \mbox{ a.e.},\\
    -&\frac{\partial \mathcal{L}}{\partial U} =  \gamma  = \dot \Gamma \; \mbox{ a.e.},\\
    &\frac{\partial \mathcal{L}}{\partial {\dot q} } =   \mu \leq  0,\quad 
    \mbox{ $\mu(\theta)=0$ if $q$ is strictly increasing at $\theta$},\footnotemark
    \tag{IC-Mon: dual}\\
    &  \gamma(\theta) \geq 0, \quad  \gamma(\theta)  [\theta q - c(q) - U] =0, \tag{IC-Env\textsuperscript{R}: dual}\\
    &  \Gamma(\lbar \theta)  \leq 0,  \quad \Gamma(\lbar \theta)U(\lbar \theta) =0; \quad \mu(\lbar \theta)\leq 0,  \quad \mu(\lbar \theta) q(\lbar \theta)=0,  \\
    &  \Gamma(\bar \theta)   = 0, \quad \mu(\bar\theta)=0.
\end{align}
\footnotetext{
    An increasing function $q$ is \emph{strictly} increasing at $\theta$ if $q(\theta+\varepsilon)-q(\theta-\varepsilon)>0$ for all $\varepsilon>0$.
     }

\subparagraph*{2. Jumps in costates at junctions.} 
Because (IC-Env\textsuperscript{R}) is a pure state constraint involving $(U,q)$, their costates $\Gamma$ and $\mu$ can be discontinuous at junctions between intervals on which it is binding and not binding.
At a junction point $t_j$,  
\begin{align}
    & \Gamma(t_j{+}) - \Gamma(t_j{-}) = \chi (t_j) \geq0, \quad   [t_j q(t_j) - c(q(t_j) ) -U(t_j)] \chi (t_j)=0, \\
    & \mu(t_j{+}) - \mu(t_j{-}) = (c'(q(t_j)) - t_j) \chi (t_j) .\end{align}
If $t_j = \bar\theta$, the right limits are understood as
the terminal values at $\bar\theta$.

\subparagraph*{3. Jumps in the state variable.}
At a jump $t_0\in(0,\bar\theta)$ of the state variable $q$, 
\footnote{
See \citet[Theorem 3.7, pp.~196–7]{SeierstadSydsaeter1987}.
In the current formulation, this condition is already implied by (IC-Mon: dual) and condition 2.
If $\theta_0\in\T$, a necessary condition is
\begin{equation}
     H(t_0{+}) - H(t_0{-})  =  q_i(t_0) f(t_0) + \Gamma(t_0{+})q_i(t_0)
    \begin{cases}
        =0, &\text{if }t_0>\lbar\theta\\
        \geq 0, &\text{if }t_0=\lbar\theta 
    \end{cases}
    \label{eqn:H}
\end{equation}
which can be obtained from the first-order condition for $t_0$ provided that $f$ is continuous. }
\begin{align}
    \Gamma(t_0{+}) = \Gamma(t_0-) \; \mbox{ (by continuity of $U$)}, \qquad 
    \mu(t_0{+})  = \mu(t_0-).
\end{align}
\subsubsection*{Construction of the multipliers}
I propose the following multipliers.
    \begin{equation*}\label{Gamma} 
        \Gamma (\theta) = 
        \begin{cases}
           - A(\theta_0^*) ,  &\text{if } \theta\in [0,{\theta_{c}}(\theta_0^*))\\
           - f(\theta), &\text{if }  \theta\in [{\theta_{c}}(\theta_0^*),\bar\theta)  \\
            0, &\text{if } \theta=\bar\theta  
        \end{cases}
        \qquad
                \gamma(\theta) = \dot\Gamma(\theta) =     
        \begin{cases}
           0 ,  &\text{if } \theta\in [0,{\theta_{c}}(\theta_0^*))\\
           - f'(\theta), &\text{if }  \theta\in [{\theta_{c}}(\theta_0^*),\bar\theta)  
        \end{cases}
    \end{equation*}
        \begin{equation*} 
            \mu (\theta) =  
            \begin{cases}
             -  (F(\theta)-F(\theta_0^*)) + (\theta-\theta_0^*)  {A}(\theta_0^*) , &\text{if }  \theta\in [0,  {\theta_{c}}(\theta_0^*)) \\
                0, &\text{if } \theta \in [{\theta_{c}}(\theta_0^*),\bar \theta]
            \end{cases}
        \end{equation*}
        and define $\Gamma(\bar\theta) = \mu(\bar\theta)=0$ if $\theta_c(\theta_0)>\bar\theta$. By construction, the multipliers satisfy
        \[
          \dot \mu(\theta) = - (  f(\theta) + \gamma(\theta)  (\theta-c'(q^*(\theta)))+\Gamma(\theta)) 
        \]
        Condition \eqref{S} implies that, if $\theta_c(\theta_0^*)\leq \bar\theta$, then $\mu(\theta_c(\theta_0)-)=0$, so that $\mu$ is continuous at $\theta_c(\theta_0^*)$.
        On the other hand, $\Gamma$ is discontinuous at $\theta_c(\theta_0^*)$ and $\bar\theta$.
        
        In particular, in the fully revealing region where $q(\theta)=q_f(\theta)$, because $c'(q_f(\theta)) =\theta$ (and $\dot q_f(\theta)>0$), the multipliers $\mu=0$ and $\Gamma(\theta) = -{v}_q(q_f(\theta),\theta) f(\theta)$ are constructed to satisfy the necessary condition.

    \subsubsection*{Applying Sufficiency Theorem}
    To apply Theorem 6.2 in \cite[pp.~361--362]{SeierstadSydsaeter1987},  
    I now verify conditions 1--3, as well as the concavity condition for sufficiency.

   \emph{1.} Costate and control equations and transversality conditions hold by construction of the multipliers.
       In particular,  the transversality conditions at $\lbar\theta$ follow from $\Gamma(\lbar\theta)=-A(\theta_0^*)\leq0$ and $U^*(\lbar\theta)=0$;  and $\mu(\lbar\theta)\leq0$ (by condition~\eqref{S}) and $q^*(\lbar\theta)=0$ if $\mu(\lbar\theta)<0$ (which implies $\theta_0^*>\lbar\theta$).

    \emph{1a. (IC-Mon: dual).}
    On $[0,\theta_c(\theta_0^*)]$, condition \eqref{S} implies $\mu(\theta)\leq 0$.
    On $[\theta_c(\theta_0^*),\bar\theta]$, $\mu(\theta)=0$ by definition.
    In addition, $q$ is strictly increasing only if $\theta=\theta_0^*$ or $\theta\in[\theta_c(\theta_0^*),\bar\theta]$ (and $\theta_c(\theta_0^*)<\bar\theta$).
    At $\theta=\theta_0^*$, we have $\mu(\theta_0^*)=0$ by the definition of $A(\theta_0^*)$.

    \emph{1b. (IC-Env\textsuperscript{R}: dual).}
    In pooling regions, $\gamma=0$.
    In the fully revealing region, by condition~\eqref{C}, $f$ is decreasing, so $\gamma = -f'(\theta)\geq0$.

    \emph{2. Jumps in costates.}
   Due to pure state constraint (IC-Env\textsuperscript{R}),  the costates $\Gamma$ and $\mu$ can be discontinuous at $\bar\theta$ and $\theta_c(\theta)$.

   \emph{2.1: Jump at $\bar\theta$.} First, at $\bar\theta$, the jump of $\Gamma(\theta)$ is nonnegative.
        If $\theta_c(\theta_0^*)<\bar\theta$, we have
       \begin{align*}
        \Gamma(\bar\theta) - \Gamma(\bar\theta-)   =  f(\bar\theta)= \chi (\bar\theta) 
        \geq0,\qquad 
        \mu(\bar\theta) - \mu(\bar\theta{-})  = ( c'(q(\bar\theta) ) -\bar\theta )\chi (\bar\theta)  = 0,
       \end{align*}
        because  $c'(q^*(\theta) )= \theta$ on the fully revealing region $[\theta_c(\theta_0^*),\bar\theta]$.
Otherwise, if $\theta_c(\theta_0^*)\geq\bar\theta$, we have
\begin{align*}
   \Gamma(\bar\theta) - \Gamma(\bar\theta-) & = A(\theta_0^*) =\chi (\bar\theta)  \geq0,\\ 
   \mu(\bar\theta) - \mu(\bar\theta{-})&=
     (1-F(\theta_0^*)) - (\bar\theta-\theta_0^*)  {A}(\theta_0^*)
   = (\theta_c(\theta_0^*)-\bar\theta)  A(\theta_0^*)  = (c'(q(\bar\theta) )-\bar\theta)  \chi (\bar\theta) 
\end{align*}
   because $\theta_c(\theta_0^*)\geq\bar\theta$ implies 
   $F(\theta_c(\theta_0^*))=1$ and
   $q(\bar\theta) = q_i(\theta_0^*)$.
    The complementary-slackness condition 
 holds because (IC-Env\textsuperscript{R}) binds.

     \emph{2.2: Jump at  $\theta_c(\theta_0^*)$.}
     Assume without loss that $\theta_c(\theta_0^*)<\bar\theta$; otherwise, this junction point is handled in 2.1.
              By condition~\eqref{S}, 
            \[
            \chi(\theta_c(\theta_0^*)) = \Gamma(\theta_c(\theta_0^*)+) -  \Gamma(\theta_c(\theta_0^*)-)
            = A(\theta_0^*) - f(\theta_c(\theta_0^*))
            \geq0
            \]
            Because $\mu$ is continuous at $\theta_c(\theta_0^*)$ and $c'(q(\theta_c(\theta_0^*)) )= \theta_c(\theta_0^*)$, we have
            \[
            \mu(\theta_c(\theta_0^*)+) -  \mu(\theta_c(\theta_0^*)-) =  [  c'(q(\theta_c(\theta_0^*)))-\theta_c(\theta_0^*)] \chi (\theta_c(\theta_0^*))  = 0.
            \]
Again, the  complementary-slackness condition holds because (IC-Env\textsuperscript{R}) binds.

    \emph{3. Jump in state variable.}
    At $\theta_0^*$ where $q^*$ is discontinuous, we have $\Gamma(\theta_0^*-) = \Gamma(\theta_0^*+) = -A(\theta_0^*)$, and condition~\eqref{S} (and equation~\eqref{eqn:FOC}) implies Condition~\eqref{eqn:H}.

\emph{4. Concavity.}
The pure state constraint $\theta q - c(q)-U\geq 0$ is convex since its left-hand side is concave in $(q,U)$.
Moreover, $H$ is affine and hence concave in the state and control variables $(q,U,{\dot q})$.
Hence, Theorems 6.2 (mixed and pure state constraints) and 3.8 (jumps in state variables) in \citet[pp.~361--362, 198--199]{SeierstadSydsaeter1987} imply that $q^*$ is optimal. 
\end{proof} 
\begin{remark}\label{rmk:Theorem1}
Technically, the sufficiency theorems in \cite{SeierstadSydsaeter1987} require the state variable $q$ to be piecewise smooth with finitely many jump discontinuities.
\footnote{
\cite{Hellwig2008} only provides necessary conditions and does not allow for pure state constraints.
}
\cite{Xiao2023a} (see Theorem~\ref{thm:certificate-D} in \supple*{sec:certificate}) shows that these conditions remain sufficient when $q$ is an arbitrary right-continuous, increasing function of bounded variation.
Further, this result does not involve relaxing the state equality constraint (IC-Env\textsuperscript{D}) to an inequality.
It also applies to general $v(q,\theta)$ that need not satisfy Condition~\ref{LD}.
\end{remark}

\allowdisplaybreaks

\subsection{Proofs of Section~\ref{general}}
\allowdisplaybreaks
\label{Hamiltonian:general}

\begin{proof}[Proof of Lemma~\ref{lem:existence}]
The admissible set is nonempty because $q=w=0$ is admissible. Since $v$ is
bounded on the compact set $Q\times\Theta$, the supremum $V^*$ is finite.
Choose an admissible sequence $(q_n,w_n)$ whose objective values converge
to $V^*$.
Both $q_n$ and $w_n$ are increasing by Lemma~\ref{lem:IC} and $Q$-valued by definition.

By Helly's selection theorem, there is a common
subsequence converging at every \(\theta\in\T\) to increasing,
\(Q\)-valued functions \(q\) and \(w\). In particular,
by continuity of $c$,
\(
    \underline U_n
    =
    \underline\theta w_n(\underline\theta)
    -
    c(q_n(\underline\theta))
    \to
    \underline U.
\)
Pointwise convergence and continuity of $c$ preserve the pointwise
terms in \textup{(IR)} and \textup{(IC-Env)}, while bounded convergence preserves the integral terms in
\textup{(IC-Env)} and \textup{(MPS)}. Hence, $(q,w)$ is admissible.

Finally, continuity of $v$ on $Q\times\Theta$ and dominated convergence give
\[
    \int_{\Theta}v(q_n(\theta),\theta)\d F(\theta)
    \to
    \int_{\Theta}v(q(\theta),\theta)\d F(\theta).
\]
Thus, $(q,w)$ attains $V^*$.
\end{proof}

\subsubsection{Setting up the Optimal Control Problem}

Define $D(\theta) = \int_{\underline \theta}^{ \theta} (w(x) -  q(x))\d F(x)$  and $U(\theta) =  \int_{\underline \theta}^\theta w(x) {\d x} + \underline U$.
Rewrite the constraints in \ref{prob:P} as
\allowdisplaybreaks
\begin{align}
    & \theta w(\theta)-c( q(\theta) ) -U(\theta)=0 \\
    & D(\theta)\geq 0  \;  \mbox{ (MPS)}, \quad \dot D =  ( w(\theta)-q(\theta) ) f(\theta)\\
    & \dot U = w(\theta), \quad \mbox{$U, D$ absolutely continuous},\\
    & \mbox{$w$ increasing}\\
    & U(\underline  \theta), w(\underline \theta) \geq  0, \quad D(\underline\theta)=D(\bar \theta) = 0,\\
    & G(U(\bar\theta),w(\bar\theta))\equiv \bar\theta w(\bar \theta) -c(w(\bar\theta)) - U(\bar\theta)\geq0.
\end{align}  
The final boundary condition follows because $D(\theta)\geq0$ (MPS) implies $q(\bar\theta)\geq w(\bar\theta)$ and hence $\bar\theta w(\bar \theta) -c(w(\bar\theta)) - U(\bar\theta)\geq0$.

    Define the Hamiltonian and the Lagrangian by
\begin{align}
    H &=  {v}(q(\theta), \theta) f(\theta) 
        + \Lambda(\theta) [ w(\theta)-q(\theta)] f(\theta) 
        +\Gamma (\theta) w(\theta) + \mu(\theta) {\dot w}(\theta) \\
    \mathcal{L} &= H + \gamma(\theta)  [\theta w(\theta)-c( q(\theta) ) -U(\theta)] 
    + \lambda(\theta) D(\theta) 
\end{align}
where $D,U,w$ are state variables; $q $ is the control variable;
$\Lambda,\Gamma,\mu$ are the costates associated with $D, U, w$, respectively;
$\gamma,\lambda$ are the Lagrangian multipliers on (IC-Env) and (MPS), respectively.

Slightly abusing notation, the exposition above uses $\dot w$ as a control variable and represents the pure state constraint (MPS) via a Lagrangian term $\lambda D$, which implies $\frac{\partial \mathcal{L}}{\partial \dot w} = \mu$ and $\dot\Lambda = -\frac{\partial\mathcal{L}}{\partial D} = -\lambda$ {whenever $\dot w$ and $\dot \Lambda$ exist}.
The sufficiency theorem below, however, dispenses with $\dot w$ and $\lambda$ altogether; hence, it does \emph{not} require $w$ or $\Lambda$ to be absolutely continuous.

\paragraph*{Sufficient conditions.} 
The following theorem gives sufficient conditions for optimality.
Notably, the theorem does not use $\dot w$ as a control variable, and instead allows $w$ to be an arbitrary right-continuous, increasing function of bounded variation.
\footnote{
    In this sense, this theorem complements the necessity result in \citet[Theorem~3.1]{Hellwig2008} by establishing sufficiency and by incorporating a pure state constraint.
}
It also extends the sufficiency theorems of \citet[Theorems~6.2 and~6.14, pp.~361--362 and 393--394]{SeierstadSydsaeter1987} to optimal control problems with an equality constraint (IC-Env),
\footnote{
Their method would require representing the equality as two inequality constraints, which would in turn require both $\theta w-c(q)-U$ and $-(\theta w-c(q)-U)$ to be quasiconcave---an impossibility.
} and relaxes the terminal transversality condition of \citet[Theorem~10.3.2]{LeonardLong1992}.
Additionally, it incorporates the pure state constraint (MPS) without imposing further assumptions on the endogenous costate $\Lambda$, allowing an arbitrary right-continuous function of bounded variation.

\begin{theorem}[Certificate of Optimality]
  \label{thm:certificate}
  Suppose that $(q^*,w^*)$ satisfies all constraints of
  problem~\ref{prob:P}.
  Then, $(q^*,w^*)$ is a solution to problem~\ref{prob:P} if there exist absolutely continuous functions
  $\Gamma$ and $\mu$, a right-continuous
function $\Lambda$ of bounded variation, 
and a measurable function $\gamma$,
satisfying the following conditions.
  \begin{enumerate}
    \item Costate and control equations holds almost everywhere:
    \begin{align}
    -&\frac{\partial \mathcal{L}}{\partial U} =  \gamma = \dot \Gamma  \\
    -&\frac{\partial \mathcal{L}}{\partial w} = -(\theta \gamma + \Lambda f + \Gamma) =\dot\mu  \\
    &\frac{\partial \mathcal{L}}{\partial q} =  (v_q(q^*,\theta) - \Lambda)  f -\gamma c'(q^*)  = 0 
\end{align} 
\item Dual feasibility conditions and complementary slackness for (IC-Mon) and (MPS): 
\begin{align*}
      &  \mu \leq  0,\quad \int_{(\lbar \theta, \bar \theta]} \mu \d w^*=0\\
      & \Lambda \text{ is decreasing}, \quad  \int_{(\lbar \theta, \bar \theta]} D^* \d \Lambda =0.\footnotemark
\end{align*}
\footnotetext{
Equivalently, the nonnegative measure $-\d\Lambda$ is supported on
$\{\theta:D^*(\theta)=0\}$.  
If $\Lambda$ is absolutely continuous, this is equivalent to $-\dot \Lambda = \frac{\partial \mathcal{L}}{\partial D} = \lambda \geq0$ and $\lambda(\theta) D^*(\theta)=0$ a.e.  
}
\item Transversality conditions:
\begin{align*}
      &  \Gamma(\lbar \theta)  \leq 0,  \quad \Gamma(\lbar \theta)U^*(\lbar \theta) =0,     \\
    &  \Gamma(\bar \theta) [U(\bar\theta)-U^*(\bar \theta)]
    + \mu(\bar\theta) [w(\bar\theta)-w^*(\bar \theta)]  \geq 0 \mbox{ for all feasible } U(\bar\theta), w(\bar\theta),   \\
    & \mu(\lbar \theta)\leq 0,\quad \mu(\lbar \theta) w^*(\lbar \theta)=0.
\end{align*}
\item Concavity: 
define $
\kappa(\theta)= \inf_{q\in (0,q_{\max}]} \left\{ \frac{-v_{qq}(q,\theta)}{c''(q)} \right\}$; then
\[
 \Gamma(\theta) + \int_{\lbar\theta}^{\theta} \kappa(s) f(s) \d s \text{ is increasing}.
\]
\end{enumerate}
\end{theorem}
\begin{remark}
Since every admissible mechanism satisfies $G(U(\bar\theta),w(\bar\theta))\geq0$
and $G$ is concave in $(U,w)$,  
a convenient sufficient condition for the upper-endpoint transversality condition is the
following: 
 there exists $\eta({\bar\theta})\geq0$ such that
\begin{equation} \label{eqn:trans:barUw}
    \begin{aligned}
    &\bigl(\Gamma(\bar\theta),\mu(\bar\theta)\bigr)
     =\eta({\bar\theta}) \nabla G\bigl(U^*(\bar\theta),w^*(\bar\theta)\bigr)
     =\eta({\bar\theta})  \left( -1,\bar\theta-c'(w^*(\bar\theta))\right) ,\\
    &\eta(\bar\theta)\geq0, \quad \eta({\bar\theta}) G\bigl(U^*(\bar\theta),w^*(\bar\theta)\bigr) = 0. 
    \end{aligned}
\end{equation}
\end{remark}

\begin{proof}[Proof of Theorem~\ref{thm:certificate}]
Fix any admissible $(q,w)$ that satisfies the constraints.
Write $\Delta z=z-z^*$ for each state $(w,U,D)$ or control $q$.  Define
\[
    \mathcal J(q,w,U)
    \equiv v(q,\theta) f+\Lambda(w-q)f+\Gamma w
      +\gamma\bigl[\theta w-c(q)-U\bigr].
\]
Condition 3 and $\dot\Gamma=\gamma$ imply $\gamma(\theta)+\kappa(\theta)f(\theta)\geq0$.
Moreover, by the definition of \(\kappa(\theta)\),
$v_{qq}(q,\theta)+\kappa(\theta)c''(q)\leq0$ for all \(q\in Q\).
Therefore,
\[
\mathcal J_{qq} 
 =v_{qq}(q,\theta) f(\theta)-\gamma(\theta)c''(q)
 =
\left[v_{qq}(q,\theta)+\kappa(\theta)c''(q)\right] f(\theta)-
\left[\gamma(\theta)+\kappa(\theta)f(\theta)\right]c''(q)
\le0.
\]
Thus, for each \(\theta\in\Theta\), \(\mathcal J\) is concave in \(q\) and affine in \(U\) and $w$. 
Consequently, concavity and the costate and control equations in condition 1 imply
\begin{align*}
    \mathcal J-\mathcal J^*
      \leq
      (\Lambda f+\Gamma+\theta\gamma)\Delta w
      -\gamma\Delta U
      +\left[(v_q(q^*,\theta)-\Lambda)f-\gamma c'(q^*)\right]\Delta q =-\dot\mu\Delta w-\dot\Gamma\Delta U.
\end{align*}
Therefore, by absolute continuity of $\mu$ and $\Gamma$,
\[
    \int_{\lbar\theta}^{\bar\theta}\left(\mathcal J - \mathcal J^*\right) \d\theta
    \leq
    - \int_{(\lbar\theta,\bar\theta]}\dot\mu\Delta w\d\theta
    - \int_{(\lbar\theta,\bar\theta]}\dot\Gamma\Delta U\d\theta
    =- \int_{(\lbar\theta,\bar\theta]} \Delta w\d\mu
    - \int_{(\lbar\theta,\bar\theta]}\Delta U\d\Gamma
\]

By the definition of $\mathcal J$ and $\theta w - c(q) - U = 0$ (IC-Env), and because $\d(\Delta D) = (\Delta w - \Delta q) f \d\theta $ and $\d(\Delta U) = \Delta w\d\theta$, we have
\[
   \Delta V := \int_{\lbar\theta}^{\bar\theta}\left(v(q,\theta)-v(q^*,\theta)\right)f\d\theta
    =\int_{\lbar\theta}^{\bar\theta}
       (\mathcal J-\mathcal J^*)\d\theta
      -\int_{(\lbar\theta,\bar\theta]}\Lambda\d(\Delta D)
      -\int_{(\lbar\theta,\bar\theta]}\Gamma\d(\Delta U).
\]
Combining these two equations and applying Stieltjes integration by parts yields
\begin{align*}
   \Delta V
    \leq 
      -\bigl[\Gamma\Delta U\bigr]_{\lbar\theta}^{\bar\theta}
      -\bigl[\mu\Delta w\bigr]_{\lbar\theta}^{\bar\theta}
      -\bigl[\Lambda\Delta D\bigr]_{\lbar\theta}^{\bar\theta} 
     +\int_{(\lbar\theta,\bar\theta]}\mu\d(\Delta w)
      +\int_{(\lbar\theta,\bar\theta]}\Delta D\d\Lambda.
\end{align*}
This integration-by-parts formula is valid for bounded-variation
functions and therefore allows the state variable $w$ and the
costate $\Lambda$ to have jumps.

Every term on the right-hand side is nonpositive. Because $D(\lbar\theta)=D(\bar\theta)=0$ for any feasible $D$, we have $ [\Lambda\Delta
D ]_{\lbar\theta}^{\bar\theta}=0$. 
The
transversality and complementary-slackness conditions give
\[
    -\bigl[\Gamma\Delta U\bigr]_{\lbar\theta}^{\bar\theta}   
    -\bigl[\mu\Delta w\bigr]_{\lbar\theta}^{\bar\theta}
      =\Gamma(\lbar\theta)U(\lbar\theta)  + \mu(\lbar\theta)w(\lbar\theta)
      - \left[  \Gamma(\bar\theta)\Delta U(\bar\theta) + \mu(\bar\theta)\Delta w(\bar\theta)   \right]\leq0.
\]
Moreover, monotonicity of $w$ and condition 1 imply
\[
    \int_{(\lbar\theta,\bar\theta]}\mu\d(\Delta w)
      =\int_{(\lbar\theta,\bar\theta]}\mu\d w-\int_{(\lbar\theta,\bar\theta]}\mu\d w^* = \int_{(\lbar\theta,\bar\theta]}\mu\d w\leq0.
\]
Finally, $D\geq0$, nonincreasing $\Lambda$, and condition 2 imply
\[
    \int_{(\lbar\theta,\bar\theta]}\Delta D\d\Lambda
      =\int_{(\lbar\theta,\bar\theta]} D\d\Lambda-\int_{(\lbar\theta,\bar\theta]} D^*\d\Lambda=\int_{(\lbar\theta,\bar\theta]} D\d\Lambda\leq0.
\]
Thus, 
\(
 \Delta V \leq 0.
\)
Hence, $(q^*,w^*)$ is optimal.
\end{proof}

\paragraph*{Necessary conditions.} 

Let $I$ be an interval on which the optimal mechanism satisfies $q(\theta) = w(\theta) = q_f(\theta)$.
Then, the maximum principle in \citet[Theorem~3.1]{BocciaDePinhoVinter2016} (see also \citet[Theorem 10.3.1]{LeonardLong1992}) implies that the costate and control equations in condition~1, as well as condition 2, of Theorem~\ref{thm:certificate} are also necessary.
\footnote{
    After substituting $q = c^{-1}(\theta w - U)$ to eliminate the mixed equality (IC-Env), the constraint qualifications for $D\geq 0$ and $\dot w\geq0$ are straightforward to verify.
    This elimination does not change the necessary conditions, after substituting $\gamma = \frac{(v_q(q^*,\theta)-\Lambda)f}{c'(q^*)}$ if $q^* > 0$ and $\gamma = 0$ if $q^* = 0$,
    so that $\frac{\partial \mathcal L}{\partial q} = 0$.
}
Because $c'(q_f) = \theta$ and $q_f$ is strictly increasing, the necessary conditions imply $\mu=0$ on $I$ and thus

\begin{align*} 
    -\dot \mu &=  {v}_q(q_f(\theta),\theta) f(\theta)  + \Gamma(\theta)  = 0 \\
    -\Lambda &= [\theta \Gamma(\theta)]'/f(\theta).
\end{align*}
It follows that $\Gamma(\theta) = -{v}_q(q_f(\theta),\theta) f(\theta)$ and  
    \[ \Lambda(\theta) 
    =  \frac{[ v_q(q_f(\theta),\theta) \cdot \theta f(\theta)]'}{f(\theta)}
    \equiv\rho(\theta).
    \]
Dual feasibility for (MPS) then requires $\Lambda(\theta)$ to be decreasing on $I$.

\begin{lemma}[Necessary Condition in Fully Revealing Regions] \label{lem:rho}
    If $(q^*,w^*)$ is optimal, then $\rho$ is decreasing on any interval $I$ where $q^*(\theta) = w^*(\theta) = q_f(\theta)$.
    \footnote{
       An alternative proof of Lemma~\ref{lem:rho} based on perturbations is in \supple*{sec:deferredproofs}.
    }
\end{lemma}

\subsubsection{Proof of Proposition~\ref{prop:optimalstochastic}}
\begin{proof}[Proof of Proposition~\ref{prop:optimalstochastic}]

    Suppose that $f$ is single-peaked at $\theta_m$ and $\epsilon_f=-\frac{\theta f'(\theta)}{f(\theta)}$ is increasing on $[\theta_m,\bar\theta]$.
By Lemma~\ref{lemma:single-peaked}, there exists some $\theta_0^*\in[0,\theta_m]$ that satisfies Conditions~\eqref{S} and \eqref{C}.
     Consider the candidate $(q^*,w^*)$, where $q^*=w^*$ are given by equation~\eqref{eqn:q*},
    which satisfies all constraints in \ref{prob:P}.
    Thus, it suffices to construct multipliers and use Theorem~\ref{thm:certificate} to certify optimality.

\subsubsection*{Construction of the multipliers}

For
\(\theta\in[\theta_0^*,\theta_c(\theta_0^*)]\), define the average density between \(\theta\) and \(\theta_c(\theta_0^*)\) by
\begin{equation}
       R(\theta)=     \frac{
    F(\theta_c(\theta_0^*))-F(\theta)
    }{
        \theta_c(\theta_0^*)-\theta
    }, 
    \quad
R(\theta_c(\theta_0^*))
=
f(\theta_c(\theta_0^*)).
\end{equation}
Let $\tau=F(\theta)$, $\tau_0^*=F(\theta_0^*)$, and $\tau_c^*=F(\theta_c(\theta_0^*))$.
Define \(K\) on \([\tau_0^*,\tau_c^*]\) by
\begin{align}
K(F(\theta))
=
\theta R(\theta)
-
\theta_0^*A(\theta_0^*) =
F(\theta)-F(\theta_0^*)
+
\theta_c(\theta_0^*)
\left[
R(\theta)-A(\theta_0^*)
\right].
\label{eq:K-multiplier}
\end{align}
Let \(\widehat K=\operatorname{cav} K\) be the concavification of \(K\), with $\widehat K(\tau_0^*)=K(\tau_0^*)$ and $\widehat K(\tau_c^*)=K(\tau_c^*)$.

\subparagraph*{The costate for $D$.}
Define the right-continuous function
\begin{equation}
    \Lambda(\theta)=
    \begin{cases}
        1,
        &\text{if }\theta\in[0,\theta_0^*),
        \\ 
        \widehat K'_+(F(\theta)),
        &\text{if }\theta\in
        [\theta_0^*,\theta_c(\theta_0^*)),
        \\ 
        1-\epsilon_f(\theta),
        &\text{if }\theta\in
        [\theta_c(\theta_0^*),\bar\theta),
    \end{cases}
    \label{eq:Lambda-multiplier}
\end{equation}
where $\widehat K'_+$ denotes the right-derivative, and $\epsilon_f(\theta) = -\theta\frac{f'(\theta)}{f(\theta)}$.
Define \(\Lambda(\bar\theta)=\Lambda(\bar\theta-).\)

At the lower boundary $\theta_0^*$, 
Condition~\eqref{S} and equality at
\(\theta_c(\theta_0^*)\) gives \(R(\theta)\leq A(\theta_0^*)\) for all \(\theta\in[\theta_0^*,\theta_c(\theta_0^*)]\).
Hence, by the definition of \(K\) in \eqref{eq:K-multiplier},
\[
K(F(\theta))\leq F(\theta)-F(\theta_0^*).
\]
Therefore,   the affine function \(\tau\mapsto\tau-\tau_0^*\) is a concave majorant of \(K\).
Since \(\widehat K\) is the least concave
majorant and \(\widehat K(\tau_0^*)=K(\tau_0^*)=0\),  
\[
    \widehat K(\tau) -  \widehat K(\tau_0^*) \leq \tau-\tau_0^*, \quad \text{for all }\tau\in[\tau_0^*,\tau_c^*].
\]
Taking the right derivative at \(\tau_0^*\) gives
$\widehat K'_+(\tau_0^*)\leq1.$
Thus, $\Lambda$ jumps (weakly) downward at $\theta_0^*$.
Moreover, concavity of \(\widehat K\) implies that
\(\widehat K'_+\) is decreasing and therefore
$\Lambda(\theta)\leq1$ on 
$[\theta_0^*,\theta_c(\theta_0^*))$.

If $\theta_c(\theta_0^*)\geq \bar\theta$, then $\Lambda$ is continuous at $\bar\theta$ by definition. 
Now assume $\theta_c(\theta_0^*)<\bar\theta$ and consider the jump of $\Lambda$ at $\theta_c(\theta_0^*)$.
Since
\(
\theta_0^*
\leq
\theta_m
\leq
\theta_c(\theta_0^*),
\)
single-peakedness and IED imply
\[
\epsilon_f(\theta)
\leq
\epsilon_f(\theta_c(\theta_0^*))
\qquad
\text{for every }
\theta\in[\theta_0^*,\theta_c(\theta_0^*)].
\]
Indeed,  on the increasing part of \(f\), \(\epsilon_f(\theta)\leq0\leq \epsilon_f(\theta_c(\theta_0^*))\); on the decreasing part,
the inequality follows from IED.

For every
\(\theta\in[\theta_0^*,\theta_c(\theta_0^*))\), integration by parts gives
\begin{align}
R(\theta)-f(\theta_c(\theta_0^*))
&=
\frac{1}{
\theta_c(\theta_0^*)-\theta
}
\int_\theta^{\theta_c(\theta_0^*)}
\left[
f(x)-f(\theta_c(\theta_0^*))
\right]\d x \notag\\
&=
\frac{1}{
\theta_c(\theta_0^*)-\theta
}
\int_\theta^{\theta_c(\theta_0^*)} (x-\theta)
\frac{\epsilon_f(x)}{x}f(x)
 \d x  \; \mbox{(because \(-f'(x)=\frac{\epsilon_f(x)}{x} f(x)\))}\notag\\
&\leq
\frac{
\epsilon_f(\theta_c(\theta_0^*))
}{
\theta_c(\theta_0^*)
}
\left[
F(\theta_c(\theta_0^*))-F(\theta)
\right] \; \mbox{(because \( {x-\theta}
\leq
\frac{x[\theta_c(\theta_0^*)-\theta]}{\theta_c(\theta_0^*)} \))}.
\label{eq:direct-average-density-bound}
\end{align}

It follows from \eqref{eq:K-multiplier} that
\begin{align}
\frac{
K(\tau_c^*)-K(F(\theta))
}{
\tau_c^*-F(\theta)
}
=
1-
\theta_c(\theta_0^*)
\frac{
R(\theta)-f(\theta_c(\theta_0^*))
}{
F(\theta_c(\theta_0^*))-F(\theta)
}
 \geq
1-\epsilon_f(\theta_c(\theta_0^*)).
\label{eq:direct-terminal-chord}
\end{align}
Therefore, the affine function
\(
\tau
\mapsto
\left(
1-\epsilon_f(\theta_c(\theta_0^*))
\right)(\tau-\tau_c^*) + K(\tau_c^*)
\)
is a concave majorant of \(K\). Since \(\widehat K\) is the least concave
majorant and \(K(\tau_c^*) = \widehat K(\tau_c^*)\), we have
\[
    \widehat K(\tau)-\widehat K(\tau_c^*)
    \leq
    \left(
    1-\epsilon_f(\theta_c(\theta_0^*))
    \right)(\tau-\tau_c^*),
    \qquad
    \text{for every }\tau\in[\tau_0^*,\tau_c^*].
\]
Taking the left derivative gives
\[
\lim_{\tau\uparrow\tau_c^*} \widehat K'_+(\tau)  
\equiv \lim_{\tau\uparrow\tau_c^*} \frac{\widehat K(\tau_c^*) - \widehat K(\tau)}{\tau_c^* - \tau} 
\geq
1-\epsilon_f(\theta_c(\theta_0^*)).
\]
Thus, \(\Lambda\) jumps (weakly) downward at
\(\theta_c(\theta_0^*)\). Concavity of \(\widehat K\) makes
\(\Lambda\) decreasing in the pooling region, while IED makes
\(1-\epsilon_f\) decreasing in the fully revealing region.
Consequently, \(\Lambda\) is decreasing on \(\Theta\).

\subparagraph*{The costate for \(U\).}
Define
\begin{equation}
    \Gamma(\theta)=
    \begin{cases}
        -A(\theta_0^*),
        &\text{if }\theta\in[0,\theta_0^*),
        \\ 
        -f(\theta_c(\theta_0^*))
        -
        \dfrac{1}{\theta_c(\theta_0^*)}
        \displaystyle
        \int_{\theta}^{\theta_c(\theta_0^*)}
        [1-\Lambda(s)]\d F(s),
        &\text{if }\theta\in
        [\theta_0^*,\theta_c(\theta_0^*)),
        \\ 
        -f(\theta),
        &\text{if }\theta\in
        [\theta_c(\theta_0^*),\bar\theta].
    \end{cases}
    \label{eq:Gamma-multiplier}
\end{equation}
The costate $\Gamma$ is continuous at \(\theta_c(\theta_0^*)\) by construction. It is also
continuous at \(\theta_0^*\), because
\begin{align*}
    \int_{\theta_0^*}^{\theta_c(\theta_0^*)}
    [1-\Lambda(s)]\d F(s)
     =
    F(\theta_c(\theta_0^*))
    -
    F(\theta_0^*)
    -
    \widehat K(\tau_c^*)+\widehat K(\tau_0^*)
     =
    \theta_c(\theta_0^*)
    \left[
    A(\theta_0^*)
    -
    f(\theta_c(\theta_0^*))
    \right]
\end{align*}
implies $\Gamma(\theta_0^{*}-) = -A(\theta_0^*)= \Gamma(\theta_0^{*}+)$.
Because $\Gamma$ is absolutely continuous in each region and is continuous at the boundaries, it is absolutely continuous on $[0,\bar\theta]$.

In the exclusion region, $\Gamma$ is constant.
In the bunching region, $\Lambda\leq1$ implies that $\Gamma$ is increasing.
In the fully revealing region, Condition~\eqref{C} implies that
$\Gamma=-f$ is increasing.
Hence, $\Gamma$ is increasing on $[0,\bar\theta]$.

\subparagraph*{The Lagrangian multiplier on (IC-Env).}
Define
\begin{equation}
    \gamma(\theta)=  \dot \Gamma(\theta) = 
    \begin{cases}
        0,
        &\text{if }\theta\in[0,\theta_0^*),
        \\ 
        \dfrac{
        [1-\widehat K'_+(F(\theta))]f(\theta)
        }{
        \theta_c(\theta_0^*)
        },
        &\text{if }\theta\in
        [\theta_0^*,\theta_c(\theta_0^*)),
        \\ 
        -f'(\theta),
        &\text{if }\theta\in
        [\theta_c(\theta_0^*),\bar\theta).
    \end{cases}
    \label{eq:gamma-multiplier}
\end{equation}
which satisfies first-order condition for the control \(q\):
\begin{equation} \label{eq:q-FOC}
        \gamma(\theta)c'(q^*(\theta)) =  [1-\Lambda(\theta)]f(\theta)
\end{equation}

\subparagraph*{The costate for \(w\).}
Define
\begin{equation}
    \mu(\theta)=
    \begin{cases}
        F(\theta_0^*)-F(\theta)
        -
        (\theta_0^*-\theta)A(\theta_0^*),
        &\text{if }\theta\in[0,\theta_0^*),
        \\ 
        \dfrac{
        \theta_c(\theta_0^*)-\theta
        }{
        \theta_c(\theta_0^*)
        }
        [
        K(F(\theta))-\widehat K(F(\theta))
         ],
        &\text{if }\theta\in
        [\theta_0^*,\theta_c(\theta_0^*)),
        \\ 
        0,
        &\text{if }\theta\in
        [\theta_c(\theta_0^*),\bar\theta],
    \end{cases}
    \label{eq:mu-multiplier}
\end{equation}
Endpoint contact of \(\widehat K\) implies that \(\mu\) is continuous
at both junctions. Since \(K\circ F\) is continuously differentiable
and \(\widehat K\circ F\) is absolutely continuous, \(\mu\) is
absolutely continuous. Direct differentiation gives $\dot\mu(\theta)=-[\theta\gamma(\theta)+\Lambda(\theta)f(\theta)+\Gamma(\theta)]$ almost everywhere.

Condition~\eqref{S} gives \(\mu\leq0\) in the exclusion region, while
\(\widehat K\geq K\) gives \(\mu\leq0\) in the pooling region.
Moreover, \(\mu=0\) at \(\theta_0^*\) and throughout the fully
revealing region. Since \(w^*\) is constant in the exclusion and
pooling regions and can jump only at \(\theta_0^*\), where
\(\mu(\theta_0^*)=0\), we obtain $\int \mu \d w^* = 0$.

 \subsubsection*{Applying Sufficiency Theorem}
    Now I verify the conditions in Theorem~\ref{thm:certificate}.
    We have verified the costate and control equations, the absolute
continuity of \(\Gamma\) and \(\mu\), the monotonicity of
\(\Lambda\), the condition \(\mu\leq0\), and the concavity condition.

 \emph{1.  Costate and control equations and transversality conditions.} 
 We have verified the costate and control equations, and (Mon: dual).
It remains to verify the transversality conditions.
At $\lbar\theta$, they follow from $\Gamma(\lbar\theta)=-A(\theta_0^*)\leq0$ and $U^*(\lbar\theta)=0$; and $\mu(\lbar\theta)\leq0$ and $w^*(\lbar\theta)=0$ if $\mu(\lbar\theta)<0$ (which implies $\theta_0^*>\lbar\theta$).

At $\bar\theta$, if $\theta_c(\theta_0^*)\leq\bar\theta$, conditions~\eqref{eqn:trans:barUw} are satisfied with $\eta(\bar\theta) = f(\bar\theta)\geq0$:
\begin{align*}
    \Gamma(\bar\theta) =-f(\bar\theta) = -\eta(\bar\theta)\leq0, \qquad 
    \mu(\bar\theta) =  \eta(\bar\theta) [\bar\theta -c'(w^*(\bar\theta))] = 0
\end{align*}
because $c'(w^*(\bar\theta))=\bar\theta$.
If $\theta_c(\theta_0^*)>\bar\theta$, then the transversality conditions also hold with $\eta(\bar\theta) = 0$ because
$\Gamma(\bar\theta)=0$ (by $f(\theta_c(\theta_0^*))=0$) and $\mu(\bar\theta)=  0$ (by $K(1)=\widehat K(1)$).

    \emph{2. Dual feasibility and complementary slackness (MPS).}
We have verified that $\Lambda$ is decreasing on $[0,\bar\theta]$.
Moreover, $w^*=q^*$ implies $D\equiv 0$ and hence $\int D \d \Lambda=0$.

\emph{3. Concavity.}
Because $\Gamma$ is increasing, the concavity condition in
Theorem~\ref{thm:certificate} is satisfied.

All the conditions of Theorem~\ref{thm:certificate} are therefore
satisfied. Thus, $(q^*,w^*)$ solves problem~\ref{prob:P}, which can be implemented by lower censorship with a minimum standard $q_i(\theta_0^*)$.

\medskip
Finally, log-concavity implies single-peakedness and increasing
elasticity on the decreasing part of \(f\). If \(f\) is increasing,
the optimal cutoff satisfies  
\(\theta_c(\theta_0^*)\geq\bar\theta\), so the fully revealing region is empty, and the same multiplier construction applies. The resulting allocation can also be implemented by a pass/fail test with a minimum standard $q_i(\theta_0^*)$.

For the final claim in the proposition, the subgradient
condition is precisely Condition~\eqref{S} on \(\Theta\) with $\theta_c(\theta_0^*)\geq\bar\theta$.
The fully revealing region is therefore empty, and the same multiplier construction applies.
\end{proof}

\subsubsection{Proof of Proposition~\ref{prop:improve}}
The proof follows immediately from Proposition~\ref{prop:optimalrating2} as $\rho(\theta) = 1 - \epsilon_f(\theta)$.

\subsection{Proofs of Section~\ref{lineardelegation}}

\begin{proof}[Proof of Proposition~\ref{prop:lineardelegation}]
    Using the virtual-density representation, replace the density $f$ with the virtual density $\phi$ and the lower bound $\lbar\theta$ with $0$, so that $U(0)=q(0)=0$.
    Redefine the average density $A(\theta)$ in terms of $\phi$ as in equation~\eqref{eqn:A}, with $A(0)=\lim_{\theta\downarrow 0} A(\theta) =\phi(0)$.
    The proof then proceeds exactly as in Proposition~\ref{prop:quality-maximization}, in which $q^*$ is already extended to $[0,\bar\theta]$ to accommodate this generalization, under which $\theta_0^*<\lbar\theta$ is possible.
\end{proof}

\begin{proof}[Proof of Example~\ref{ex:alpha}]
Let \(V(t;\alpha)\) denote the principal's expected payoff under lower
censorship with cutoff type \(t\).
By the virtual-density
representation,
\[
V(t;\alpha)
=
\int_0^{\bar\theta}q(\theta;t)\phi (\theta;\alpha)\d\theta,
\qquad
\phi (\theta;\alpha)
=
(1-\alpha)f(\theta)+\alpha
 \left( F_\zeta(\bar\theta)-F_\zeta(\theta) \right) .
\]
Therefore,
\(V(t;\alpha)=(1-\alpha)V(t;0)+\alpha V(t;1)\).

When $\alpha=1$, the virtual density $\phi(\theta;1)$ is decreasing in $\theta$, so we have
\[
\frac{\partial V(t;1)}{\partial t}
=
q_i(t)
\left[
\frac{1}{\theta_c(t)-t}
\int_t^{\theta_c(t)}\phi (\theta;1)\,\d\theta
-
\phi(t;1)
\right]
\leq0.
\]
Thus, \(V(t;1)\) is decreasing in \(t\).

Now fix \(0\leq\alpha<\alpha'<1\) and \(t'>t\). Linearity implies
\begin{align*}
(1-\alpha)
\left( V(t';\alpha')-V(t;\alpha') \right)
=
(1-\alpha')
\left(  V(t';\alpha)-V(t;\alpha) \right)  +
(\alpha'-\alpha)
\left(  V(t';1)-V(t;1)  \right).
\end{align*}
Because the last term is nonpositive,
\[
V(t';\alpha)\leq V(t;\alpha)
\quad\Longrightarrow\quad
V(t';\alpha')\leq V(t;\alpha').
\]
Let \(t^*(\alpha)=\arg\max_t V(t;\alpha)\) denote the optimal cutoff type. The preceding
single-crossing property implies that
\(\min t^*(\alpha)\) and \(\max t^*(\alpha)\) are decreasing
in \(\alpha\). 
\end{proof}

\begin{proof}[Proof of Proposition \ref{prop:optimalrating2}]
Under Condition~\ref{LD}, we have $\rho(\theta)=
\beta(\theta)\zeta(\theta) + \theta \frac{\phi'(\theta)}{f(\theta)}$, so  Lemma~\ref{lem:rho} implies that the optimal deterministic rating scheme is not optimal among all rating schemes.
By Lemma~\ref{lem:existence}, the optimal rating scheme exists and therefore must be nondeterministic.
\end{proof}

\section{A sufficiency theorem for optimal control problems with monotonicity and state constraints}
\label{sec:certificate}
I present a sufficiency theorem for optimal control problems with monotonicity constraints and a pure state equality constraint, which has applications in optimal delegation, particularly with participation constraints (see also \cite{Xiao2023a}).
The method is related to the Lagrangian approach of \cite{AmadorBagwell2022} for delegation with participation constraints.
In \citet{AmadorBagwell2022}, the Lagrangian method is applied to a
truncated problem, and the optimal allocation characterized on the
truncated type domain is continuous.
By contrast, the sufficiency theorem presented here applies to the original problem and allows for jump discontinuities in the monotone variable.

Compared to existing sufficiency theorems in the optimal control literature, this theorem handles the monotonicity constraint (IC-Mon\textsuperscript{D}) without assuming piecewise absolute continuity of the monotone variable or using its derivative as a control variable, thereby allowing for jumps in the monotone variable.
In this sense, it complements the necessity result of \citet[Theorem~3.1]{Hellwig2008} by establishing sufficiency while also incorporating a pure state constraint (IC-Env\textsuperscript{D}).

When the state variable is piecewise absolutely continuous with finitely many jumps, it is plausible to use its derivative as a control variable, provided that jump conditions for the state variable are included.
In this case, the theorem extends the sufficiency theorems of \citet[Theorems~6.2, pp.~361--362]{SeierstadSydsaeter1987} and \citet[Theorem~10.3.2]{LeonardLong1992} by incorporating a pure state equality constraint (IC-Env\textsuperscript{D}).
\footnote{
Using their methods would require representing the state equality constraint as two inequality constraints, which in turn would require both $\theta q-c(q)-U$ and $-\theta q+c(q)+U$ to be quasiconcave---a contradictory requirement when $c$ is strictly convex.
}

Assume $\theta\in\Theta=[\underline \theta, \bar \theta] \subseteq \reals_+$ is a random variable with distribution $F$ and continuous density $f=F'>0$.
Assume $c\colon \reals_+ \to \reals_+$ is a twice continuously differentiable, strictly increasing, and strictly convex function that satisfies $c(0)=c'(0)=0$ and $\lim_{q\to \infty} c'(q)>\bar \theta$.
Assume $v\colon \reals_+\times \Theta \to \reals$ is a twice continuously differentiable function that is concave in $q$.

Assume $q\colon\Theta\to \reals_+$ is a right-continuous function of bounded variation. 
Consider the following problem:
\begin{align*}
\max_{q\colon \Theta\to \reals_+} \quad  &\int_{\underline \theta}^{\bar \theta} {v}(q(\theta), \theta) \d F(\theta)
\tag*{[D]} 
\label{prob:D}\\
\text{subject to}\quad 
    & \underline{U} = \lbar\theta q(\underline\theta) - c(q(\underline\theta))\geq0 \tag*{(IR\textsuperscript{D})}\\
    & q(\theta) \mbox{ increasing},  \tag*{(IC-Mon\textsuperscript{D})}\\
    & \theta q(\theta) - c(q(\theta) ) =  \int_{\underline \theta}^\theta q(x) \d x + \underline U, \mbox{ for all $\theta\in\T$.}  \tag*{(IC-Env\textsuperscript{D})} 
\end{align*}
Define $U(\theta) =  \int_{\underline \theta}^\theta q(x) \d x + \underline U$.
Rewrite constraints as 
\begin{align}
    & \theta q(\theta) - c(q(\theta) ) - U(\theta) = 0 \tag{IC-Env\textsuperscript{D}} \\
    & \dot U = q(\theta), \mbox{ $U$ absolutely continuous},\\
    & \mbox{$q$ increasing}\\ 
    & U(\underline  \theta), q(\underline \theta) \geq  0,\quad  U(\bar \theta), q(\bar \theta) \mbox{ free.}
\end{align}  
Define the Hamiltonian and the Lagrangian by
\begin{align} 
   H   &= v(q(\theta),\theta) f(\theta) 
         + \Gamma (\theta) q(\theta)  ,\\
   \mathcal{L}   &= H + \gamma(\theta)  [\theta q(\theta) - c(q(\theta) ) -U(\theta)],
\end{align}

The following theorem gives sufficient conditions for optimality.

\begin{theorem}[Certificate of Optimality]
\label{thm:certificate-D}
Suppose that \((q^*,U^*)\) satisfies all constraints of
problem~\ref{prob:D}. Then \((q^*,U^*)\) is a solution to
problem~\ref{prob:D} if there exist right-continuous, piecewise absolutely continuous functions
\(\Gamma\) and \(\mu\), with possible
jumps at a set of junction points \(\{t_j\}\), and a measurable function $\gamma$, satisfying the following
conditions. 

\begin{enumerate}
\item Costate and control equations and transversality conditions: 
\begin{align}
    -&\frac{\partial\mathcal L}{\partial U}
    =
    \gamma
    =
    \dot\Gamma \; \mbox{ a.e.},
    \label{eq:cert-Gamma}
    \\
    -&\frac{\partial\mathcal L}{\partial q}
    =
    -\left[
      v_q(q^*,\theta)  f+\Gamma
        +\gamma\bigl(\theta-c'(q^*)\bigr)
    \right]
    =
    \dot\mu \; \mbox{ a.e.},
    \label{eq:cert-mu}
    \\
    &
    \mu\leq0,
    \quad
    \int_{(\lbar\theta,\bar\theta]}
        \mu(\theta-)\d q^*(\theta)
    =     0,
    \label{eq:cert-mon}
    \\
    &\Gamma(\lbar\theta)
    \leq0,
    \quad
    \Gamma(\lbar\theta)U^*(\lbar\theta)=0,
    \label{eq:cert-lower-U}
    \\
    &\mu(\lbar\theta)
    \leq0,
    \quad
    \mu(\lbar\theta)q^*(\lbar\theta)=0,
    \label{eq:cert-lower-q}
    \\
    &\Gamma(\bar\theta) = \mu(\bar\theta)= 0.
\end{align}

\item Jumps in the costates. At every junction point \(t_j\) where $\Gamma$ or $\mu$ jumps,
there exists a scalar \(\chi(t_j)\geq0\) such that
\begin{align}
    \Gamma(t_j+)-\Gamma(t_j-)
    &=
    \chi(t_j) \geq0,
    \label{eq:cert-Gamma-jump}
    \\
    \mu(t_j+)-\mu(t_j-)
    &=
    \bigl(c'(q^*(t_j))-t_j\bigr)\chi(t_j).
    \label{eq:cert-mu-jump}
\end{align}
If \(t_j=\bar\theta\), the right limits are understood as
the terminal values \(\Gamma(\bar\theta)\) and \(\mu(\bar\theta)\).

\item Concavity: define 
$Q_\theta=\{q\ge 0: \theta q-c(q)\ge 0\}$ and
$$\kappa(\theta) = \sup\left\{
k:
v_{qq}(q,\theta)+k c''(q)\le 0 \text{ for all $q\in Q_\theta$}
\right\};$$
then
\[
 \Gamma(\theta) + \int_{\lbar\theta}^{\theta} \kappa(s) f(s) \d s \text{ is increasing}.
\]
\end{enumerate}
\end{theorem}

\begin{proof}[Proof of Theorem~\ref{thm:certificate-D}]
Fix any admissible \((q,U)\) satisfying the constraints, and write
\[
    \Delta q=q-q^*,
    \qquad
    \Delta U=U-U^*.
\]
Define
\[
    g(q,U;\theta)
    :=
    \theta q-c(q)-U
\]
and
\[
    \mathcal J(q,U;\theta)
    \equiv
    v(q(\theta),\theta) f(\theta)
    +
    \Gamma(\theta)q(\theta)
    +
    \gamma(\theta)g(q(\theta),U(\theta),\theta).
\]

Condition~3 and \eqref{eq:cert-Gamma} imply
\[
    \gamma(\theta)+\kappa(\theta)f(\theta)\geq0
    \quad\text{a.e.}
\]
Moreover, by the definition of \(\kappa(\theta)\),
\[
    v_{qq}(q,\theta)+\kappa(\theta)c''(q)\leq0
    \quad\text{for every }q\in Q_\theta.
\]
Therefore,
\[
\mathcal J_{qq}(q,U;\theta)
 =v_{qq}(q,\theta) f(\theta)-\gamma(\theta)c''(q)
 =
\left[v_{qq}(q,\theta)+\kappa(\theta)c''(q)\right] f(\theta)-
\left[\gamma(\theta)+\kappa(\theta)f(\theta)\right]c''(q)
\le0.
\]
Thus, for each \(\theta\in\Theta\), \(\mathcal J\) is concave in \(q\) on
\(Q_\theta\) and affine in \(U\). Since every feasible allocation
satisfies
\[
    U(\theta)=\theta q(\theta)-c(q(\theta))\geq0,
\]
both \(q(\theta)\) and \(q^*(\theta)\) belong to \(Q_\theta\).
Hence, at every point that is not a junction, concavity of \(\mathcal J\) in \(q\) and the costate and control equations in condition 1 imply
\begin{align}
    \mathcal J-\mathcal J^*
    &\leq
    \left[
        v_q(q^*,\theta) f+\Gamma
        +\gamma\bigl(\theta-c'(q^*)\bigr)
    \right]\Delta q
    -
    \gamma\Delta U=
    -\dot\mu\Delta q
    -
    \dot\Gamma\Delta U.
    \label{eq:cert-continuous-bound}
\end{align}

Now consider the junction points where the costates jump. Since \(F\) has
a continuous density, $K(\theta) = \int_{\lbar\theta}^{\theta}\kappa(s)\d F(s)$
is continuous. Hence, Condition~3 implies that every jump of
\(\Gamma\) is nonnegative. In accordance with Condition~2, write
\[
    \Gamma(t_j+)-\Gamma(t_j-)
    =
    \chi(t_j)\geq0.
\]
Since \(c\) is convex,
\[
    g(q,U;t_j)=t_jq-c(q)-U
\]
is concave in \((q,U)\). Therefore,
\((q,U)\mapsto\chi(t_j)g(q,U;t_j)\) is concave. Moreover,
(IC-Env\textsuperscript{D}) implies
\[
    g(q(t_j),U(t_j);t_j)
    =
    g(q^*(t_j),U^*(t_j);t_j)
    =
    0.
\]
The concavity inequality at \(t_j\) consequently gives
\begin{align}
    0
    &=
    \chi(t_j)
    \left[
        g(q(t_j),U(t_j);t_j)
        -
        g(q^*(t_j),U^*(t_j);t_j)
    \right]
    \notag\\
    &\leq
    \chi(t_j)
    \left[
        \bigl(t_j-c'(q^*(t_j))\bigr)\Delta q(t_j)
        -
        \Delta U(t_j)
    \right]
    \notag\\
    &=
    -\bigl[\mu(t_j+)-\mu(t_j-)\bigr]\Delta q(t_j)
    -
    \bigl[\Gamma(t_j+)-\Gamma(t_j-)\bigr]\Delta U(t_j),
    \label{eq:cert-jump-bound}
\end{align}
where the last equality follows from
\eqref{eq:cert-Gamma-jump}--\eqref{eq:cert-mu-jump}.

Integrating \eqref{eq:cert-continuous-bound} over all continuity
intervals and adding the jump inequalities
\eqref{eq:cert-jump-bound} yields
\begin{equation}
    \int_{\lbar\theta}^{\bar\theta}
        \bigl(\mathcal J-\mathcal J^*\bigr)\d\theta
    \leq
    -
    \int_{(\lbar\theta,\bar\theta]}
        \Delta q\d\mu
    -
    \int_{(\lbar\theta,\bar\theta]}
        \Delta U\d\Gamma.
    \label{eq:cert-BV-bound}
\end{equation}
The Stieltjes integrals in \eqref{eq:cert-BV-bound} include both the
absolutely continuous parts and the jumps of the costates.

By definition of $\mathcal{J}$, state equations, and $g(q,U;\theta)=0$ (IC-Env\textsuperscript{D}), we have
\begin{equation}
   \Delta V : = \int_{\lbar\theta}^{\bar\theta}
        \left(  v(q,\theta) - v(q^*,\theta) \right)f \d\theta
    =
        \int_{\lbar\theta}^{\bar\theta}
        \bigl(\mathcal J-\mathcal J^*\bigr)\d\theta
    -
    \int_{\lbar\theta}^{\bar\theta}
        \Gamma \Delta q \d\theta
    =
    \int_{\lbar\theta}^{\bar\theta}
        \bigl(\mathcal J-\mathcal J^*\bigr)\d\theta
    -
    \int_{\lbar\theta}^{\bar\theta}
        \Gamma\d(\Delta U).
    \label{eq:cert-objective}
\end{equation}
Combining \eqref{eq:cert-BV-bound} and
\eqref{eq:cert-objective} gives
\begin{align}
   \Delta V 
    \leq 
    -
    \int_{(\lbar\theta,\bar\theta]}
        \Delta q\d\mu
    -
    \int_{(\lbar\theta,\bar\theta]}
        \Delta U\d\Gamma
    -
    \int_{\lbar\theta}^{\bar\theta}
        \Gamma\d(\Delta U).
    \label{eq:cert-before-ibp}
\end{align}

Using integration by parts for bounded-variation functions,
\begin{align}
    \Delta V
    \leq 
    -
    \bigl[\Gamma\Delta U\bigr]_{\lbar\theta}^{\bar\theta}
    -
    \bigl[\mu\Delta q\bigr]_{\lbar\theta}^{\bar\theta}
    +
    \int_{(\lbar\theta,\bar\theta]}
        \mu(\theta-)\d(\Delta q)(\theta).
    \label{eq:cert-master}
\end{align}
Here, the Stieltjes integral uses the left limit \(\mu(\theta-)\). This convention
ensures that the bounded-variation integration-by-parts formula holds
without an additional cross-jump term.

Every term on the right-hand side of
\eqref{eq:cert-master} is nonpositive. By the lower- and upper-endpoint transversality conditions, we have
\begin{align*}
  -
    \bigl[\Gamma\Delta U\bigr]_{\lbar\theta}^{\bar\theta}
    -
    \bigl[\mu\Delta q\bigr]_{\lbar\theta}^{\bar\theta}
    =
    \Gamma(\lbar\theta)\Delta U(\lbar\theta)
    +
    \mu(\lbar\theta)\Delta q(\lbar\theta)
    -
        \Gamma(\bar\theta)\Delta U(\bar\theta)
     -
        \mu(\bar\theta)\Delta q(\bar\theta)
        \leq 0
\end{align*}

Finally, monotonicity of \(q\) and
\eqref{eq:cert-mon} imply
\begin{align*}
    \int_{(\lbar\theta,\bar\theta]}
        \mu(\theta-)\d(\Delta q)(\theta)
 =
    \int_{(\lbar\theta,\bar\theta]}
        \mu(\theta-)\d q(\theta)
    -
    \int_{(\lbar\theta,\bar\theta]}
        \mu(\theta-)\d q^*(\theta)
=
    \int_{(\lbar\theta,\bar\theta]}
        \mu(\theta-)\d q(\theta)
    \leq0.
\end{align*}
Thus, $\Delta V \leq 0$ for every feasible \((q,U)\).
Hence, \((q^*,U^*)\) is optimal.
\end{proof}

\section{Optimality Certificate for Stochastic Rating Schemes}
\label{app:stochastic-certificate}

This subsection applies Theorem~\ref{thm:certificate} to the candidate
direct mechanism in Example~\ref{ex:stochastic}.  We state the free-boundary
conditions that define the candidate, construct the costates and multipliers
required by the theorem, and show that any exact solution satisfying the stated
feasibility and sign conditions meets all of the theorem's hypotheses.  We then
use validated ball arithmetic and a Krawczyk operator to isolate an exact root
and certify the required inequalities on their full intervals.  This yields a
formal computer-assisted proof of the candidate's global optimality.

The multiplier construction and numerical calculations were developed with the
assistance of ChatGPT-5.6 Sol.

\subsection{Candidate and Free-Boundary System}
\label{sec:ratings-cert-candidate}

In the example,
\[
    \lbar\theta=0.01,
    \qquad
    \bar\theta=1,
    \qquad
    v(q,\theta)=q,
    \qquad
    c(q)=\frac{q^2}{2}.
\]
The density and distribution function are
\begin{align}
    f(\theta)
    &=C\theta^{-1/2}e^{\theta/2},
        \label{eq:ratings-cert-density}\\
    F(\theta)
    &=\frac{
        \erfi(\sqrt{\theta/2})-\erfi(\sqrt{\lbar\theta/2})
      }{
        \erfi(1/\sqrt 2)-\erfi(\sqrt{\lbar\theta/2})
      },
        \label{eq:ratings-cert-cdf}\\
    C
    &=\frac{1}{
        \sqrt{2\pi}\bigl[
          \erfi(1/\sqrt 2)-\erfi(\sqrt{\lbar\theta/2})
        \bigr]
      }
      \approx0.4567082818.
        \label{eq:ratings-cert-normalization}
\end{align}

For arbitrary constants $(\Lambda,B)$, define
\begin{align}
    M_{\Lambda,B}(\theta)
    &\equiv
      \Lambda\theta f(\theta)+B-\Lambda F(\theta),
        \label{eq:ratings-cert-M-family}\\
    \widetilde q_{\Lambda,B}(\theta)
    &\equiv
      \frac{(\Lambda-1)\theta^2f(\theta)}
           {M_{\Lambda,B}(\theta)},
        \label{eq:ratings-cert-q-family}\\
    \widehat\Gamma_{\Lambda,B}(\theta)
    &\equiv
      \frac{B-\Lambda F(\theta)}{\theta}.
        \label{eq:ratings-cert-Gammahat-family}
\end{align}
The hat emphasizes that $\widehat\Gamma_{\Lambda,B}$ is an auxiliary
function.  It will equal the costate $\Gamma$ on the separating interval, but
not on the lower pooling interval.  Its derivative and the derivative of
$\widetilde q_{\Lambda,B}$ are
\begin{align}
    \widehat\Gamma_{\Lambda,B}'(\theta)
    &=-\frac{M_{\Lambda,B}(\theta)}{\theta^2},
        \label{eq:ratings-cert-Gammahat-prime}\\
    \widetilde q_{\Lambda,B}'(\theta)
    &=
    \widetilde q_{\Lambda,B}(\theta)
    \left[
       \frac{\theta+3}{2\theta}
       -\frac{\Lambda f(\theta)(\theta-1)}
             {2M_{\Lambda,B}(\theta)}
    \right].
        \label{eq:ratings-cert-qprime-family}
\end{align}

For a candidate quadruple $(\Lambda,B,\theta_L,\theta_H)$, let
\begin{equation}
    q_L\equiv\widetilde q_{\Lambda,B}(\theta_L),
    \qquad
    q_H\equiv\widetilde q_{\Lambda,B}(\theta_H),
    \qquad
    w_L\equiv\frac{q_L^2}{2\lbar\theta}.
    \label{eq:ratings-cert-endpoints-family}
\end{equation}
The four constants solve
\begin{align}
    0={}&
      q_L\bigl[
        \widehat\Gamma_{\Lambda,B}(\theta_L)
        -\widehat\Gamma_{\Lambda,B}(\lbar\theta)
      \bigr]
      -(1-\Lambda)F(\theta_L),
      \tag{FB-L}\label{eq:ratings-cert-FB-L}\\
    0={}&
      \widehat\Gamma_{\Lambda,B}(\theta_H)
      -\frac{1-F(\theta_H)}{\theta_H-q_H},
      \tag{FB-$\Gamma$}\label{eq:ratings-cert-FB-Gamma}\\
    0={}&
      w_L+
      \int_{\theta_L}^{\theta_H}
        \frac{
          \widetilde q_{\Lambda,B}(x)
          \widetilde q_{\Lambda,B}'(x)
        }{x}\d x
      -q_H,
      \tag{C-W}\label{eq:ratings-cert-C-W}\\
    0={}&
      w_L F(\theta_H)-q_L F(\theta_L)
      -\int_{\theta_L}^{\theta_H}
        \widetilde q_{\Lambda,B}(x)f(x)\d x \notag\\
      &\quad+
      \int_{\theta_L}^{\theta_H}
        \frac{
          \widetilde q_{\Lambda,B}(x)
          \widetilde q_{\Lambda,B}'(x)
        }{x}
        \bigl[F(\theta_H)-F(x)\bigr] \d x.
      \tag{C-D}\label{eq:ratings-cert-C-D}
\end{align}
The lower free-boundary equation will be seen below to be precisely the
condition that the costates match at $\theta_L$.  The last two equations
impose $w^*(\theta_H)=q_H$ and $D^*(\theta_H)=0$.

The computer-assisted verification below proves that
\eqref{eq:ratings-cert-FB-L}--\eqref{eq:ratings-cert-C-D} has a unique exact
solution in the stated interval box.  Denote it by
$(\Lambda_s,B,\theta_L,\theta_H)$.  Numerically,
\begin{align}
    \Lambda_s&\approx0.6484032082813592,
    &B&\approx-0.03127595936631157,\notag\\
    \theta_L&\approx0.03194209760055742,
    &\theta_H&\approx0.8445620637795093.
    \label{eq:ratings-cert-root}
\end{align}
Henceforth, set
\begin{equation}
    M(\theta)\equiv M_{\Lambda_s,B}(\theta),
    \qquad
    \widetilde q(\theta)
      \equiv\widetilde q_{\Lambda_s,B}(\theta),
    \qquad
    \widehat\Gamma_s(\theta)
      \equiv\widehat\Gamma_{\Lambda_s,B}(\theta).
    \label{eq:ratings-cert-fixed-notation}
\end{equation}
The implied endpoint values are
\begin{equation}
    q_L\approx0.03787871728,
    \qquad
    q_H\approx1.0086814119,
    \qquad
    w_L\approx0.07173986114.
    \label{eq:ratings-cert-endpoint-values}
\end{equation}

The candidate solutions are
\begin{align}
    q^*(\theta)
    &=
    \begin{cases}
        q_L,
          &\lbar\theta\leq\theta\leq\theta_L,\\[2pt]
        \widetilde q(\theta),
          &\theta_L<\theta<\theta_H,\\[2pt]
        q_H,
          &\theta_H\leq\theta\leq1,
    \end{cases}
    \label{eq:ratings-cert-qstar}\\
    w^*(\theta)
    &=
    \begin{cases}
        w_L,
          &\lbar\theta\leq\theta\leq\theta_L,\\[4pt]
        w_L+
        \displaystyle\int_{\theta_L}^{\theta}
          \frac{\widetilde q(x)\widetilde q'(x)}{x}\d x,
          &\theta_L<\theta<\theta_H,\\[10pt]
        q_H,
          &\theta_H\leq\theta\leq1.
    \end{cases}
    \label{eq:ratings-cert-wstar}
\end{align}

\begin{figure}[htb]
\centering
\includegraphics[width=0.45\textwidth]{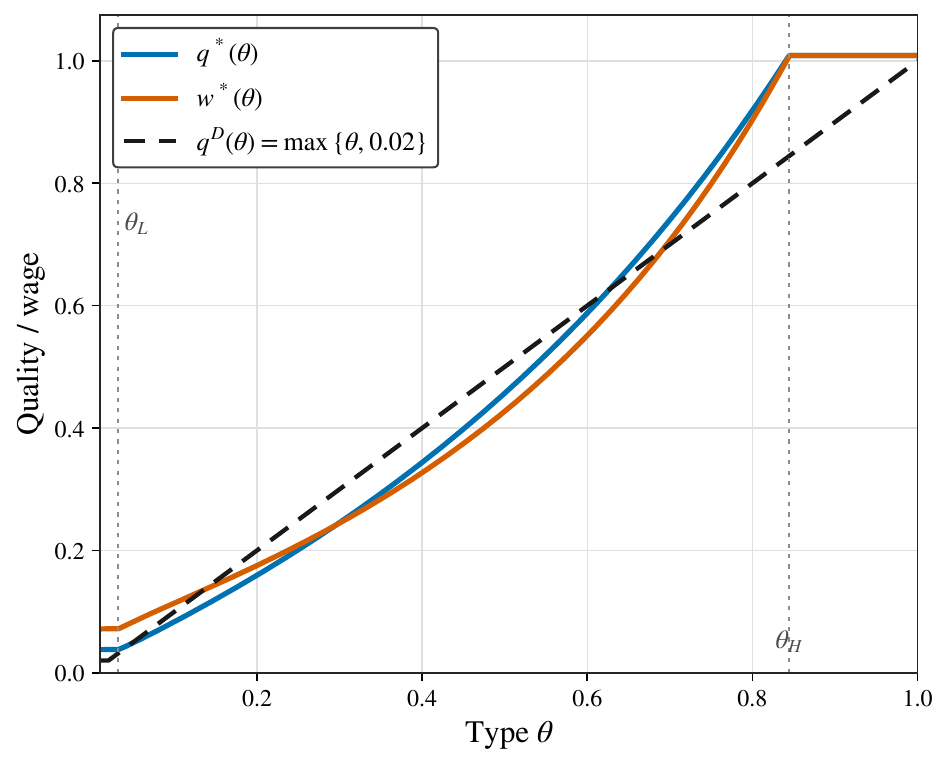}
\caption{The optimal $(q^*,w^*)$}
\end{figure}

Define
\begin{align}
    U^*(\theta)
    &\equiv
      \theta w^*(\theta)-\frac{q^*(\theta)^2}{2},
      \label{eq:ratings-cert-Ustar}\\
    D^*(\theta)
    &\equiv
      \int_{\lbar\theta}^{\theta}
        \bigl[w^*(x)-q^*(x)\bigr]f(x)\d x.
      \label{eq:ratings-cert-Dstar}
\end{align}
Then $U^*(\lbar\theta)=0$, $U^{*\prime}=w^*$, and
\begin{equation}
    w^{*\prime}(\theta)
    =\frac{q^*(\theta)q^{*\prime}(\theta)}{\theta}
    \quad\text{a.e.}
    \label{eq:ratings-cert-wprime}
\end{equation}
The contact equations give
\begin{equation}
    D^*(\lbar\theta)=D^*(\theta_H)=0,
    \qquad
    D^*(\theta)=0
      \quad\text{for }\theta\in[\theta_H,1].
    \label{eq:ratings-cert-D-contact}
\end{equation}

\subsection{Costates and Multipliers}
\label{sec:ratings-cert-multipliers}

We now construct the costates $(\Lambda,\Gamma,\mu)$ and the multiplier
$\gamma$ required by Theorem~\ref{thm:certificate}.

For the upper contact interval define
\begin{equation}
    R(\theta)\equiv\frac{1-F(\theta)}{f(\theta)},
    \qquad
    \epsilon_f(\theta)
      \equiv-\frac{\theta f'(\theta)}{f(\theta)}
      =\frac{1-\theta}{2}.
    \label{eq:ratings-cert-upper-definitions}
\end{equation}
On $[\theta_H,1]$, set
\begin{align}
    \Gamma_u(\theta)
    &\equiv
      -\frac{1-F(\theta)}{q_H-\theta},
      \label{eq:ratings-cert-Gamma-upper}\\
    \gamma_u(\theta)
    &\equiv
      \Gamma_u'(\theta)
      =\frac{f(\theta)\bigl[q_H-\theta-R(\theta)\bigr]}
             {(q_H-\theta)^2},
      \label{eq:ratings-cert-gamma-upper}\\
    \Lambda_u(\theta)
    &\equiv
      1-\frac{q_H\gamma_u(\theta)}{f(\theta)},
      \label{eq:ratings-cert-Lambda-upper}\\
    \lambda_u(\theta)
    &\equiv
      -\Lambda_u'(\theta)
      =\frac{q_H}{(q_H-\theta)^3}
      \left\{
        2\bigl[q_H-\theta-R(\theta)\bigr]
        -\frac{\epsilon_f(\theta)}{\theta}
          (q_H-\theta)R(\theta)
      \right\}.
      \label{eq:ratings-cert-lambda-upper}
\end{align}

For the lower pooling interval, define the auxiliary switching function by
\begin{equation}
    P(\theta)
    \equiv
    q_L\bigl[
      \widehat\Gamma_s(\theta)
      -\widehat\Gamma_s(\lbar\theta)
    \bigr]
    -(1-\Lambda_s)F(\theta),
    \qquad
    \lbar\theta\leq\theta\leq\theta_L.
    \label{eq:ratings-cert-P}
\end{equation}
The costates and multipliers are
\begin{align}
    \Lambda(\theta)
    &=
    \begin{cases}
      \Lambda_s,
        &\lbar\theta\leq\theta\leq\theta_H,\\
      \Lambda_u(\theta),
        &\theta_H\leq\theta\leq1,
    \end{cases}
    \label{eq:ratings-cert-Lambda}\\
    \Gamma(\theta)
    &=
    \begin{cases}
      \widehat\Gamma_s(\lbar\theta)
        +\dfrac{1-\Lambda_s}{q_L}F(\theta),
        &\lbar\theta\leq\theta\leq\theta_L,\\[9pt]
      \widehat\Gamma_s(\theta),
        &\theta_L\leq\theta\leq\theta_H,\\[2pt]
      \Gamma_u(\theta),
        &\theta_H\leq\theta\leq1,
    \end{cases}
    \label{eq:ratings-cert-Gamma}\\
    \gamma(\theta)
    &=
    \begin{cases}
      \dfrac{(1-\Lambda_s)f(\theta)}{q_L},
        &\lbar\theta\leq\theta\leq\theta_L,\\[9pt]
      -\dfrac{M(\theta)}{\theta^2},
        &\theta_L\leq\theta\leq\theta_H,\\[9pt]
      \gamma_u(\theta),
        &\theta_H\leq\theta\leq1,
    \end{cases}
    \label{eq:ratings-cert-gamma}\\
    \mu(\theta)
    &=
    \begin{cases}
      \dfrac{\theta}{q_L}P(\theta),
        &\lbar\theta\leq\theta\leq\theta_L,\\[9pt]
      0,
        &\theta_L\leq\theta\leq1.
    \end{cases}
    \label{eq:ratings-cert-mu}
\end{align}
These formulas give the costates and multiplier required by
Theorem~\ref{thm:certificate}.

The identities needed at the lower junction are transparent from
\eqref{eq:ratings-cert-FB-L}:
\begin{equation}
    P(\lbar\theta)=P(\theta_L)=0,
    \qquad
    \Gamma(\theta_L^-)=\Gamma(\theta_L^+),
    \qquad
    \mu(\theta_L)=0.
    \label{eq:ratings-cert-lower-matching}
\end{equation}
Moreover, $\widetilde q(\theta_L)=q_L$ and
\[
    -\frac{M(\theta)\widetilde q(\theta)}{\theta^2}
    =(1-\Lambda_s)f(\theta),
\]
so $\gamma$ is also continuous at $\theta_L$.  At the upper junction, the
free-boundary conditions imply
\begin{equation}
    \widehat\Gamma_s(\theta_H)=\Gamma_u(\theta_H),
    \qquad
    -\frac{M(\theta_H)}{\theta_H^2}=\gamma_u(\theta_H),
    \qquad
    \Lambda_s=\Lambda_u(\theta_H).
    \label{eq:ratings-cert-upper-matching}
\end{equation}
Hence the costates are continuous at both junctions.

In particular, the constructed costate $\Lambda$ is absolutely continuous.
Hence, only for this candidate, the measure multiplier has the density
\begin{align}
    \lambda(\theta)
    &\equiv-\dot\Lambda(\theta)
    =
    \begin{cases}
      0,
        &\lbar\theta\leq\theta<\theta_H,\\
      \lambda_u(\theta),
        &\theta_H\leq\theta\leq1,
    \end{cases}
    \quad\text{a.e.},
    \label{eq:ratings-cert-lambda}\\
    -\d\Lambda&=\lambda(\theta)\d\theta.\notag
\end{align}
The sign condition below verifies that this density is nonnegative.

The endpoint conditions are
\begin{equation}
    \Gamma(\lbar\theta)=\frac{B}{\lbar\theta}<0,
    \qquad
    \mu(\lbar\theta)=0,
    \qquad
    \Gamma(1)=\mu(1)=0.
    \label{eq:ratings-cert-endpoints}
\end{equation}
Thus the lower transversality and complementary-slackness conditions hold
because $U^*(\lbar\theta)=0$ and $w^*(\lbar\theta)>0$.  No endpoint
restriction is imposed on $\Lambda$ because both endpoints of $D$ are fixed.
The domain conditions ensure that $q^*$ is interior in $Q$, so the
quality-control equation has no bound multiplier.

\subsection{Conditional Optimality Certificate}
\label{sec:ratings-cert-sufficiency}

\begin{proposition}[Conditional optimality certificate]
\label{prop:ratings-cert-optimality}
Assume $v(q,\theta)=q$ and $c(q)=q^2/2$.
Suppose the free-boundary system
\eqref{eq:ratings-cert-FB-L}--\eqref{eq:ratings-cert-C-D}
has an exact solution $(\Lambda_s,B,\theta_L,\theta_H)$ such that
\begin{equation}
    \begin{gathered}
        \lbar\theta<\theta_L<\theta_H<1<q_H<q_{\max},
        \qquad 0<q_L\leq q_H,\\
        0<\Lambda_s<1,
        \qquad B<0,
        \qquad M(\lbar\theta)<0.
    \end{gathered}
    \label{eq:ratings-cert-domain-conditions}
\end{equation}
Suppose further that, on their respective intervals,
\begin{equation}
    \widetilde q'\geq0,
    \qquad
    D^*\geq0,
    \qquad
    \lambda_u\geq0,
    \qquad
    P\leq0.
    \label{eq:ratings-cert-sign-conditions}
\end{equation}
Then the mechanism $(q^*,w^*)$ in
\eqref{eq:ratings-cert-qstar}--\eqref{eq:ratings-cert-wstar}
globally solves problem~\ref{prob:P}.
\end{proposition}

\begin{proof}
Because
\[
    M'(\theta)=\frac{\Lambda_s f(\theta)(\theta-1)}{2}<0
    \qquad\text{on }[\lbar\theta,\theta_H],
\]
the condition $M(\lbar\theta)<0$ implies that $M<0$ throughout this interval.
Thus $\widetilde q$ is well defined and positive.  The condition
$\widetilde q'\geq0$, together with
\eqref{eq:ratings-cert-qstar}--\eqref{eq:ratings-cert-wprime}, implies that
$q^*$ and $w^*$ are nondecreasing and satisfy the stated bounds.  Equations
\eqref{eq:ratings-cert-Ustar}--\eqref{eq:ratings-cert-D-contact}, together with
$D^*\geq0$, verify (IR), (IC-Env), and (MPS).  Hence the candidate is feasible
for problem~\ref{prob:P}.

The multiplier formulas and the matching conditions
\eqref{eq:ratings-cert-lower-matching}--\eqref{eq:ratings-cert-upper-matching}
give absolutely continuous costates $\Lambda$, $\Gamma$, and $\mu$.  Direct
differentiation of
\eqref{eq:ratings-cert-Lambda}--\eqref{eq:ratings-cert-mu} verifies the costate
and control equations in Theorem~\ref{thm:certificate}; here
$v(q,\theta)=q$ and $c'(q)=q$.  Equation~\eqref{eq:ratings-cert-endpoints}
gives the required transversality conditions.

Moreover, $P\leq0$ implies $\mu\leq0$.  Since $w^*$ is constant wherever
$\mu$ may be negative and $\mu=0$ wherever $w^*$ increases,
\[
    \int_{(\lbar\theta,1]}\mu\d w^*=0.
\]
For this candidate, $-\d\Lambda=\lambda(\theta)\d\theta$, and
$\lambda_u\geq0$ implies $\lambda\geq0$.  Thus $\Lambda$ is nonincreasing.
Moreover, $\lambda=0$ below $\theta_H$ and $D^*=0$ on $[\theta_H,1]$.
Consequently, $-\d\Lambda$ is supported on
$\{\theta:D^*(\theta)=0\}$ and
\[
    \int_{(\lbar\theta,1]}D^*\d\Lambda=0.
\]
Finally, $\Lambda_s<1$ and monotonicity of $\Lambda$ imply
$\Lambda(\theta)<1$ everywhere, so the quality-control equation gives
\[
    \gamma(\theta)
    =\frac{[1-\Lambda(\theta)]f(\theta)}{q^*(\theta)}>0.
\]
Since $\dot\Gamma=\gamma$, the costate $\Gamma$ is increasing.  All hypotheses
of Theorem~\ref{thm:certificate} are therefore satisfied, so
$(q^*,w^*)$ globally solves problem~\ref{prob:P}.
\end{proof}

\subsection{Formal Computer-Assisted Verification}
\label{sec:ratings-cert-computer-proof}

The accompanying script \texttt{certify\_stochastic\_example.py} uses
80-decimal-digit Arb balls, validated quadrature, and an analytically
differentiated interval Jacobian.  Let ${\mathcal R}$ collect the four residuals in
\eqref{eq:ratings-cert-FB-L}--\eqref{eq:ratings-cert-C-D}, let
\[
 x_0=
 \begin{pmatrix}
  0.6484032082813591704\\
 -0.03127595936631156913\\
  0.0319420976005574207\\
  0.8445620637795093244
 \end{pmatrix},
 \qquad
 X=x_0+[-10^{-18},10^{-18}]^4,
\]
and let $J(X)$ be the resulting interval enclosure of
$D{\mathcal R}(X)$.  For a fixed nonsingular midpoint preconditioner $Y$, the
program evaluates the Krawczyk operator
\[
 K(X)=x_0-Y{\mathcal R}(x_0)+[I-YJ(X)](X-x_0)
\]
and verifies the strict inclusion $K(X)\subset\operatorname{int}X$.  More
precisely, its outward-rounded output gives
\begin{align*}
 K_1(X)&\subset
 [0.6484032082813591703750514606697\pm4.56\mathord\times10^{-32}],\\
 K_2(X)&\subset
 [-0.03127595936631156912786385957346\pm6.52\mathord\times10^{-33}],\\
 K_3(X)&\subset
 [0.0319420976005574207023168427295\pm3.06\mathord\times10^{-32}],\\
 K_4(X)&\subset
 [0.8445620637795093243961336875765\pm8.70\mathord\times10^{-32}].
\end{align*}
The Krawczyk inclusion proves existence and uniqueness of an exact root in
$X$.

It remains to certify the inequalities that hold on continua of types.  The
endpoint-safe reductions used by the program are recorded here.  Define
\begin{align*}
 K_0(\theta)&\equiv\Lambda_sF(\theta)-B,
 &N(\theta)&\equiv K_0(\theta)-\Lambda_s\theta f(\theta)=-M(\theta),\\
 H_q(\theta)&\equiv(\theta+3)K_0(\theta)
                    -4\Lambda_s\theta f(\theta),
 &S_q(\theta)&\equiv\theta f(\theta)-K_0(\theta),\\
 \tau&\equiv2\Lambda_s-1.
\end{align*}
Direct differentiation gives
\begin{equation}
 \frac{\widetilde q'(\theta)}{\widetilde q(\theta)}
   =\frac{H_q(\theta)}{2\theta N(\theta)},
 \qquad
 H_q'(\theta)=K_0(\theta)+\Lambda_s(1-\theta)f(\theta)>0,
 \label{eq:ratings-cert-q-shape}
\end{equation}
and
\begin{equation}
 P'(\theta)
   =(1-\Lambda_s)f(\theta)
      \left[\frac{q_L}{\widetilde q(\theta)}-1\right],
 \qquad
 S_q'(\theta)
   =f(\theta)\left[\frac{\theta+1}{2}-\Lambda_s\right].
 \label{eq:ratings-cert-sign-reductions}
\end{equation}
Moreover, $K_0>0$ and
$N'(\theta)=\Lambda_s(1-\theta)f(\theta)/2>0$.  Thus the certified
inequality $N(\lbar\theta)=-M(\lbar\theta)>0$ keeps every denominator away
from zero.
The interval calculation certifies, with strict outward-rounded bounds,
\begin{gather}
 -M(\lbar\theta)>0.0015,
 \quad H_q(\theta_L)>0.0224,
 \quad \widetilde q(\lbar\theta)-q_L>0.0686,
 \quad w_L-q_L>0.0338,
 \label{eq:ratings-cert-certified-lower-margins}\\
 \theta_L<\tau<\theta_H,
 \quad S_q(\theta_L)>0.0045,
 \quad S_q(\tau)<-0.0226,
 \quad S_q(\theta_H)>0.0366.
 \label{eq:ratings-cert-certified-middle-margins}
\end{gather}
Equation~\eqref{eq:ratings-cert-q-shape} therefore proves
$\widetilde q'>0$ on $[\theta_L,\theta_H]$.  It also shows that
$\widetilde q$ has at most one turning point on
$[\lbar\theta,\theta_L]$.  Since
$\widetilde q(\lbar\theta)>q_L=\widetilde q(\theta_L)$ and
$P(\lbar\theta)=P(\theta_L)=0$, $\widetilde q$ cannot remain above $q_L$
throughout the open interval: otherwise~\eqref{eq:ratings-cert-sign-reductions}
would give $P(\theta_L)<P(\lbar\theta)$.  It therefore crosses $q_L$ once
before reaching its unique minimum and then increases to $q_L$ at
$\theta_L$.  Equation~\eqref{eq:ratings-cert-sign-reductions} now shows that
$P'$ first is negative and then positive.  Hence $P\leq0$ on the full lower
interval, including its two exact endpoint zeros.

For the MPS slack, let $g\equiv w^*-q^*$.  On the lower pool,
$D^*=(w_L-q_L)F>0$ away from $\lbar\theta$.  On the separating interval,
\[
 g'(\theta)=\widetilde q'(\theta)
     \left[\frac{\widetilde q(\theta)}{\theta}-1\right],
 \qquad
 \operatorname{sgn}\left[\frac{\widetilde q(\theta)}{\theta}-1\right]
   =\operatorname{sgn}S_q(\theta).
\]
The signs in~\eqref{eq:ratings-cert-certified-middle-margins} imply that $g$
increases, then decreases, and then increases.  Since $g(\theta_L)>0$,
(C-W) gives $g(\theta_H)=0$.  The final increasing arc is therefore negative
before $\theta_H$, so $g$ has exactly one sign-changing zero in
$(\theta_L,\theta_H)$ before its terminal zero at $\theta_H$.  Since
$D^{*\prime}=gf$, while (C-D) gives $D^*(\theta_H)=0$, $D^*$ first increases
and then decreases to zero.  Thus $D^*>0$ on $(\lbar\theta,\theta_H)$ and
$D^*=0$ on the upper contact interval.

Finally, put $Q=q_H$, $d(\theta)=Q-\theta$,
$a_u(\theta)=(1-\theta)/(2\theta)$, and $R=(1-F)/f$.  The expression in
braces in~\eqref{eq:ratings-cert-lambda-upper} is
\[
 L(\theta)=2[d(\theta)-R(\theta)]-a_u(\theta)d(\theta)R(\theta),
 \qquad
 \lambda_u(\theta)=\frac{Q}{d(\theta)^3}L(\theta).
\]
Using $R'=-1+a_uR$ gives
\begin{align*}
 L'(\theta)
 &=a_ud-\bigl[a_u+d(a_u'+a_u^2)\bigr]R,\\
 4\theta^2\bigl[a_u+d(a_u'+a_u^2)\bigr]
 &=-\theta^3+Q\theta^2+(3-2Q)\theta-Q.
\end{align*}
The program certifies
\begin{equation}
 \frac{3(\theta_H+1)}2-Q>1.758,
 \qquad
 2(1-\Lambda_s)
 -\frac{1-\theta_H}{2\theta_H}
   [\theta_H+\Lambda_s(Q-\theta_H)]>0.615.
 \label{eq:ratings-cert-certified-upper-margins}
\end{equation}
The first inequality makes the cubic in the preceding display increasing on
$[\theta_H,1]$; its value at $1$ is $-2(Q-1)<0$, so it is negative throughout.
Since $a_u,d>0$ and $R\geq0$ below the upper endpoint, this gives $L'>0$ on
$[\theta_H,1)$.  By
(FB-$\Gamma$) and the defining identity
$Q=\widetilde q(\theta_H)$, the second
inequality is equivalent to $L(\theta_H)>0$.  Therefore
$\lambda_u>0$ on $[\theta_H,1]$.

The same calculation certifies all remaining strict domain conditions in
\eqref{eq:ratings-cert-domain-conditions}; in particular,
$0<\Lambda_s<1$, $B<0$, $q_L>0.0378$, and
$1.0086<q_H<1.0087<2=q_{\max}$.  Consequently every hypothesis of
Proposition~\ref{prop:ratings-cert-optimality} holds.

\begin{corollary}[Certified optimality]
\label{cor:ratings-certified-optimality}
The mechanism $(q^*,w^*)$ in
\eqref{eq:ratings-cert-qstar}--\eqref{eq:ratings-cert-wstar} globally solves
problem~\ref{prob:P}.
\end{corollary}

\subsection{Certified Objective Comparison}
\label{sec:ratings-cert-objective-comparison}

Under the restriction to deterministic ratings, the optimal quality schedule
for this example is
\[
    q^D(\theta)=\max\{\theta,2\lbar\theta\}
    =
    \begin{cases}
        0.02,&\lbar\theta\leq\theta\leq0.02,\\
        \theta,&0.02<\theta\leq1.
    \end{cases}
\]
Its expected-quality value is
\begin{align}
    J_D
    &\equiv\int_{\lbar\theta}^{1}q^D(\theta)\d F(\theta)\notag\\
    &=0.02F(0.02)+\int_{0.02}^{1}\theta f(\theta)\d\theta\notag\\
    &=0.02F(0.02)
      +C\left[
          2\sqrt{\theta}e^{\theta/2}
          -\sqrt{2\pi}\,
           \erfi\!\left(\sqrt{\frac{\theta}{2}}\right)
        \right]_{0.02}^{1}\notag\\
    &\in
      [0.4143711206552709,\,0.4143711206552711].
    \label{eq:ratings-cert-deterministic-value}
\end{align}
At the exact root isolated above, the candidate's expected-quality value is
\begin{align}
    J^*
    &\equiv q_LF(\theta_L)
      +\int_{\theta_L}^{\theta_H}
         \widetilde q(\theta)f(\theta)\d\theta
      +q_H[1-F(\theta_H)]\notag\\
    &\in
      [0.4159053676869892,\,0.4159053676869894].
    \label{eq:ratings-cert-candidate-value}
\end{align}
Validated propagation of the root box through both objectives gives
\begin{align}
    J^*-J_D
    &\in[0.0015342470317183,\,0.0015342470317185],\notag\\
    \frac{J^*-J_D}{J_D}
    &\in[0.0037025916026487,\,0.0037025916026489].
    \label{eq:ratings-cert-objective-gain}
\end{align}
In particular, the candidate's value is strictly larger, and the relative gain
is approximately $0.3703\%$.

\subsection{A Feasible Construction for a Larger Gain}
\label{sec:ratings-cert-large-gain-construction}

This subsection establishes only the existence and value of a feasible
stochastic rating scheme that satisfies all constraints of problem~\ref{prob:P}.
It does not use the multipliers in
Theorem~\ref{thm:certificate}. Nor does it claim that the
constructed mechanism is optimal.

Retain $c(q)=q^2/2$, $v(q,\theta)=q$, and
$\Theta=[\lbar\theta,1]$, where $\lbar\theta=0.01$.  Consider the density
\begin{equation}
 f(\theta)
 =Z\left[\varepsilon+\exp\{-k(\theta-\lbar\theta)\}\right],
 \qquad
 \varepsilon=0.005,
 \qquad
 k=100,
 \label{eq:ratings-cert-large-gain-density}
\end{equation}
where
\[
 Z=
 \left[
   \varepsilon(1-\lbar\theta)
   +\frac{1-\exp\{-k(1-\lbar\theta)\}}{k}
 \right]^{-1}
 \approx66.8896321070.
\]

\begin{proposition}[A feasible $10.69\%$ improvement]
\label{prop:ratings-cert-large-gain}
For the density in~\eqref{eq:ratings-cert-large-gain-density}, there exists a
feasible stochastic rating scheme whose expected quality exceeds that of the
optimal deterministic rating scheme by more than $10.69\%$.
\end{proposition}

\begin{proof}
Set $\theta_L=0.134$.  For $b>\theta_L$, define
\begin{equation}
 K(b)
 =
 \frac{b^2}
 {\displaystyle
   \frac{\theta_L^4}{2\lbar\theta}
   +\frac23(b^3-\theta_L^3)}.
 \label{eq:ratings-cert-large-gain-K}
\end{equation}
For the moment, fix $b$ and write $K=K(b)$.  Consider the direct mechanism
\begin{equation}
 q_b(\theta)=
 \begin{cases}
  K\theta_L^2,
    &\lbar\theta\leq\theta\leq\theta_L,\\
  K\theta^2,
    &\theta_L<\theta<b,\\
  Kb^2,
    &b\leq\theta\leq1,
 \end{cases}
 \label{eq:ratings-cert-large-gain-q}
\end{equation}
and
\begin{equation}
 w_b(\theta)=
 \begin{cases}
  \displaystyle\frac{K^2\theta_L^4}{2\lbar\theta},
    &\lbar\theta\leq\theta\leq\theta_L,\\[2mm]
  \displaystyle K^2\left[
    \frac{\theta_L^4}{2\lbar\theta}
    +\frac23(\theta^3-\theta_L^3)
  \right],
    &\theta_L<\theta<b,\\[2mm]
  Kb^2,
    &b\leq\theta\leq1.
 \end{cases}
 \label{eq:ratings-cert-large-gain-w}
\end{equation}
Equation~\eqref{eq:ratings-cert-large-gain-K} makes both schedules continuous
at $b$ because $w_b(b)=q_b(b)=Kb^2$.

\begin{figure}[htb]
\centering
\includegraphics[width=0.5\textwidth]{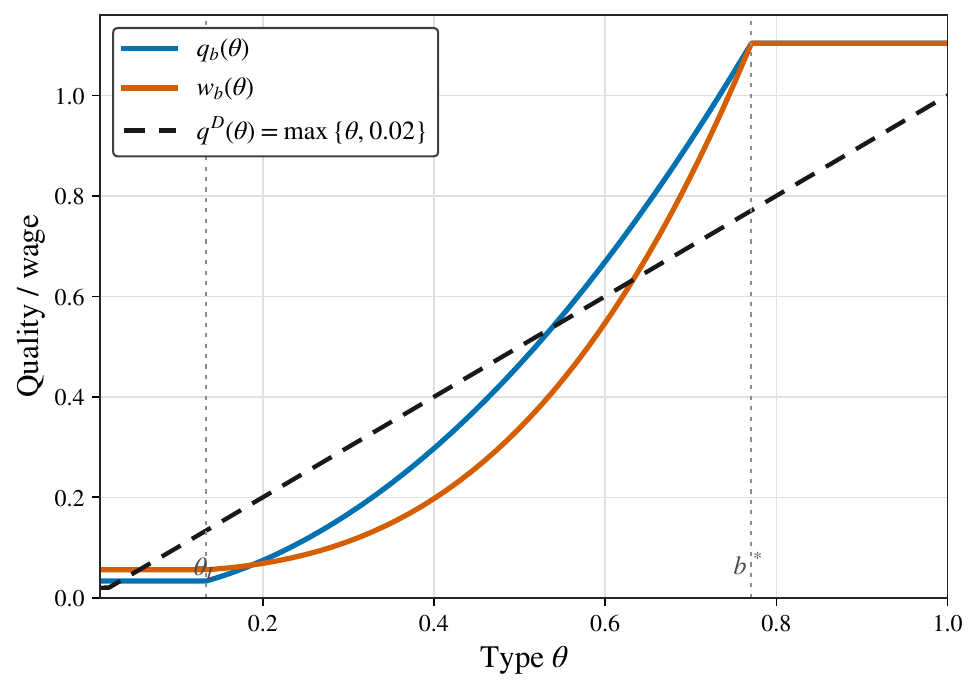}
\caption{A feasible $(q,w)$ with a $10.69\%$ gain over $q^D$.}
\end{figure}

We next select $b$ so that Bayesian plausibility holds.  Put
\[
 g(\theta)=\varepsilon+\exp\{-k(\theta-\lbar\theta)\},
 \qquad
 I_j(x,y)=\int_x^y\theta^j g(\theta)\d\theta,
\]
and define the unnormalized terminal MPS slack
\[
 \widetilde\Phi(b)
 =\int_{\lbar\theta}^{b}
    [w_b(\theta)-q_b(\theta)]g(\theta)\d\theta.
\]
Writing
$C_0=\theta_L^4/(2\lbar\theta)-2\theta_L^3/3$, direct integration gives
\begin{align}
 \widetilde\Phi(b)
 = 
 \left(\frac{K^2\theta_L^4}{2\lbar\theta}-K\theta_L^2\right)
 I_0(\lbar\theta,\theta_L)
 +K^2C_0I_0(\theta_L,b) +\frac23K^2I_3(\theta_L,b)-KI_2(\theta_L,b).
 \label{eq:ratings-cert-large-gain-Phi}
\end{align}
Every term is elementary.  In particular,
\begin{align*}
 I_j(x,y)
 &=\frac{\varepsilon}{j+1}(y^{j+1}-x^{j+1})
   +A_j(y)-A_j(x),\\
 A_j(t)
 &=-e^{-k(t-\lbar\theta)}
   \sum_{r=0}^{j}
   \frac{j!}{(j-r)!}\frac{t^{j-r}}{k^{r+1}}.
\end{align*}
Thus $\widetilde\Phi$ is continuous.  An outward-rounded $80$-digit Arb
evaluation of~\eqref{eq:ratings-cert-large-gain-Phi} gives, for
\[
 b_-=0.770734451623,
 \qquad
 b_+=0.770734451626,
\]
the strict enclosures
\begin{align*}
 \widetilde\Phi(b_-)
 &\in[3.1357,3.1359]\mathord\times10^{-15},\\
 \widetilde\Phi(b_+)
 &\in[-2.7969,-2.7967]\mathord\times10^{-15}.
\end{align*}
The intermediate value theorem therefore gives some
$b^*\in(b_-,b_+)$ such that $\widetilde\Phi(b^*)=0$.  Numerically,
\begin{equation}
 b^*\approx0.7707344516245857,
 \qquad
 K(b^*)\approx1.857835556986241.
 \label{eq:ratings-cert-large-gain-parameters}
\end{equation}

Fix such a $b^*$ and suppress the subscript $b^*$.  The schedules $q$ and
$w$ are continuous and increasing.  On the separating interval,
\[
 q'(\theta)=2K\theta,
 \qquad
 w'(\theta)=2K^2\theta^2,
 \qquad
 \theta w'(\theta)=q(\theta)q'(\theta),
\]
while both derivatives vanish on the two pooling intervals.  Hence, for
$U(\theta)=\theta w(\theta)-q(\theta)^2/2$,
\[
 U'(\theta)=w(\theta),
 \qquad
 U(\lbar\theta)
 =\lbar\theta\frac{K^2\theta_L^4}{2\lbar\theta}
  -\frac{(K\theta_L^2)^2}{2}
 =0.
\]
Thus (IC-Env), (IC-Mon), and (IR) hold.

It remains to verify (MPS).  Let $h=w-q$.  On the lower pool,
\[
 h
 =K\theta_L^2
   \left(\frac{K\theta_L^2}{2\lbar\theta}-1\right)>0.
\]
On $(\theta_L,b^*)$,
\begin{equation}
 h'(\theta)=2K\theta(K\theta-1).
 \label{eq:ratings-cert-large-gain-hprime}
\end{equation}
The same interval calculation certifies
\[
 K\theta_L^2-2\lbar\theta>0.0133,
 \qquad
 \theta_L<\frac1K<b^*,
 \qquad
 1<K\,(b^*)^2<2.
\]
Moreover, $h(b^*)=0$ by construction.  Equation
\eqref{eq:ratings-cert-large-gain-hprime} therefore implies that $h$ is
initially positive, crosses zero exactly once, and is negative until it
returns to zero at $b^*$.  Consequently,
\[
 D(\theta)
 =\int_{\lbar\theta}^{\theta}[w(x)-q(x)]\d F(x)
\]
first increases and then decreases to
$D(b^*)=Z\widetilde\Phi(b^*)=0$.  Since $w=q=K\,(b^*)^2$ above $b^*$,
$D(\theta)=0$ there.  Hence $D\geq0$ everywhere with $D(1)=0$, proving
(MPS).  The quality bound also holds because
$q(1)=K\,(b^*)^2\approx1.1036130190<2=q_{\max}$.

For completeness, this feasible direct mechanism has an explicit stochastic
rating implementation.  Let $\theta_m$ be the unique interior zero of $h$ and
write $q_L=K\theta_L^2$, $q_H=K\,(b^*)^2$, and
\[
 q^m=q(\theta_m)=w(\theta_m)
 \approx0.06525166079,
 \qquad
 \theta_m\approx0.18740973633.
\]
Because $w$ is constant on every quality fiber, define
$\widehat w(q(\theta))=w(\theta)$.  For every on-path
$q\in[q_L,q_H]$, reveal $q$ with probability
\begin{equation}
 p(q)=\frac{\widehat w(q)-q^m}{q-q^m},
 \qquad
 p(q^m)=\widehat w'(q^m)=K\theta_m,
 \label{eq:ratings-cert-large-gain-p}
\end{equation}
and send a common signal $\mathsf{pass}$ otherwise.  The single-crossing of
$\widehat w(q)$ and $q$ implies that $\widehat w(q)$ lies between $q$ and
$q^m$, so $p(q)\in[0,1]$.  Moreover,
\[
 (1-p(q))(q-q^m)=q-\widehat w(q),
\]
and $D(1)=0$ implies that the posterior mean following $\mathsf{pass}$ is
exactly $q^m$. Complete the scheme off path by sending \(\mathsf{fail}\) for
\(q<q_L\) and revealing \(q\) for \(q>q_H\). In the supporting PBE,
$\mu_{\mathsf{fail}} = \delta_0$ and hence \(\omega(\mathsf{fail})=0\). Individual rationality
therefore rules out profitable downward deviations, while $q_H>1=\bar\theta$ makes every
full-revelation payoff decreasing above $q_H$.  Thus the rating scheme
implements~\eqref{eq:ratings-cert-large-gain-q}.

Finally, because the density is decreasing, the optimal deterministic
schedule is $q^D(\theta)=\max\{\theta,2\lbar\theta\}$.  In terms of the
moments above,
\begin{align*}
 V^S
 &=Z\left[
   K\theta_L^2I_0(\lbar\theta,\theta_L)
   +KI_2(\theta_L,b^*)
   +K\,(b^*)^2I_0(b^*,1)
 \right],\\
 V^D
 &=Z\left[
   2\lbar\theta I_0(\lbar\theta,2\lbar\theta)
   +I_1(2\lbar\theta,1)
 \right].
\end{align*}
Outward-rounded evaluation on $[b_-,b_+]$ gives
\begin{align*}
 V^S&\approx0.2026476391389674,
 &V^D&\approx0.1830627387369327,\\
 \frac{V^S-V^D}{V^D}
 &\in[0.1069846355,0.1069846357].
\end{align*}
In particular,
$V^S-1.1069V^D>1.54\mathord\times10^{-5}$, so this feasible stochastic
rating scheme raises expected quality by more than $10.69\%$ relative to the
optimal deterministic benchmark.
\end{proof}

\section{Deferred Proofs}
\label{sec:deferredproofs}

\subsection{Single-peakedness and Conditions (S) and (C)}

Use the maintained
assumptions on $c$ and $f$, and extend $F=f=0$ below
$\underline\theta$ and $F=1$, $f=0$ above $\bar\theta$. Recall that
$\theta_c(t)=c'(q_i(t))$, where $c(q_i(t))=tq_i(t)$. Set
$\theta_c(0)=0$ and define
\[
    A(t):=\frac{F(\theta_c(t))-F(t)}{\theta_c(t)-t},
    \qquad 0<t\leq\bar\theta,
    \qquad A(0):=0.
\]
The value at zero is the continuous extension: since
$\theta_c(t)\to0$ and $\underline\theta>0$, the numerator vanishes
for all sufficiently small positive $t$.

\begin{figure}[htb]
    \centering
    \begin{subfigure}[b]{0.425\textwidth}
        \centering
        \includegraphics[width=\textwidth]{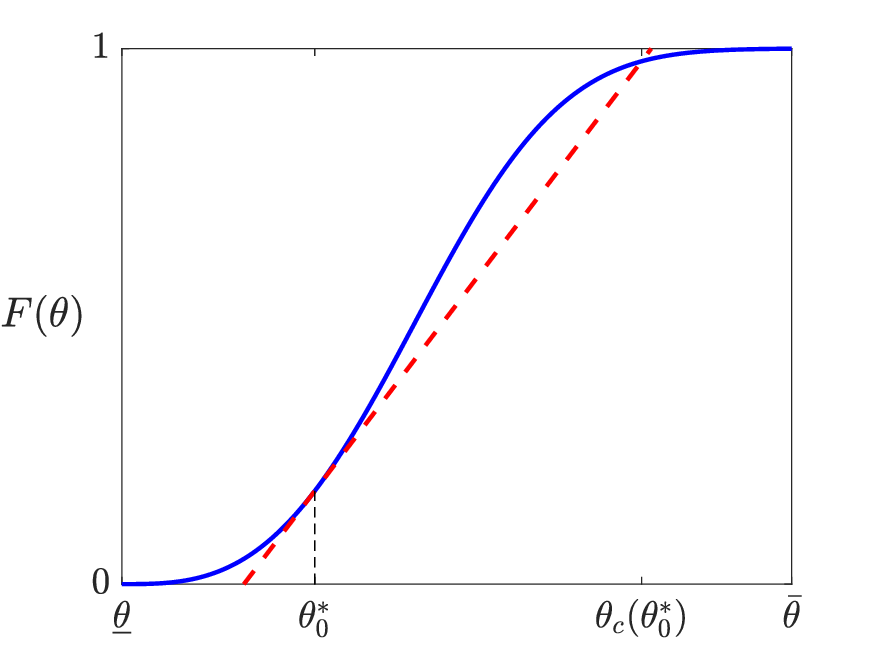}
        \caption{Unimodal $f$ with $\theta_c(\theta_0^*)<\bar\theta$}
    \end{subfigure}
    \hspace{10pt}
    \begin{subfigure}[b]{0.425\textwidth}
        \centering
        \includegraphics[width=\textwidth]{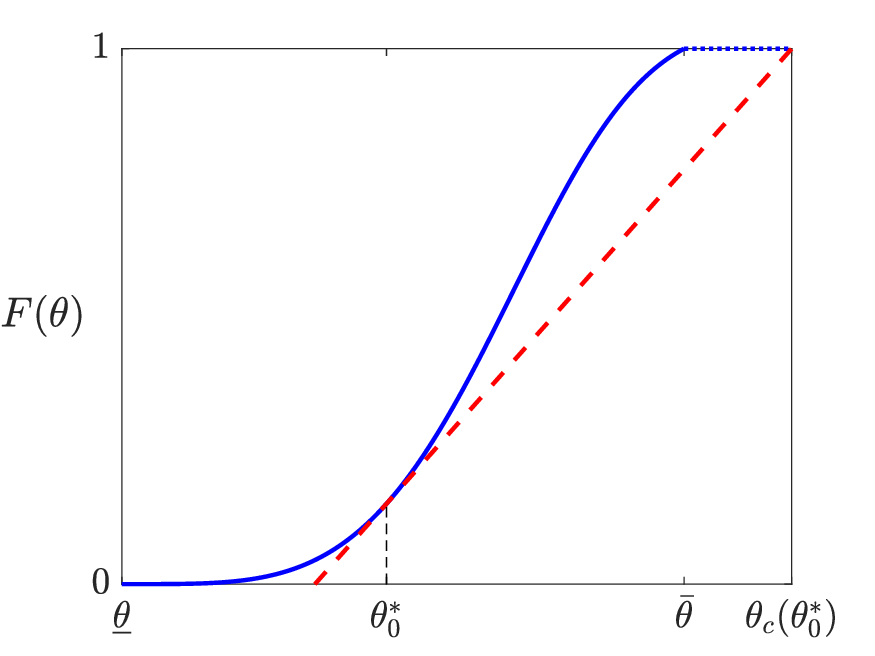}
        \caption{Unimodal $f$ with $\theta_c(\theta_0^*)>\bar\theta$}
        \label{fig:F:d}
    \end{subfigure}

    \medskip
    \begin{subfigure}[b]{0.425\textwidth}
        \centering
        \includegraphics[width=\textwidth]{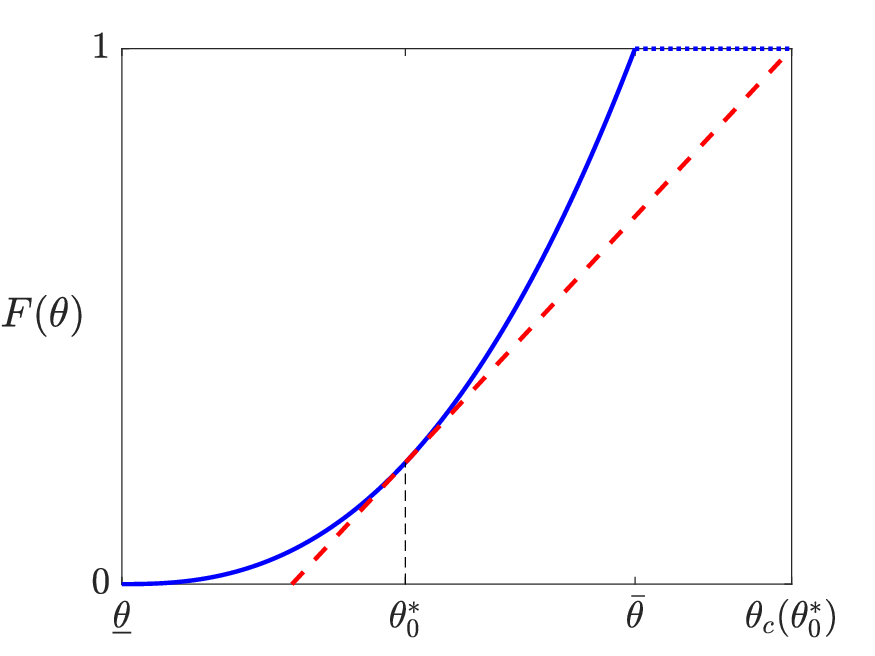}
        \caption{Increasing $f$}
    \end{subfigure}
    \hspace{10pt}
    \begin{subfigure}[b]{0.425\textwidth}
        \centering
        \includegraphics[width=\textwidth]{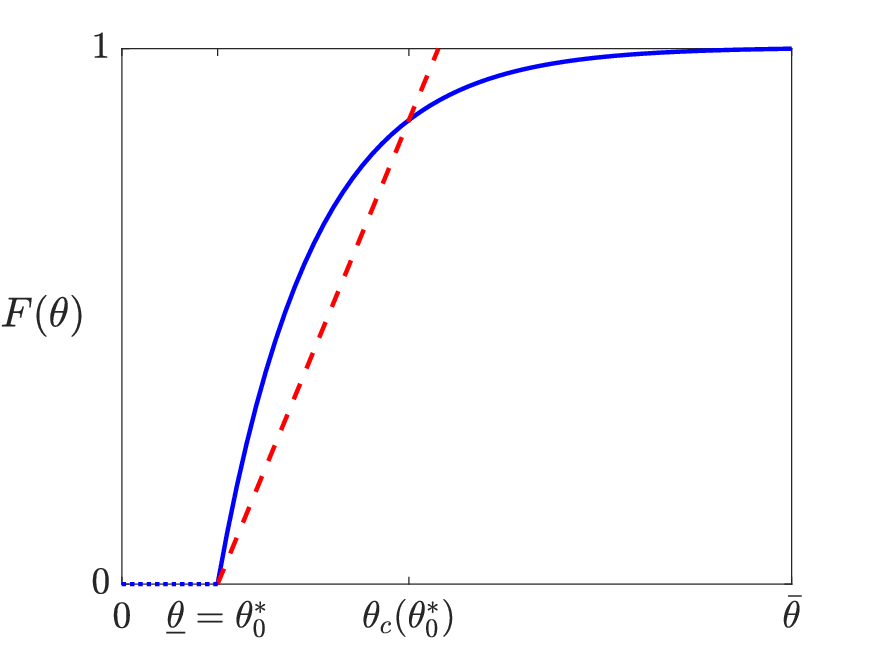}
        \caption{Decreasing $f$}
    \end{subfigure}
    \caption{A single-peaked density admits a cutoff satisfying
    Conditions~\eqref{S2} and~\eqref{C2}.}
    \label{fig:F}
\end{figure}

\begin{lemma}\label{lemma:single-peaked2}
If $f$ is single-peaked, then there exists
$\theta_0^*\in[0,\bar\theta)$ satisfying the following conditions:
\begin{description}
    \nameditem{S}{S2}
    $F(\theta)-F(\theta_0^*)\geq
    (\theta-\theta_0^*)A(\theta_0^*)$
    for every $\theta\in[0,\theta_c(\theta_0^*)]$,
    with equality at $\theta=\theta_c(\theta_0^*)$.

    \nameditem{C}{C2}
    $f$ is nonincreasing on
    $[\theta_c(\theta_0^*),\bar\theta]$.
\end{description}
If the mode $\theta_m\in(\underline\theta,\bar\theta]$ is unique,
then the solution is unique and satisfies
$\theta_c^{-1}(\theta_m)<\theta_0^*<\theta_m$.
If $f$ is nonincreasing, then $\theta_0^*=\underline\theta$
satisfies both conditions.
\end{lemma}

\begin{proof}[Proof of Lemma~\ref{lemma:single-peaked2}]
Let $\theta_m$ be a mode of $f$. The extended distribution $F$ is
convex on $[0,\theta_m]$ and concave on $[\theta_m,\infty)$.
We first construct a line that supports the convex part from below
and meets the concave part at $\theta_c(\theta_0^*)$.

The cost assumptions imply that $\theta_c$ is continuous and
strictly increasing, with $\theta_c(t)>t$ for $t>0$,
$\theta_c(0)=0$, and $\theta_c(\bar\theta)>\bar\theta$.
Thus $a:=\theta_c^{-1}(\theta_m)$ is well-defined and satisfies
$0<a<\theta_m$. Put
$s:=\max\{a,\underline\theta\}$ and
$\Delta(t):=A(t)-f(t)$ for $t\in[\underline\theta,\bar\theta]$.
Since $A(\theta_m)$ averages the nonincreasing density to the right
of the mode, $\Delta(\theta_m)\leq0$.

If $s=\underline\theta$ and $\Delta(s)\leq0$, choose
$\theta_0^*=\underline\theta$. Otherwise, $\Delta(s)\geq0$:
this is immediate when $s=\underline\theta$, while for
$s=a>\underline\theta$ it follows from
\[
    A(a)=\frac{1}{\theta_m-a}\int_a^{\theta_m}f(x)\d x
    \geq f(a).
\]
Continuity of $\Delta$ then gives a root
$\theta_0^*\in[s,\theta_m]$, so $A(\theta_0^*)=f(\theta_0^*)$.
The selected cutoff lies below $\bar\theta$: this is immediate if
$\theta_m<\bar\theta$, and if $\theta_m=\bar\theta$ it follows from
$\Delta(\bar\theta)=-f(\bar\theta)<0$.

In either case,
$\underline\theta\leq\theta_0^*\leq\theta_m
\leq\theta_c(\theta_0^*)$, and the line
\[
    \ell(\theta)
    :=F(\theta_0^*)+A(\theta_0^*)(\theta-\theta_0^*)
\]
lies below $F$ on $[0,\theta_m]$. Indeed, at an interior cutoff it
is a tangent to the convex function $F$. At
$\theta_0^*=\underline\theta$, its slope lies between zero and
$f(\underline\theta)$, the left and right slopes of $F$ at the
support boundary, so it is again a supporting line.

By definition of $A$, $F-\ell$ equals zero at
$\theta_c(\theta_0^*)$. On
$[\theta_m,\theta_c(\theta_0^*)]$, this difference is concave,
nonnegative at the left endpoint, and zero at the right endpoint.
It is therefore nonnegative throughout. This proves
Condition~\eqref{S2}. Since $\theta_c(\theta_0^*)\geq\theta_m$,
Condition~\eqref{C2} also holds; its interval is empty if
$\theta_c(\theta_0^*)>\bar\theta$.
If $f$ is nonincreasing, take $\theta_m=\underline\theta$.
The construction then selects $\theta_0^*=\underline\theta$,
proving the last claim.

We next prove uniqueness when the mode
$\theta_m>\underline\theta$ is unique. Let $t\in[0,\bar\theta)$
satisfy Conditions~\eqref{S2} and~\eqref{C2}. First,
Condition~\eqref{C2} implies $\theta_c(t)\geq\theta_m$.
Otherwise, $f$ would be both nondecreasing and nonincreasing on
$[\max\{\theta_c(t),\underline\theta\},\theta_m]$, making the mode
nonunique. In particular, $t\geq a>0$. If $t<\underline\theta$,
then $A(t)>0$ because $\theta_c(t)\geq\theta_m>\underline\theta$,
whereas Condition~\eqref{S2} at $\underline\theta$ would give
$0\geq(\underline\theta-t)A(t)>0$. Hence $t\geq\underline\theta$.

The one-sided limits of Condition~\eqref{S2} at $t$ give
\[
    A(t)\leq f(t),
    \qquad A(t)=f(t)\ \text{if }t>\underline\theta.
\]
We claim that $t<\theta_m<\theta_c(t)$. If $t=\theta_m$, the
average density over $[t,\theta_c(t)]$ is strictly below $f(t)$
because the mode is unique, contradicting $A(t)=f(t)$.
If $t>\theta_m$, Condition~\eqref{S2} at $\theta_m$ requires
\[
    \frac{1}{t-\theta_m}\int_{\theta_m}^{t}f(x)\d x
    \leq A(t)=f(t).
\]
But weak decrease and uniqueness of the mode make the average on
the left strictly greater than $f(t)$. Thus $t<\theta_m$.
Similarly, $\theta_c(t)=\theta_m$ would give $A(t)>f(t)$ by
averaging the nondecreasing density over $[t,\theta_m]$, again a
contradiction. Consequently, every solution satisfies
\[
    \underline\theta\leq t<\theta_m<\theta_c(t).
\]

Suppose there are two solutions $t_1<t_2$, and define
\[
    \ell_i(\theta):=F(t_i)+A(t_i)(\theta-t_i),
    \qquad i=1,2.
\]
Since $t_2>\underline\theta$ and $f$ is nondecreasing before the
mode,
\[
    A(t_2)=f(t_2)\geq f(t_1)\geq A(t_1).
\]
Condition~\eqref{S2} for $t_1$ also gives
$\ell_2(t_2)=F(t_2)\geq\ell_1(t_2)$. Thus $\ell_2$ starts weakly
above $\ell_1$ at $t_2$ and has a weakly larger slope. In particular,
\[
    \ell_2(\theta_c(t_1))
    \geq\ell_1(\theta_c(t_1))
    =F(\theta_c(t_1)).
\]

On the other hand,
\[
    F(\theta_m)-\ell_2(\theta_m)
    =\int_{t_2}^{\theta_m}[f(x)-f(t_2)]\d x>0.
\]
The integrand is nonnegative by weak increase, and the unique mode
makes it strictly positive near $\theta_m$ by continuity.
The function $F-\ell_2$ is concave on
$[\theta_m,\theta_c(t_2)]$, is strictly positive at $\theta_m$,
and is zero at $\theta_c(t_2)$. Since strict monotonicity of
$\theta_c$ gives
$\theta_m<\theta_c(t_1)<\theta_c(t_2)$, concavity implies
$F(\theta_c(t_1))>\ell_2(\theta_c(t_1))$, a contradiction.
The solution is therefore unique. Its strict location implies
$\theta_c^{-1}(\theta_m)<\theta_0^*<\theta_m$, as claimed.
\end{proof}

\subsection{Necessary Condition in Fully Revealing Regions}

\begin{lemma}[Necessary Condition in Fully Revealing Regions]
\label{lem:rho2}
If $(q^*,w^*)$ is optimal, then on any interval $I$ where
$q^*(\theta)=w^*(\theta)=q_f(\theta)$,
\[
    \rho(\theta)
    =\frac{[\theta v_q(q_f(\theta),\theta)f(\theta)]'}{f(\theta)}
\]
is nonincreasing, and is constant if \emph{(MPS)} is slack on $I$,
that is, if $D^*(\theta)>0$ for every $\theta\in I$.
\end{lemma}

\begin{proof}[Proof of Lemma~\ref{lem:rho2}]
Work first on an open fully revealing interval
$I\subseteq\operatorname{int}\Theta$. Since $w^*=q^*$ on $I$, the
MPS state $D^*$ is constant there; write $D^*=D_0\geq0$.
We compare two types by perturbing the mechanism in a small
neighborhood of each. The perturbation creates MPS slack near the
lower type and removes it near the higher type. Adjusting the size
of the latter perturbation restores the terminal equality.

Fix $\theta_1<\theta_2$ in $I$. Choose nonnegative
$a,b\in C_c^\infty(I)$ supported in small neighborhoods of
$\theta_1$ and $\theta_2$, respectively, with the support of $a$
strictly to the left of that of $b$, and normalize them so that
\[
    \int_I a\d F=\int_I b\d F=1.
\]
Take $b$ to be a normalized square of a nonzero smooth function.
Then $(b')^2\leq C_b b$ for some finite constant $C_b$; this bound
will ensure that the right perturbation removes slack throughout
its support. Extend $a$ and $b$ by zero outside $I$, and choose a
compact interval $J\Subset I$ containing both supports in its
interior.

For small parameters $(\varepsilon,t)$, put
$g=\varepsilon a-tb$ and define
\[
\begin{aligned}
    U_{\varepsilon,t}(\theta)
        &:=U^*(\theta)+\theta g(\theta),\\
    w_{\varepsilon,t}(\theta)
        &:=w^*(\theta)+(\theta g(\theta))',\\
    q_{\varepsilon,t}(\theta)
        &:=c^{-1}\!\left(c(q^*(\theta))+\theta^2g'(\theta)\right).
\end{aligned}
\]
These schedules coincide with the candidate outside $J$ and
satisfy $U_{\varepsilon,t}'=w_{\varepsilon,t}$ almost everywhere and
$\theta w_{\varepsilon,t}-c(q_{\varepsilon,t})=U_{\varepsilon,t}$
exactly. Thus \emph{(IC-Env)} holds, and the lower-endpoint
\emph{(IR)} constraint is unchanged.

The remaining range and monotonicity constraints are also preserved
for sufficiently small parameters. Indeed, $q_f(J)$ lies in the
interior of $Q$, and $q_f'=1/c''(q_f)$ is bounded away from zero on
$J$. Hence $w_{\varepsilon,t}'=q_f'+(\theta g)''>0$ on $J$,
while both schedules remain in $Q$. Differentiating the exact
envelope identity gives
$c'(q_{\varepsilon,t})q_{\varepsilon,t}'
=\theta w_{\varepsilon,t}'>0$ on $J$, so quality is increasing as
well. Since the perturbations vanish near the ends of $J$, both
schedules remain nondecreasing on all of $\Theta$.

We now choose $t$ to preserve the terminal MPS equality. Define
\[
    G(\varepsilon,t)
    :=\int_I(w_{\varepsilon,t}-q_{\varepsilon,t})\d F.
\]
The terminal equality holds precisely when $G(\varepsilon,t)=0$,
because $w^*=q^*$ on $I$ and the mechanism is unchanged elsewhere.
All parameter dependence is confined to $J$, where
$c'(q_f(\theta))=\theta>0$, so $G$ is $C^2$ near the origin.
Using this identity and the normalization of $a$ and $b$ gives
\[
    G(0,0)=0,\qquad
    G_\varepsilon(0,0)=\int_I a\d F=1,\qquad
    G_t(0,0)=-\int_I b\d F=-1.
\]
The implicit-function theorem therefore yields a $C^2$ function
$t(\varepsilon)$, defined for both signs of sufficiently small
$\varepsilon$, such that
\[
    G(\varepsilon,t(\varepsilon))=0,
    \qquad t(\varepsilon)=\varepsilon+O(\varepsilon^2).
\]
In particular, $t(\varepsilon)>0$ for small $\varepsilon>0$.
Write $(q_\varepsilon,w_\varepsilon)
:=(q_{\varepsilon,t(\varepsilon)},w_{\varepsilon,t(\varepsilon)})$
and $g_\varepsilon:=\varepsilon a-t(\varepsilon)b$.

Concavity and Taylor expansion of $c^{-1}$ give, uniformly on $J$,
\[
    q_\varepsilon=q_f+\theta g_\varepsilon'-R_\varepsilon,
    \qquad 0\leq R_\varepsilon\leq C(g_\varepsilon')^2.
\]
Consequently, $w_\varepsilon-q_\varepsilon
=g_\varepsilon+R_\varepsilon$ on $J$. For $\varepsilon>0$, this
gap is nonnegative on the support of $a$. On the support of $b$,
putting $t=t(\varepsilon)>0$ gives
\[
    w_\varepsilon-q_\varepsilon
    \leq-tb+Ct^2(b')^2
    \leq-tb(1-CC_b t)\leq0
\]
for sufficiently small $\varepsilon$. Thus, as $\theta$ increases,
the cumulative change $D_\varepsilon-D^*$ first rises over the left
support, remains constant between the supports, and then falls
over the right support. It returns to zero because
$G(\varepsilon,t(\varepsilon))=0$. It therefore stays nonnegative
throughout, and \emph{(MPS)} holds exactly.

If $D_0>0$, the same construction is feasible for both signs of
sufficiently small $\varepsilon$. Indeed,
$\|D_\varepsilon-D^*\|_\infty=O(|\varepsilon|)$, and this change
vanishes outside $J$. The positive slack $D_0$ on $J$ therefore
preserves all cumulative inequalities also when $\varepsilon<0$.

Finally, $t'(0)=1$ gives the uniform expansion
$q_\varepsilon=q^*+\varepsilon\theta(a'-b')+O(\varepsilon^2)$.
Optimality and integration by parts imply
\begin{align}
    0&\geq
    \left.\frac{\d}{\d\varepsilon}
        \int_\Theta v(q_\varepsilon(\theta),\theta)\d F(\theta)
    \right|_{\varepsilon=0+}\nonumber\\
    &=\int_I\theta v_q(q_f(\theta),\theta)f(\theta)
        [a'(\theta)-b'(\theta)]\d\theta\nonumber\\
    &=\int_I\rho(\theta)[b(\theta)-a(\theta)]\d F(\theta).
    \label{eq:rho-variational-inequality}
\end{align}
The boundary term vanishes because $a$ and $b$ have compact support
in $I$. If $D_0>0$, feasibility for both signs of $\varepsilon$
makes the derivative equal to zero.

Letting the normalized supports of $a$ and $b$ shrink to
$\theta_1$ and $\theta_2$, respectively, and using continuity of
$\rho$, we obtain
\[
    \rho(\theta_2)\leq\rho(\theta_1),
    \qquad\text{with equality if }D_0>0.
\]
Since the two types were arbitrary, $\rho$ is nonincreasing on $I$
and constant when \emph{(MPS)} is slack there. Continuity extends
both conclusions to any included endpoints.
\end{proof}

\subsection{Proof of the Claim in Example~\ref{ex:alpha} (Partial Cost Internalization)}
\label{app:alpha-all-ratings}
\begin{proposition}
    Assume that $v(q,\theta)=q-\alpha\frac{c(q)}{\theta}$, where $\alpha \in [0,1)$. 
    If $f$ satisfies IED on $\Theta$, then lower censorship is optimal among all ratings.
\end{proposition}
\begin{proof}Fix \(\alpha\in[0,1)\), and define
\[
T(\theta)
:=
F_{\zeta}(\bar\theta)-F_{\zeta}(\theta),
\qquad
\phi(\theta)
=
(1-\alpha)f(\theta)+\alpha T(\theta),
\qquad
\Phi(\theta)
=
\int_0^\theta\phi(x)\d x.
\]
Thus \(T'(\theta)=-f(\theta)/\theta\) almost everywhere on
\(\Theta\). Recall that
\[
\epsilon_f(\theta)
=
-\frac{\theta f'(\theta)}{f(\theta)}.
\]
On \(\Theta\),
\begin{align}
\phi'(\theta)
&=
-\frac{f(\theta)}{\theta}
\left[
\alpha+(1-\alpha)\epsilon_f(\theta)
\right],
\label{eq:alpha-phi-prime}\\
\rho(\theta)
&:=
1+\frac{\theta\phi'(\theta)}{f(\theta)}
=
(1-\alpha)\left[1-\epsilon_f(\theta)\right].
\label{eq:alpha-rho}
\end{align}
Consequently, IED implies that \(\phi'\) changes sign at most once,
from positive to negative, and that \(\rho\) is decreasing.

By Proposition~\ref{prop:lineardelegation} and its supporting-line
construction, there is an optimal deterministic lower-censorship
rating with cutoff \(t\). Let
\[
r:=\theta_c(t),
\qquad
s:=\max\{t,\lbar\theta\},
\qquad
A:=
\frac{\Phi(r)-\Phi(t)}{r-t}.
\]
The cutoff satisfies
\begin{equation}
\Phi(\theta)-\Phi(t)
\geq
(\theta-t)A
\quad
\text{for every }\theta\in[0,r],
\qquad
\Phi(r)-\Phi(t)=(r-t)A,
\label{eq:alpha-support}
\end{equation}
and \(\phi\) is decreasing on
\([r,\bar\theta]\) whenever \(r<\bar\theta\).

Because \(\alpha<1\), the virtual density has a strict upward jump at
the lower boundary:
\[
\phi(\lbar\theta+)-\phi(\lbar\theta-)
=
(1-\alpha)f(\lbar\theta)>0.
\]
Hence the deterministic optimum can be selected with \(s<r\).
Indeed, if \(r\leq\lbar\theta\), the induced allocation is fully
revealing. For a minimum standard slightly above
\(q_f(\lbar\theta)\), the cutoff remains below \(\lbar\theta\), while
a positive interval of the lowest types increases its quality.
This strictly raises the principal's payoff because
\[
v_q(q_f(\theta),\theta)=1-\alpha>0,
\]
contradicting optimality of full revelation.

Let \((q^*,w^*)\) denote the lower-censorship candidate. It satisfies
\(q^*=w^*\), and hence \(D^*\equiv0\). Moreover,
\[
c'(q^*(\theta))
=
\begin{cases}
0,
& \lbar\theta\leq\theta<s,\\
r,
& s\leq\theta<\min\{r,\bar\theta\},\\
\theta,
& r\leq\theta\leq\bar\theta.
\end{cases}
\]

It is convenient to construct a transformed costate
\(\widetilde\Gamma\) and then set
\begin{equation}
\Gamma(\theta)
=
\widetilde\Gamma(\theta)+\alpha T(\theta),
\qquad
\gamma(\theta)
=
\widetilde\Gamma'(\theta)
-\alpha\frac{f(\theta)}{\theta}.
\label{eq:alpha-transform}
\end{equation}
Since \(v_q(q,\theta)=1-\alpha c'(q)/\theta\), the control and costate
equations in Theorem~\ref{thm:certificate} then reduce to
\begin{align}
[1-\Lambda(\theta)]f(\theta)
&=
c'(q^*(\theta))\widetilde\Gamma'(\theta),
\label{eq:alpha-q-foc}\\
\mu'(\theta)
&=
-\left[
\phi(\theta)+\widetilde\Gamma(\theta)
+
\bigl(\theta-c'(q^*(\theta))\bigr)
\widetilde\Gamma'(\theta)
\right].
\label{eq:alpha-mu-costate}
\end{align}
Furthermore, because \(\kappa(\theta)=\alpha/\theta\),
\begin{equation}
\Gamma(\theta)
+
\int_{\lbar\theta}^{\theta}
\kappa(x)f(x)\d x
=
\widetilde\Gamma(\theta)+\alpha T(\lbar\theta).
\label{eq:alpha-concavity-transform}
\end{equation}
Thus the concavity condition in Theorem~\ref{thm:certificate} is
equivalent to \(\widetilde\Gamma\) being increasing.

\medskip
\noindent
\emph{Case 1: \(r<\bar\theta\).}
For \(\theta\in[s,r]\), define
\[
R_\phi(\theta)
:=
\frac{\Phi(r)-\Phi(\theta)}{r-\theta},
\qquad
R_\phi(r):=\phi(r),
\]
and define \(K\) in actual-\(F\) quantiles by
\begin{equation}
K(F(\theta))
:=
F(\theta)-F(s)
+
r\left[R_\phi(\theta)-A\right].
\label{eq:alpha-K}
\end{equation}
If \(s=t\), then \(R_\phi(s)=A\). If \(s=\lbar\theta>t\), the cutoff
first-order condition gives
\[
A=\phi(t)=\alpha T(\lbar\theta).
\]
Because \(\phi\) is constant on \([t,\lbar\theta)\), the definition of
\(A\) again gives \(R_\phi(s)=A\). Thus
\[
K(F(s))=0.
\]

Let \(\widehat K=\operatorname{cav}K\) be the least concave majorant
of \(K\) on \([F(s),F(r)]\). It contacts \(K\) at both endpoints.
Define the following right-continuous multipliers:
\begin{equation}
\Lambda(\theta)
=
\begin{cases}
1,
& \lbar\theta\leq\theta<s,\\[1mm]
\widehat K'_+(F(\theta)),
& s\leq\theta<r,\\[1mm]
\rho(\theta),
& r\leq\theta\leq\bar\theta,
\end{cases}
\label{eq:alpha-Lambda}
\end{equation}
\begin{equation}
\widetilde\Gamma(\theta)
=
\begin{cases}
-A,
& \lbar\theta\leq\theta<s,\\[1mm]
-\phi(r)
-\dfrac{
F(r)-F(\theta)
-\widehat K(F(r))+\widehat K(F(\theta))
}{r},
& s\leq\theta<r,\\[4mm]
-\phi(\theta),
& r\leq\theta\leq\bar\theta,
\end{cases}
\label{eq:alpha-Gamma}
\end{equation}
and
\begin{equation}
\mu(\theta)
=
\begin{cases}
\Phi(t)-\Phi(\theta)-(t-\theta)A,
& \lbar\theta\leq\theta<s,\\[1mm]
\dfrac{r-\theta}{r}
\left[
K(F(\theta))-\widehat K(F(\theta))
\right],
& s\leq\theta<r,\\[4mm]
0,
& r\leq\theta\leq\bar\theta.
\end{cases}
\label{eq:alpha-mu}
\end{equation}
Finally, define \(\Gamma\) and \(\gamma\) by
\eqref{eq:alpha-transform}.

Endpoint contact of \(\widehat K\) implies that
\(\widetilde\Gamma\) and \(\mu\) are continuous at \(s\) and \(r\).
They are absolutely continuous within each region, so
\(\Gamma\) and \(\mu\) are absolutely continuous on \(\Theta\).
Moreover, almost everywhere in the pooling region,
\[
\widetilde\Gamma'(\theta)
=
\frac{[1-\Lambda(\theta)]f(\theta)}{r},
\]
while \(\widetilde\Gamma'=0\) in the exclusion region and
\(\widetilde\Gamma'=-\phi'\) in the fully revealing region.
Together with \eqref{eq:alpha-rho}, these identities verify
\eqref{eq:alpha-q-foc}. Direct differentiation of
\eqref{eq:alpha-mu}, using \eqref{eq:alpha-K} and
\(\widehat K(F(r))=K(F(r))\), verifies
\eqref{eq:alpha-mu-costate}. Hence all the original costate and
control equations hold.

It remains to verify the sign and monotonicity conditions.
Condition~\eqref{eq:alpha-support} gives
\[
R_\phi(\theta)\leq A
\qquad
\text{for every }\theta\in[s,r].
\]
Therefore,
\[
K(F(\theta))
\leq
F(\theta)-F(s).
\]
The affine function \(\tau\mapsto\tau-F(s)\) is thus a concave
majorant of \(K\). Since it contacts \(\widehat K\) at \(F(s)\),
\[
\widehat K'_+(F(s))\leq1.
\]
Concavity of \(\widehat K\) then implies
\(\Lambda\leq1\) throughout the pooling region. Consequently,
\(\widetilde\Gamma'\geq0\) in that region. It is also nonnegative in
the exclusion region and, because \(\phi\) is decreasing, in the
fully revealing region. Hence \(\widetilde\Gamma\) is increasing,
which verifies the concavity condition through
\eqref{eq:alpha-concavity-transform}.

The same supporting-line inequality and
\(\widehat K\geq K\) imply
\[
\mu(\theta)\leq0
\qquad
\text{for every }\theta\in\Theta.
\]
Moreover, \(w^*\) is constant wherever \(\mu\) may be negative, and
\(\mu=0\) at the cutoff and throughout the fully revealing region.
Thus
\[
\int_{(\lbar\theta,\bar\theta]}\mu(\theta)\d w^*(\theta)=0.
\]

We next verify the upper splice of \(\Lambda\). Define
\[
m(\theta)
:=
\alpha+(1-\alpha)\epsilon_f(\theta).
\]
IED implies that \(m\) is increasing. Since \(\phi\) is decreasing
at and above \(r\), \(m(r)\geq0\). For every \(\theta\in[s,r)\),
\begin{align}
R_\phi(\theta)-\phi(r)
&=
\frac{1}{r-\theta}
\int_\theta^r
\frac{x-\theta}{x}m(x)f(x)\d x \notag\\
&\leq
\frac{m(r)}{r}
\left[F(r)-F(\theta)\right].
\label{eq:alpha-upper-bound}
\end{align}
It follows from \eqref{eq:alpha-K} that
\begin{align}
\frac{
K(F(r))-K(F(\theta))
}{
F(r)-F(\theta)
}
&=
1-
r\frac{
R_\phi(\theta)-\phi(r)
}{
F(r)-F(\theta)
}
\notag\\
&\geq
1-m(r)
=
\rho(r).
\label{eq:alpha-terminal-chords}
\end{align}
Thus the affine function through \(K(F(r))\) with slope \(\rho(r)\)
majorizes \(K\). By minimality of the concave majorant, it also
majorizes \(\widehat K\), with equality at \(F(r)\). Therefore,
\[
\widehat K'_-(F(r))
\geq
\rho(r).
\]
Hence \(\Lambda\) has a weakly downward jump at \(r\). It also has a
weakly downward jump at \(s\), because its left limit is \(1\) and
\(\widehat K'_+(F(s))\leq1\). Within the pooling region it is
decreasing by concavity of \(\widehat K\), and within the revealing
region it is decreasing by \eqref{eq:alpha-rho} and IED. Thus
\(\Lambda\) is decreasing on \(\Theta\). Since \(D^*\equiv0\),
complementary slackness for (MPS) follows immediately.

For completeness, consider the endpoint conditions. The
supporting-line construction gives
\[
A\geq\phi(\lbar\theta-)=\alpha T(\lbar\theta),
\]
with equality whenever \(t<\lbar\theta\). Since
\(\widetilde\Gamma(\lbar\theta)=-A\),
\[
\Gamma(\lbar\theta)
=
-A+\alpha T(\lbar\theta)
\leq0.
\]
If \(t<\lbar\theta\), then
\(\Gamma(\lbar\theta)=\mu(\lbar\theta)=0\). If
\(t\geq\lbar\theta\), then \(U^*(\lbar\theta)=0\), and
\(\mu(\lbar\theta)w^*(\lbar\theta)=0\) follows from
\eqref{eq:alpha-mu}. Hence all lower-endpoint conditions hold.

At the upper endpoint,
\[
\Gamma(\bar\theta)
=
-\phi(\bar\theta)
=
-(1-\alpha)f(\bar\theta),
\qquad
\mu(\bar\theta)=0.
\]
Because \(c'(w^*(\bar\theta))=\bar\theta\), the upper transversality
condition \eqref{eqn:trans:barUw} holds with
\[
\eta(\bar\theta)=(1-\alpha)f(\bar\theta)\geq0.
\]

\medskip
\noindent
\emph{Case 2: \(r\geq\bar\theta\).}
There is no fully revealing region. Define
\begin{equation}
\Lambda(\theta)=1,
\qquad
\widetilde\Gamma(\theta)=-A,
\qquad
\mu(\theta)
=
(\theta-t)A-[\Phi(\theta)-\Phi(t)]
\label{eq:alpha-no-revelation}
\end{equation}
on \(\Theta\), and again set
\[
\Gamma=\widetilde\Gamma+\alpha T,
\qquad
\gamma=\widetilde\Gamma'
-\alpha\frac{f}{\theta}.
\]
Equations \eqref{eq:alpha-q-foc} and
\eqref{eq:alpha-mu-costate} follow immediately.
Condition~\eqref{eq:alpha-support} gives \(\mu\leq0\), and
\(\widetilde\Gamma\) is constant. Thus the monotonicity,
complementary-slackness, and concavity conditions all hold. The
lower-endpoint conditions are verified exactly as in Case~1.

Since \(\Phi\) is constant above \(\bar\theta\), equality in
\eqref{eq:alpha-support} at \(r\) gives
\[
\Phi(\bar\theta)-\Phi(t)=(r-t)A.
\]
Consequently,
\[
\Gamma(\bar\theta)=-A,
\qquad
\mu(\bar\theta)=(\bar\theta-r)A.
\]
Moreover,
\(c'(w^*(\bar\theta))=c'(q_i(t))=r\). Therefore,
\eqref{eqn:trans:barUw} holds with
\(\eta(\bar\theta)=A\geq0\).

In both cases, all the hypotheses of
Theorem~\ref{thm:certificate} are satisfied. Thus the
lower-censorship candidate solves problem~\ref{prob:P}. Every rating
scheme induces a feasible point of that problem, while the candidate
is itself implemented by a deterministic lower-censorship rating.
It is therefore optimal among all, possibly stochastic, rating
schemes.
\end{proof}

\section{Optimal Rating Design with Transfers}
\label{transfer}

\subsection{Transfers Contingent on the Rating Result or Quality}
\label{transfer:contingent}

Suppose first that the principal can charge a transfer $T(s)\in\reals$ that
depends on the rating result $s\in S$, where a positive $T(s)$ denotes a payment from
the agent to the principal. If the principal observes quality, she may instead
charge a quality-contingent certification fee. 
They can both be implemented by a menu of test-fee structures $\{\pi(\cdot|\theta),t(\theta)\}_{\theta\in\Theta}$.

For a rating scheme $\pi$, let
\[
    \hat w_{\pi}(q)
    \equiv
    \E_{s\sim\pi(q)}[\E[\tilde q\mid s]]
    \quad\text{and}\quad
    \hat T_{\pi}(q)
    \equiv
    \E_{s\sim\pi(q)}[T(s)]
\]
denote, respectively, the agent's expected wage and expected transfer after
choosing quality $q$.

\begin{lemma}\label{lem:transfer-substitution}
    Consider two rating-and-transfer pairs $(\pi_1,T_1)$ and $(\pi_2,T_2)$.
    If
    \[
        \hat w_{\pi_1}(q)-\hat T_{\pi_1}(q)
        =
        \hat w_{\pi_2}(q)-\hat T_{\pi_2}(q)
        \quad\text{for every }q\in Q,
    \]
    then they generate the same quality-choice correspondence for every type.
    The same conclusion holds for quality-contingent fees.
\end{lemma}

\begin{proof}
    Under either pair, type $\theta$ chooses $q$ to maximize the same function,
    $\hat w_{\pi_i}(q)-\hat T_{\pi_i}(q)-c(q)/\theta$. Hence the two pairs
    induce the same set of optimal quality choices. The argument for
    quality-contingent fees is identical.
\end{proof}

The lemma implies that it is without loss to use a fully revealing test and
adjust the transfer schedule to provide the desired incentives. If the principal can observe quality and design quality-contingent fees, then a wide range of disclosure-fee schedules
can implement the same quality schedule, as long as the transfer scheme is calibrated
accordingly to provide the same incentives. In other words, the design of the rating
scheme becomes irrelevant.
 This perfect substitutability between ratings and
contingent transfers also appears in \citet{AlbanoLizzeri2001}, in a moral-hazard
setting, and \citet{PollrichStrausz2024}, in a pure adverse-selection setting.

Accordingly, consider a feasible direct mechanism
$(q(\theta),w(\theta),t(\theta))$, where
\[
    w(\theta)
    =\E_{s\sim\pi(q(\theta))}[\E[\tilde q\mid s]]
    \quad\text{and}\quad
    t(\theta)
    =\E_{s\sim\pi(q(\theta))}[T(s)]
\]
are the interim wage and transfer. 
By standard arguments, incentive compatibility is
equivalent to monotonicity of $q$ and the envelope condition
\begin{align}
    U(\theta)
    \equiv w(\theta)-t(\theta)-\frac{c(q(\theta))}{\theta} =U(\underline\theta)
      +\int_{\underline\theta}^{\theta}
        \frac{c(q(x))}{x^2}\,\d x,
    \label{envelopeB}\tag{IC-Env*}
\end{align}
with $U(\underline\theta)\geq0$.

Suppose that the principal maximizes expected quality and transfer revenue, so
her payoff is $q(\theta)+\lambda t(\theta)$, where $\lambda\geq0$ directly
measures the weight placed on transfer revenue relative to quality. If
$\lambda>0$, information rent is costly and individual rationality binds for
the lowest type. If $\lambda=0$, setting $U(\underline\theta)=0$ is without
loss. Thus, in either case, take $U(\underline\theta)=0$. The principal solves
\begin{equation}\label{eqn:transfer-problem}
    \max_{q,w,t}
    \int_{\underline\theta}^{\bar\theta}
        \bigl[q(\theta)+\lambda t(\theta)\bigr]
        \,\d F(\theta)
\end{equation}
subject to incentive compatibility, individual rationality, and the feasibility
condition (MPS).

Define the inverse hazard rate and the associated virtual ability by
\begin{equation}\label{eqn:virtual-ability}
    h(\theta)\equiv\frac{1-F(\theta)}{f(\theta)},
    \qquad
    \vartheta(\theta)
    \equiv
    \left(\frac{1}{\theta}+\frac{h(\theta)}{\theta^2}\right)^{-1}
    =\frac{\theta^2}{\theta+h(\theta)}.
\end{equation}
Under IFR, $h$ is decreasing, and $\vartheta$ is increasing.

Substituting \eqref{envelopeB} into the objective and using Bayesian
plausibility, $\E[w(\theta)]=\E[q(\theta)]$, reduces the problem to
a classical mechanism design problem; see
\citet{BaronMyerson1982} and \citet[Chapter~1]{LaffontTirole1993}.
\begin{equation}\label{eqn:transfer-reduced}
    \max_{q\text{ increasing}}
    \int_{\underline\theta}^{\bar\theta}
    \left\{
        (1+\lambda)q(\theta)
        -\lambda\frac{c(q(\theta))}{\vartheta(\theta)}
    \right\}\,\d F(\theta).
\end{equation}

\begin{proposition}[Expected-quality maximization]\label{prop:T*}
    Suppose that $\lambda>0$, \(c'(q_{\max})
>
\frac{1+\lambda}{\lambda}\bar\theta,\) and $F$ satisfies IFR. Then the optimal quality
    schedule is
    characterized by
    \begin{equation}\label{eqn:q*2}
        c'(q^*(\theta))
        =\frac{1+\lambda}{\lambda}\vartheta(\theta),
        \tag{FOC: $q$}
    \end{equation}
    which can be implemented by a fully revealing test and a result- or quality-contingent certification fee
    \begin{equation}\label{eqn:T*}
        T^*(s)
        =s-\frac{c(q^*(\underline\theta))}{\underline\theta}
        -\int_{q^*(\underline\theta)}^{s}
            \frac{c'(u)}{q^{*-1}(u)}\,\d u
    \end{equation}
    for all $s=[q^*(\lbar\theta),q^*(\bar\theta)]$, and $T(s)= q_{\max}$ for all
    $s\notin[q^*(\lbar\theta),q^*(\bar\theta)]$ to deter off-path deviations.
\end{proposition}
\begin{remark}\label{rmk:screening}
    The same allocation can be implemented by screening via menu of test--fee structures
    $\{\pi^*(\cdot|\theta),t^*(\theta)\}_{\theta\in\Theta}$, where
    \begin{equation}\label{eqn:t*}
        t^*(\theta)
        =q^*(\theta)-\frac{c(q^*(\theta))}{\theta}
        -\int_{\underline\theta}^{\theta}
            \frac{c(q^*(x))}{x^2}\,\d x,
    \end{equation}
    and $\pi^*(q|\theta)= q^*(\theta)$ if
    $q\geq q^*(\theta)$, and $\pi^*(q|\theta)$ returns a ``fail'' if $q<q^*(\theta)$. Conditional on selecting a contract, the agent chooses its threshold quality; the
    incentive constraints across contracts are exactly those of the direct
    mechanism.
\end{remark}
\begin{proof}
    From \eqref{envelopeB} and integration by parts
    \[
        \int_{\underline\theta}^{\bar\theta}U(\theta)\,\d F(\theta)
        =\int_{\underline\theta}^{\bar\theta}
            \frac{1-F(\theta)}{\theta^2}c(q(\theta))\,\d\theta
        =\int_{\underline\theta}^{\bar\theta}
            \frac{h(\theta)}{\theta^2}c(q(\theta))\,\d F(\theta).
    \]
    Substituting
    $t=w-c(q)/\theta-U$ into \eqref{eqn:transfer-problem} and applying (BP)
    therefore yields \eqref{eqn:transfer-reduced}. The stated pointwise
    maximizer is unique by strict convexity of $c$. Under IFR, $h$ is
    nonincreasing, so $\vartheta$ is increasing. The pointwise objective then
    has increasing differences in $(q,\theta)$, which implies that $q^*$ is
    increasing and hence feasible.  

    Under full revelation, $w(\theta)=q^*(\theta)$, so \eqref{eqn:t*} follows
    directly from \eqref{envelopeB} with $U(\underline\theta)=0$. If $q^*$ is
    strictly increasing, differentiating \eqref{eqn:t*} gives
    \[
        \frac{\d t^*(\theta)}{\d\theta}
        =q^{*\prime}(\theta)
         \left[1-\frac{c'(q^*(\theta))}{\theta}\right].
    \]
    Integrating this expression over quality and using
    $t^*(\underline\theta)=q^*(\underline\theta)
    -c(q^*(\underline\theta))/\underline\theta$ gives \eqref{eqn:T*}.
\end{proof}

When $\lambda=0$, the principal maximizes expected quality without assigning
any cost to transfers. The reduced objective is then $\E[q(\theta)]$, so the
optimal schedule is $q^*(\theta)=q_{\max}$ for every type; it can be implemented
using the interim transfer in \eqref{eqn:t*}.

For comparison, when $\lambda>0$, the full-information allocation replaces
$\vartheta(\theta)$ with $\theta$ in \eqref{eqn:q*2}. Since
$\vartheta(\theta)\leq\theta$, private information leads to underinvestment of quality. Moreover, $(1+\lambda)/\lambda$ decreases from infinity to $1$
as $\lambda$ increases from zero, so placing more weight on transfer revenue
creates an additional downward distortion. In the monopoly-certifier limit
$\lambda\to\infty$, the interior condition becomes
\begin{equation}\label{eqn:monopoly-quality}
    c'(q^*(\theta))=\vartheta(\theta).
\end{equation}

The optimal fee need not be increasing in quality. On the equilibrium range,
\[
    \begin{aligned}
        T^{*\prime}(q^*(\theta))
         =1-\frac{c'(q^*(\theta))}{\theta}=\frac{h(\theta)-\theta/\lambda}{\theta+h(\theta)},
    \end{aligned}
\]
so its slope depends on the inverse hazard rate and the principal's weight on
transfer revenue.

\subsection{Constant Testing Fees}
\label{transfer:general}

The substitutability result relies on allowing the transfer to vary with the
rating result or with quality. Once the principal is restricted to a constant
upfront testing fee, the transfer can no longer reproduce an arbitrary
net-return schedule, and the design of the rating becomes consequential again.
This restriction is also natural in applications in which certification fees
are posted before the test is administered.
Moreover, if the principal can tamper with the rating, then the restriction to a constant
testing fee is required for incentive compatibility of principals.

\citet{AlbanoLizzeri2001} show that a monopoly certifier optimally uses a
stochastic rating: it reveals quality with some probability and otherwise sends
a common, nearly uninformative signal to test participants. See also
\citet{SaeediShourideh2020}.
For objectives that place sufficiently large weight on fee revenue relative to
expected quality, \citet[Chapter~3]{Xiao2025} shows that a noisy test remains
optimal and induces quality below the full-information benchmark. In the
monopoly-certifier benchmark, a pass/fail test is revenue-maximizing within the
class of deterministic ratings.

\end{appendices}

\singlespacing

\bibliographystyle{econ-te}
\bibliography{certification.bib}

\end{document}